\documentclass[final,3p,times]{elsarticle}

\usepackage[normalem]{ulem}

\usepackage[utf8]{inputenc}
\usepackage{graphicx}
\usepackage{amsmath}
\usepackage{amssymb}
\usepackage{physics}
\usepackage{bbm}
\usepackage{bbold}
\usepackage{xspace}
\usepackage{color}
\usepackage{footnote}
\usepackage{comment}
\usepackage{placeins}
\usepackage{hyperref}
\usepackage{subfigure}

\usepackage{todonotes}
\setuptodonotes{inline,color=orange!50}
\usepackage{mathtools}
\usepackage{dsfont}

\definecolor{dgreen}{rgb}{0.0,0.5,0.0}
\definecolor{orange}{RGB}{252,77,6}
\definecolor{brown}{RGB}{200,127,50}
\definecolor{blue}{RGB}{00,000,100}
\definecolor{blue2}{RGB}{00,000,250}
\definecolor{green1}{RGB}{00,100,00}
\definecolor{green2}{RGB}{00,150,00}
\definecolor{green3}{RGB}{00,200,00}
\definecolor{green4}{RGB}{00,250,00}

\newcommand{\impLabel}{\mathrm{d}}
\newcommand{\chainLabel}{\mathrm{c}}
\newcommand{\impChainLabel}{\mathrm{cd}}
\newcommand{\impLabelMFD}{\mathrm{imp}}
\newcommand{\chainLabelMFD}{\mathrm{chain}}
\newcommand{\hermConj}{\mathrm{h.c.}}

\journal{Annals of Physics}
\date{23.12.2024}

\begin{document}

\title{Mean Field Decoupling of Single Impurity Anderson Model through Auxiliary Majorana Fermions}

\author[Hamburg]{Irakli Titvinidze}
\affiliation[Hamburg]{I. Institute of Theoretical Physics, University of Hamburg, 20355 Hamburg, Germany}
\author[Hamburg]{Julian Stobbe}
\author[Moscow1,Moscow2]{Alexey N. Rubtsov}
\affiliation[Moscow1]{Russian Quantum Center, Skolkovo Innovation City, 121205 Moscow, Russia}
\affiliation[Moscow2]{Department of Physics, Lomonosov Moscow State University, Leninskie gory 1, 119991 Moscow, Russia}
\author[London,Hamburg]{Georg Rohringer}
\affiliation[London]{Theory and Simulation of Condensed Matter, Department of Physics, King's College London, The Strand, London WC2R 2LS, United Kingdom}


\begin{abstract}
We present a method to study the time evolution of the single impurity Anderson model which exploits a mean field decoupling of the interacting impurity and the non-interacting bath (in form of a chain).
This is achieved by the introduction of a pair of auxiliary Majorana fermions between the impurity and the chain. 
After decoupling, we obtain a self-consistent set of equations for the impurity and chain. 
First, we study the behavior of the system in equilibrium at zero temperature. 
We observe a phase transition as a function of the interaction at the impurity and the coupling between the impurity and the chain between the Kondo regime, where the mean field parameters are zero and, hence, we have a well-defined spin at the impurity, to a phase where mean field parameters acquire finite values leading to a screening of the impurity spin by conduction bath electrons.
In the latter case, we observe charge and spin fluctuations at the impurity site. 
Let us note that, while the sharp equilibrium phase transition is a feature of the mean field treatment of the problem it is likely to show itself as a crossover in an exact treatment of the problem.
Starting from this equilibrium ground state at zero temperature we quench in the interaction strength at the impurity and/or the hybridization strength between the impurity and the chain and study the time evolution of the system.
We find that for quenches to weak to intermediate coupling the system converges to the equilibrium state defined by the final set of parameters after the quench. We analyze the oscillation frequency and as well as the thermalization rate during this quench.
A quench to a strong interaction value results in persistent oscillations and a trapping of the system in a non-thermal state.
We speculate that these two regimes of different long-time behavior are separated by a dynamical phase transition.
We, however, argue that, while our description of the weak to moderate correlated regimes is correct at all time scales, for large final interaction, our approximation is fully valid only at moderate time scale
whereas persistent oscillations and related sharp phase transitions are likely artifacts of the mean field treatment of the problem (as in the equilibrium case).

\end{abstract}

\maketitle

\section{Introduction}\label{Introduction}

Due to recent impressive experimental progress in various fields of condensed matter physics, such as ultracold atoms\cite{sc.ha.12, tr.ch.08, ka.ta.16, ch.ba.12, ja.zo.05},
coupled cavity arrays\cite{ha.br.08, ka.sh.19},
ultrafast laser spectroscopy\cite{pe.lo.06, he.ro.12},
spintronics\cite{ma.gm.23, li.ma.23},
quantum wires or quantum dots\cite{go.go.98, sm.ch.22, bo.sh.20, kl.to.21},
molecular junctions\cite{pa.pa.02, li.sh.02, ve.kl.06, tao.06, cu.hu.19, he.gr.06, br.ge.19},
correlated heterostructures\cite{oh.mu.02, zh.re.22}
or carbon nanotubes\cite{ch.ga.12, go.sc.18},
correlated electron systems in low (one or two) spatial dimensions out of equilibrium have attracted growing attention.
More specifically, the behavior of such systems after a perturbation (quench) can lead to fascinating insights into the dynamics of quantum matter. Of particular interest are, among many others, the thermalization behavior after a quench \cite{ri.du.08, ca.ca.07, caza.06, ly.mi.23, ba.do.23, qi.ji.21} and non-equilibrium phase transitions\cite{mi.ta.06, ch.ch.23, ka.ke.22, li.ya.16,ma.di.16} triggered by tuning selected system parameters.

On the theoretical side, the investigation of strongly correlated systems out of equilibrium is a considerable challenge as standard equilibrium techniques such as the Matsubara formalism are not applicable.
Hence, one often resorts to manageable models to study interacting quantum systems in a dynamic setting.
One of the simplest models for studying strongly correlated systems in and out of equilibrium is the single-impurity Anderson model (SIAM)\cite{ande.61}. 
It was originally developed to study magnetic impurities in metallic hosts\cite{ma.pe.60, cl.ma.62} and to investigate the Kondo effect\cite{kond.64, hews.93}. 
For the out-of-equilibrium system, it is a good model to study the dynamics of quantum dots\cite{go.go.98, sm.ch.22, bo.sh.20, kl.to.21} and molecular junctions\cite{pa.pa.02, li.sh.02, ve.kl.06, tao.06, cu.hu.19, he.gr.06, br.ge.19}.
Let us also mention, that the SIAM is one of the key building blocks of dynamical mean field theory (DMFT)\cite{ge.ko.96, me.vo.89, ao.ts.14}, a powerful method for studying strongly correlated systems in equilibrium and out-of-equilibrium in high spatial dimensions.

For this reason, various methods to tackle this system both in and out of equilibrium have been developed in the last decades. 
Among them are time dependent numerical renormalization group (NRG)\cite{an.sc.05,  ha.ro.09, ng.co.14, ng.co.17, bu.co.08, mu.we.12, le.we.16, ng.da.20},
self-energy method within time dependent NRG\cite{ng.co.21},
hybrid time dependent NRG\cite{ei.sc.12}, 
time dependent density renormalization group (tDMRG)\cite{ca.ma.02, wh.fe.04, schm.04, he.ma.09, ko.sa.22},
diagrammatic quantum Monte Carlo (QMC)\cite{schi.10},
continuous time QMC \cite{gu.mi.11},
configuration interaction method\cite{li.de.15}, 
functional renormalization group (FRG)\cite{ke.ja.12},
iterative summation of real-time path integrals\cite{we.ec.08}, 
numerically exact path integral\cite{se.mi.10}, 
noncrossing approximation (NCA)\cite{wi.me.94, no.pu.99},
time-dependent Gutzwiller approximation\cite{sc.fa.10, la.st.12}, 
bosonization\cite{lo.ke.05, he.ke.10},
slave spin approach\cite{guer.19},
flow equation techniques\cite{kehr.05},
dual fermions\cite{ju.li.12},
and many others.
In addition, methods have been developed for the investigation of steady-state behavior
such as auxiliary master equation approach\cite{do.nu.14, ti.do.15},
scattering state NRG\cite{ande.08},
QMC\cite{ha.he.99},
scattering-state Bethe ansatz\cite{me.an.06},
and
imaginary-time formulation of steady-state non-equilibrium\cite{ha.he.07}. 

While these techniques provide satisfactory results most of them are numerically highly demanding. This restricts the tractable system size (i.e., number of bath sites in the SIAM) and/or the times which can be reached within the numerical simulations.
Hence, there is a considerable need for simpler approaches which at the same time treat correlation effects at the impurity adequately.

In this work, we propose a method to study the behavior of the SIAM in equilibrium and out-of-equilibrium after a quench of the interaction parameter $U$ between electrons at the impurity and/or the hybridization $V$ between the impurity and the bath (which is represented in the form of a non-interacting chain). 
The idea of our approach is a mean field decoupling of the impurity and the chain. 
Since the coupling term in the Hamiltonian between these to parts of the system is quadratic, i.e., it consists of one fermionic creation and one annihilation operator, a straight-forward mean field treatment would lead to averages over single fermionic operators which are poorly defined.
To overcome this problem we introduce a pair of auxiliary Majorana fermions between the impurity and the chain, which take part in the decoupling procedure and, in fact, make it feasible.
Thus, after decoupling, we obtain separate equations for the impurity and the chain both coupled to the Majorana fermions via mean field parameters, which are determined from the respective opposite sub-system (i.e., the mean field parameter of the chain is obtained from the impurity and vice versa).
In equilibrium this leads to a set of coupled algebraic equations which have to be solved self-consistently.
For systems out of equilibrium, on the other hand, we obtain a set of coupled ordinary differential equations, which can be solved by standard methods.

The advantage of our technique is that it is much simpler than more advanced approaches such as DMRG, NRG, or QMC. 
This substantially reduces the numerical effort, which allows us to investigate large systems (long chains) up to very long times. 
On the downside, this method can obviously not fully capture the quantum entanglement between the impurity and the chain which is typical for mean field theories of this type.

Let us note that the coupling of impurities to Majorana fermions has already been investigated in other contexts\cite{em.ki.92, komn.09, li.ba.11, le.fl.11, le.li.13, ga.mi.14, we.wo.17, sh.ma.19, wr.we.21}. 
In particular, in Refs.~\cite{em.ki.92, komn.09} they emerge due to the mapping of the two-channel Kondo problem to a Majorana resonant-level model,
while in Refs.~\cite{li.ba.11, le.fl.11, le.li.13, ga.mi.14, we.wo.17, sh.ma.19, wr.we.21} their appearance is due to the Majorana zero modes and Majorana bound states.
In our case, on the contrary, they are introduced as mere auxiliary quantities which allow us to conduct a mean field decoupling of impurity and chain.

The paper is organized as follows: In Secs.~\ref{Model} and~\ref{Method}, we introduce the Hamiltonian of the system and our new method, respectively.
In Sec.~\ref{Resonance_Level_Model} we benchmark our model with the exactly solvable resonance level model.  
Subsequently, in Sec.~\ref{Results}, we present our results. 
In particular, in Sec.~\ref{Equilibrium} we discuss our equilibrium results, while in Sec.~\ref{Time_Evolution} we describe the behavior of the system out of equilibrium after a quench of parameters. 
Our findings are discussed in more detail in Sec.~\ref{Discussion}.
Finally, Sec.~\ref{Conclusions} is devoted to conclusions and an outlook.
The paper contains a number of appendices as well as supplemental material (SM) where we provide more detailed derivations of our methods and additional results. 
The codes for the equilibrium and time evolution calculations can be found in Ref.\cite{Link_to_Codes}.

\begin{figure}[t!]
\begin{center}
\subfigure[Impurity coupled to $r=1$ site with OPB]{
\label{Fig:SIAM_Line}
\includegraphics[width=0.475\textwidth]{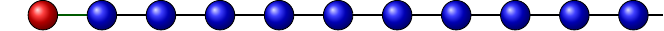}
}
\subfigure[Schematic structure for the decoupling for the {(a)}
]{
\label{Fig:SIAM_Line_Majorana}
\includegraphics[width=0.475\textwidth]{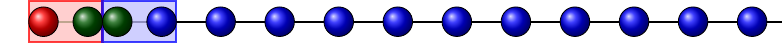}
}
\end{center}
\caption{%
(a) Schematic structure of our model where the single impurity (red) is coupled to the first site of the chain (blue), with open boundary conditions (OBC). In (b), we provide a schematic illustration of the mean field decoupling of the impurity and chain, by means of Majorana fermions (green).
}
\label{Fig:SIAM}
\end{figure}

\section{Model}\label{Model}

We consider the one-dimensional single impurity Anderson model, where an impurity, with local Hubbard interaction $U$ and spin-dependent onsite energy $\varepsilon_{\impLabel,\sigma}$, is coupled to the $r$-th site of a chain with $L$ non-interacting sites (see Fig.~\ref{Fig:SIAM} for an illustration of the case where the impurity is coupled to the first site of the chain.). 
The Hamiltonian reads as follows:
\begin{subequations}
    \label{Hamiltonian}
    \begin{align}
    \label{Hamiltonian_tot}
    &  \hat{\cal H}(t)=\hat{\cal H}_\impLabel(t) + \sum_{\sigma=\uparrow,\downarrow}\left[\hat{\cal H}_{\chainLabel,\sigma} + \hat{\cal H}_{\impChainLabel,\sigma}(t)\right]
\\
    \label{Hamiltonian_d}
    &  \hat{\cal H}_\impLabel(t)=U(t)\hat n_{\impLabel,\uparrow}\hat n_{\impLabel,\downarrow} - \sum_{\sigma=\uparrow,\downarrow} \left(\mu_\sigma-\varepsilon_{\impLabel,\sigma}(t) \right) \hat n_{\impLabel,\sigma}
\\
    \label{Hamiltonian_c_sigma}
    &  \hat {\cal H}_{\chainLabel,\sigma} =\sum_{i=1}^L \left[ J_{i,\sigma} \left(\hat c_{i,\sigma}^{\dagger}\hat c_{i+1,\sigma}^{\phantom\dagger} + \hermConj \right)
        -\left(\mu_{\sigma}-\varepsilon_{i,\sigma}\right)\hat n_{i,\sigma}
    \right]
\\
    \label{Hamiltonian_cd_sigma}
    &  \hat {\cal H}_{\impChainLabel,\sigma}(t) = V_{\sigma}(t)\left(\hat d_{\sigma}^{\dagger} \hat c_{r,\sigma}^{\phantom\dagger} + \hat c_{r,\sigma}^{\dagger} \hat d_{\sigma}^{\phantom\dagger} \right)
    \,.
\end{align}
\end{subequations}
$\hat{\cal H}_{\impLabel}(t)$, $\hat{\cal H}_{\chainLabel,\sigma}$ and $\hat{\cal H}_{\impChainLabel,\sigma}(t)$ describe the (interacting) impurity, the (non-interacting) chain, and the coupling between the impurity and the chain (for fermions with spin $\sigma$), respectively.

$\hat{d}^{(\dagger)}_\sigma$ annihilates (creates) a particle with spin $\sigma \in \{\uparrow,\downarrow \}$ at the impurity site and $\hat{n}_{d,\sigma}\!=\!\hat{d}^{\dagger}_\sigma\hat{d}^{\phantom{\dagger}}_\sigma$ is the corresponding particle density. 
$U(t)$ represents the local interaction strength between a particle with spin $\uparrow$ and a particle with spin $\downarrow$ at the impurity site, while $\varepsilon_{\rm d,\sigma}(t)$ corresponds to the onsite energy of a particle at the impurity with spin $\sigma$. 
$\mu_{\sigma}$ is the chemical potential of the impurity and the chain. 

$\hat{c}_{i\sigma}^{(\dagger)}$ annihilates (creates) a particle with spin $\sigma$ at the $i$-th site of the non-interacting chain and $\hat{n}_{i\sigma}\!=\!\hat{c}_{i\sigma}^{\dagger}\hat{c}^{\phantom{\dagger}}_{i\sigma}$ is the corresponding particle density. 
$\varepsilon_{i,\sigma}$ corresponds to onsite energy and $J_{i,\sigma}$ represents the hopping amplitude between the neighboring sites $i$ and $i+1$  of the chain. 
In the case of open boundary conditions (OBC), we have $J_{L,\sigma}=0$, while in the case of a periodic boundary condition (PBC) $J_{L,\sigma} \neq 0$ describes the hopping between the sites $1$ and $L$. 

Finally, $V_{\sigma}(t)$ denotes the spin-dependent coupling between the impurity and the $r$-th site of the non-interacting chain.

In this work, we first investigate the behavior of the system in equilibrium, in its ground state. Then, we perform a sudden change of some model parameters\cite{Quenching}:
$U(t)=U_{\rm in} \theta(-t) + U_{\rm fi}\theta(t)$, $V_\sigma(t)=V_{\sigma,{\rm in}} \theta(-t) + V_{\sigma,{\rm fi}}\theta(t)$, and $\varepsilon_{d,\sigma}(t)=\varepsilon_{\sigma,{\rm in}} \theta(-t) + \varepsilon_{\sigma,{\rm fi}}\theta(t)$.
Here, $\theta(t)$ is the Heaviside step function. 



Let us mention that, while the derivation of our mean field approach in the following Sec.~\ref{Method} is performed for a general form of the Hamiltonian in Eqs.~\eqref{Hamiltonian}, where the impurity is coupled to chain site $r$, the concrete numerical calculations are carried out for a selected set of parameters [specified at the beginning of Sec.~\ref{Results}] where the impurity is coupled to the first chain site.
{
We have restricted our calculations for the single impurity Anderson model to the half-filled system where the density can be most easily fixed to $n\!=\!1$ by setting $\varepsilon_{d,\sigma}(t)=-U(t)/2$
and $\mu\!=\!0$. In this way we avoid the technical difficulties which emerge for calculations out of half filling by the necessity for an update of the chemical potential at each time step to fix the fix the required density (see Sec.~\ref{App:mean_field_dynamics_beyond_the_half_filling}). Let us, however, stress that this is {\em not} a principal limitation of our method and will be addressed in future research work.}

\section{Method}\label{Method}

The main idea of our approach is a mean field decoupling of the hybridization term in Eq.~\eqref{Hamiltonian_cd_sigma}. 
In contrast to standard mean field theories, where such a treatment is applied to the interaction term [Eq.~\eqref{Hamiltonian_d}], our approach allows us to take into account the local interaction effects at the impurity {\em exactly}.
The method shares some similarities with the bosonic Gutzwiller mean field theory\cite{ro.ko.91} or superperturbation theory\cite{hafe.09}, where bilinear terms have been decoupled. 
Let us also point out that the dynamical mean field theory (DMFT) can be interpreted as a mean field decoupling of the non-interacting part of the Hubbard model, i.e., the hopping term\cite{ge.ko.96, me.vo.89, ao.ts.14,rubt.08}.

The standard way to perform a mean field decoupling of $\hat {\cal H}_{\impChainLabel,\sigma}$ in Eq.~\eqref{Hamiltonian_cd_sigma} would correspond to the replacement $\hat d_{\sigma}^{\dagger} \hat c_{r,\sigma}^{\phantom\dagger}\rightarrow\hat d_{\sigma}^{\dagger} \langle\hat c_{r,\sigma}^{\phantom\dagger}\rangle+\langle\hat d_{\sigma}^{\dagger}\rangle \hat c_{r,\sigma}^{\phantom\dagger}-\langle\hat d_{\sigma}^{\dagger}\rangle \langle\hat c_{r,\sigma}^{\phantom\dagger}\rangle$. However, the averages of the single fermionic operators $\langle\hat d_{\sigma}^{\dagger}\rangle$ and $\langle\hat c_{r,\sigma}^{\phantom\dagger}\rangle$ are poorly defined. To overcome this problem, we introduce a pair of auxiliary Majorana fermions $\hat\gamma_{\sigma}$ between the impurity and the chain [see Fig.~\ref{Fig:SIAM_Line_Majorana}], exploiting the property $\gamma_\sigma ^2=1$. 
{We emphasize that here the introduced Majorana fermions serve solely as auxiliary tool and, hence, do not allow specific physical interpretation. Let us stress that their introduction extends the original Hilbert space of both the impurity and the chain by a factor of $4$.}

A Majorana fermion can be represented as the sum of a Dirac fermion and its Hermitian conjugate
\begin{equation}
    \label{Definition_Majorana}
    \hat\gamma_{\sigma}=\hat a_{\sigma}^{\dagger}+\hat a_{\sigma}^{\phantom\dagger}
\end{equation}
and the following relations hold
\begin{equation}
    \label{Majorana}
    \hat\gamma_{\sigma}^{\dagger}=\hat\gamma_{\sigma}^{\phantom\dagger}\,,
    \quad
    \{\hat\gamma_{\sigma},\hat\gamma_{\sigma'}\}=2\delta_{\sigma\sigma'}\,,
    \quad
    \hat\gamma_{\sigma}^2=1\,.
\end{equation}
After taking into account these properties of Majorana fermions, we rewrite Eq.~\eqref{Hamiltonian_cd_sigma} as follows
\begin{equation}
    \label{Hamiltonian_cd_with_Mayorana}
    \hat {\cal H}_{\impChainLabel,\sigma}(t)
    = V_{\sigma}^{\phantom\dagger}(t) \hat{d}_{\sigma}^{\dagger} \hat\gamma_{\sigma}^{\phantom\dagger} \hat\gamma_{\sigma}^{\phantom\dagger} \hat c_{r,\sigma}^{\phantom\dagger}
    + V_{\sigma}^{\phantom\dagger} (t) \hat c_{r,\sigma}^{\dagger} \hat\gamma_{\sigma}^{\phantom\dagger} \hat\gamma_{\sigma}^{\phantom\dagger} \hat d_{\sigma}^{\phantom\dagger}  \,.
\end{equation}
Now we have four fermionic operators and, therefore, can perform the standard mean field decoupling
\begin{subequations}
    \label{Decoupling}
    \begin{align}
    \hat d_{\sigma}^{\dagger}\hat\gamma_{\sigma}^{\phantom\dagger} \hat\gamma_{\sigma}^{\phantom\dagger} \hat c_{r,\sigma}^{\phantom\dagger} 
    &\simeq \langle \hat\gamma_{\sigma}^{\phantom\dagger} \hat c_{r,\sigma}^{\phantom\dagger} \rangle \hat d_{\sigma}^{\dagger}\hat\gamma_{\sigma}^{\phantom\dagger}
    + \langle \hat d_{\sigma}^{\dagger}\hat\gamma_{\sigma}^{\phantom\dagger}\rangle \hat\gamma_{\sigma}^{\phantom\dagger} \hat c_{r,\sigma}^{\phantom\dagger}
    - \langle \hat\gamma_{\sigma}^{\phantom\dagger} \hat c_{r,\sigma}^{\phantom\dagger} \rangle \langle \hat d_{\sigma}^{\dagger}\hat\gamma_{\sigma}\rangle
\\
    \hat c_{r,\sigma}^{\dagger} \hat\gamma_{\sigma}^{\phantom\dagger} \hat\gamma_{\sigma}^{\phantom\dagger} \hat d_{\sigma}^{\phantom\dagger}
    &\simeq \langle \hat\gamma_{\sigma}^{\phantom\dagger} \hat d_{\sigma}^{\phantom\dagger}\rangle \hat c_{r,\sigma}^{\dagger} \hat\gamma_{\sigma}^{\phantom\dagger}
    + \langle \hat c_{r,\sigma}^{\dagger} \hat\gamma_{\sigma}^{\phantom\dagger} \rangle \hat\gamma_{\sigma}^{\phantom\dagger} \hat d_{\sigma}^{\phantom\dagger}
    -\langle \hat\gamma_{\sigma}^{\phantom\dagger} \hat d_{\sigma}^{\phantom\dagger}\rangle \langle \hat c_{r,\sigma}^{\dagger} \hat\gamma_{\sigma}^{\phantom\dagger} \rangle
    \end{align}
\end{subequations}
We obtain the following self-consistent set of equations for the mean field decoupled Hamiltonian
\begin{subequations}
    \label{Hamiltonian_Decoupled}
    \begin{align}
    \label{Hamiltonian_tot_0,Decoupled}
    &\hat{\cal H}(t) =
        \hat{\cal H}_\impLabelMFD(t) 
        + \sum_{\sigma}\left[
            \hat {\cal H}_{\chainLabelMFD,\sigma}(t)
            + {\cal C}_\sigma(t)
        \right]
\\
    \label{Hamiltonian_impurity_Decoupled}
    &\hat{\cal H}_{\impLabelMFD}(t) = 
        \hat{\cal H}_{\impLabel}(t)
        + \sum_\sigma  V_\sigma(t)\left(g_\sigma^{\phantom*}(t) \hat d_{\sigma}^{\dagger}\hat\gamma_\sigma^{\phantom\dagger}
        + g_\sigma^{*}(t) \hat\gamma_\sigma^{\phantom*} \hat d_{\sigma}^{\phantom\dagger}\right)
\\
    \label{Hamiltonian_chain_Decoupled_sigma}
    &\hat {\cal H}_{\chainLabelMFD,\sigma}(t) = 
        \hat {\cal H}_{c,\sigma} 
        + V_\sigma(t)\left(f_{\sigma}^{\phantom*}(t) \hat c_{r,\sigma}^{\dagger}\hat\gamma_\sigma^{\phantom\dagger}
        + f_{\sigma}^{*}(t)\hat\gamma_\sigma^{\phantom*} \hat c_{r,\sigma}^{\phantom\dagger}\right)
\\
    \label{Ct}
    &{\cal C}_\sigma(t) = 
        -2 V_{\sigma}(t) \Re \left(f_\sigma^{\phantom*}(t) g_\sigma^{*}(t)\right),
\end{align}
\end{subequations}
where we have defined
\begin{subequations}
    \label{Meanfield_parameters}
    \begin{align}
        \label{Meanfield_f}
        &f_{\sigma}^{\phantom*}(t) = 
            \langle \hat\gamma_{\sigma}^{\phantom\dagger} \hat d_{\sigma}^{\phantom\dagger}\rangle (t)
        &f_{\sigma}^{*}(t) = 
            \langle \hat d_{\sigma}^{\dagger}\hat\gamma_{\sigma}^{\phantom\dagger}\rangle (t) \\
        \label{Meanfield_g}
        &g_\sigma^{\phantom*}(t) = 
            \langle \hat\gamma_{\sigma}^{\phantom\dagger} \hat c_{r,\sigma}^{\phantom\dagger} \rangle (t)
        &g_\sigma^{*}(t) = 
            \langle \hat c_{r,\sigma}^{\dagger} \hat\gamma_{\sigma}^{\phantom\dagger}\rangle(t)
\end{align}
\end{subequations}
Here we have made an additional approximation by considering Majorana fermions, which are part of the impurity and chain Hamiltonians, as independent entities  (for a justification see appendix~\ref{App:Jordan-Wigner}). 
Moreover, we have assigned a time argument to indicate a possible time dependence of the mean field parameters $f_\sigma(t)$ and $g_\sigma(t)$ induced by a corresponding time dependence of the parameters $V_\sigma(t)$ and $U(t)$. Since we are interested in the behavior of the system after a quench of $V_\sigma$ and/or $U$ their time dependence will be given by a step function at $t\!=\!0$, as it was mentioned above.

It is not immediately obvious that the mean field decoupling in Eq.~\eqref{Decoupling} is justified. 
Typically, such a decoupling is performed in the dominating scattering channel corresponding to the prevailing fluctuations in the system (spin, charge, pairing, etc.). 
However, such a major physical channel cannot be straightforwardly identified in Eqs.~\eqref{Hamiltonian_cd_with_Mayorana} since the (to some extent artificially) introduced Majorana fermions have no direct physical meaning. 
Instead, the validity of the mean field treatment in Eqs.~\eqref{Decoupling} can be demonstrated by a mapping of the fermionic system via a Jordan-Wigner transform to two (interacting) spin chains, which are coupled to each other at the first (impurity) site.
In this spin Hamiltonian, the term 
    $\hat{s}_{0,\sigma}^{+} \hat{s}_{1,\sigma}^{\phantom+}$, 
which couples the spin of the impurity (site $0$) with the first spin of the chain, can be decoupled as
${\hat{s}_{0,\sigma}^{+} \hat{s}_{1,\sigma}^{\phantom+} 
    \sim 
        \langle
            \hat{s}_{0,\sigma}^{+} \rangle\hat{s}_{1,\sigma}^{\phantom+} 
            + \hat{s}_{0,\sigma}^{+} \langle\hat{s}_{1,\sigma}^{\phantom+} 
        \rangle 
        -
        \langle \hat{s}_{0,\sigma}^{+}\rangle \langle \hat{s}_{1,\sigma}\rangle}$.
It can then be demonstrated that the resulting decoupled spin Hamiltonian can be mapped back to the fermionic system, which is decoupled via Majorana fermions in Eqs.~\eqref{Hamiltonian_impurity_Decoupled}. 
More details about these mappings are given in Appendix~\ref{App:Mapping_to_Spin}.

To study the behavior of the system defined in Eqs.~\eqref{Hamiltonian_Decoupled} in the ground state as well as its time evolution, we introduce the density matrix $\rho_{ij}(t)$ for the impurity site and the nearest neighbor chain correlators
\begin{subequations}
\label{Cij_and_Kij}
\begin{align}
    \label{Cij_correlator}
    &\hat C_{ij,\sigma} =\left\{
    \begin{array}{ccc}
            \hat c_{i,\sigma}^{\dagger} \hat{c}_{j,\sigma}^{\phantom\dagger} &
            \quad & 
            1 \leq i,j \leq L
        \\
            \hat\gamma_{\sigma}^{\phantom\dagger} \hat{c}_{j,\sigma}^{\phantom\dagger} &
            \quad & 
            i=0,\,1 \leq j \leq L
        \\
            \hat c_{i,\sigma}^{\dagger} \hat\gamma_{\sigma}^{\phantom\dagger} &
            \quad &
            1 \leq i \leq L,\,  j=0
        \\
        \hat{\mathbb{1}} &
        \quad & 
        i=j=0
    \end{array}
    \right.
\\
    \label{Cij_matrix}
    & C_{ij,\sigma} (t)
        = \langle \hat C_{ij,\sigma} \rangle (t)
\\
    \label{Kij_correlator}
    & \hat K_{ij,\sigma} =
\left. 
\begin{array}{ccc}
    \hat c_{i,\sigma}^{\phantom\dagger} \hat c_{j,\sigma}^{\phantom\dagger} & \quad & 1 \leq i , j \leq L
\end{array}
\right.
\\
\label{Kij_matrix}
&K_{ij,\sigma} (t) = \langle \hat K_{ij,\sigma} \rangle (t).
\end{align}
\end{subequations}

Based on the density matrix $\rho_{ij}(t)$ and the correlation function $C_{ij,\sigma}(t)$, the mean field parameters can be determined as follows (for details see Appendix~\ref{App:Solving_the_equilibrium_problem})
\begin{subequations}
\label{Define_f_and_g}
\begin{align}
\label{Define_f_via_rho_up}
&f_{\uparrow}(t)=\rho_{13}(t)+\rho_{24}(t)
\\
\label{Define_f_via_rho_down}
&f_{\downarrow}(t)=\rho_{12}(t)+\rho_{34}(t)
\\
\label{Define_g_via_C}
&g_\sigma^{\phantom*}(t)= C_{0r,\sigma}(t) \,.
\end{align}
\end{subequations}

We will first examine the behavior of the system in equilibrium at zero temperature. 
To this end, we solve Eqs.~\eqref{Hamiltonian_Decoupled} and~\eqref{Meanfield_parameters} for $f_\sigma(t)$ and $g_\sigma(t)$ self-consistently at $t=0$.
In the next step, we use the obtained equilibrium results $f_\sigma\!\equiv\!f_\sigma(0)$ and $g_\sigma\!\equiv\!g_\sigma(0)$ as initial values for the time evolution $f_\sigma^{\phantom*}(t)$ and $g_\sigma^{\phantom*}(t)$ after a quench of the interaction parameter $U$ and/or the coupling between the chain and the impurity $V_\sigma$.
To this end, we exploit the standard equations of motion for the impurity site
\begin{equation}
    \label{equation_of_motion_Density_Operator}
    \frac{\dd}{\dd t}{\hat \rho}(t)= -i \left[\hat{\cal H}_{\impLabelMFD}, \hat\rho(t)\right]
\end{equation}
and for the chain
\begin{subequations}
    \label{equation_of_motion}
    \begin{align}
    &\frac{\dd}{\dd t}C_{ij,\sigma}(t) 
        = i \langle [\hat{\cal H}_{\chainLabelMFD,\sigma}, \hat C_{ij,\sigma}]\rangle(t)
\\
    &\frac{\dd}{\dd t}K_{ij,\sigma}(t) 
        = i \langle [\hat{\cal H}_{\chainLabelMFD,\sigma}, \hat K_{ij,\sigma}]\rangle(t).
\end{align}
\end{subequations}
This yields ${L(L+1)+10}$ coupled ordinary differential equations for $\hat{\rho}(t), C_{ij,\sigma}(t), K_{ij,\sigma}(t)$.
Let us stress that due to Eqs.~\eqref{Define_f_and_g} the mean field parameters $f_\sigma(t)$ and $g_\sigma(t)$ will be updated at each time step during the time evolution of the system.
The explicit form of the differential equations is discussed in detail in the Appendix~\ref{App:mean field_dynamics}.

\section{Resonance Level Model}\label{Resonance_Level_Model}

To benchmark our method, we first apply it to the exactly solvable resonance level model (RLM) at zero temperature ($T=0$), in which the impurity is coupled to the first site of the chain with a length of $L=1000$ sites with OBC. 
The RLM can be considered as the limiting case of the SIAM when $U=0$. 
Since in this case the spin-up and spin-down components are independent of each other, the SIAM corresponds to two equivalent copies of RLM for different spin components.
Hence, we can restrict ourselves to one spin species and suppress all spin indices in the following.
The hopping amplitudes $J_{i}$ between the neighboring chain sites $i$ and $i\!+\!1$ are site independent, that is, $J_{i}\equiv J$ and the onsite energies on the chain sites are equal to zero. 
The system is studied as a function of the hybridization between the chain and the impurity $V$ and the onsite energy at the impurity $\varepsilon_\mathrm{d}$, which will also be used as quench parameters.

\begin{figure}[t!]
\centering \includegraphics[width=0.45\textwidth]{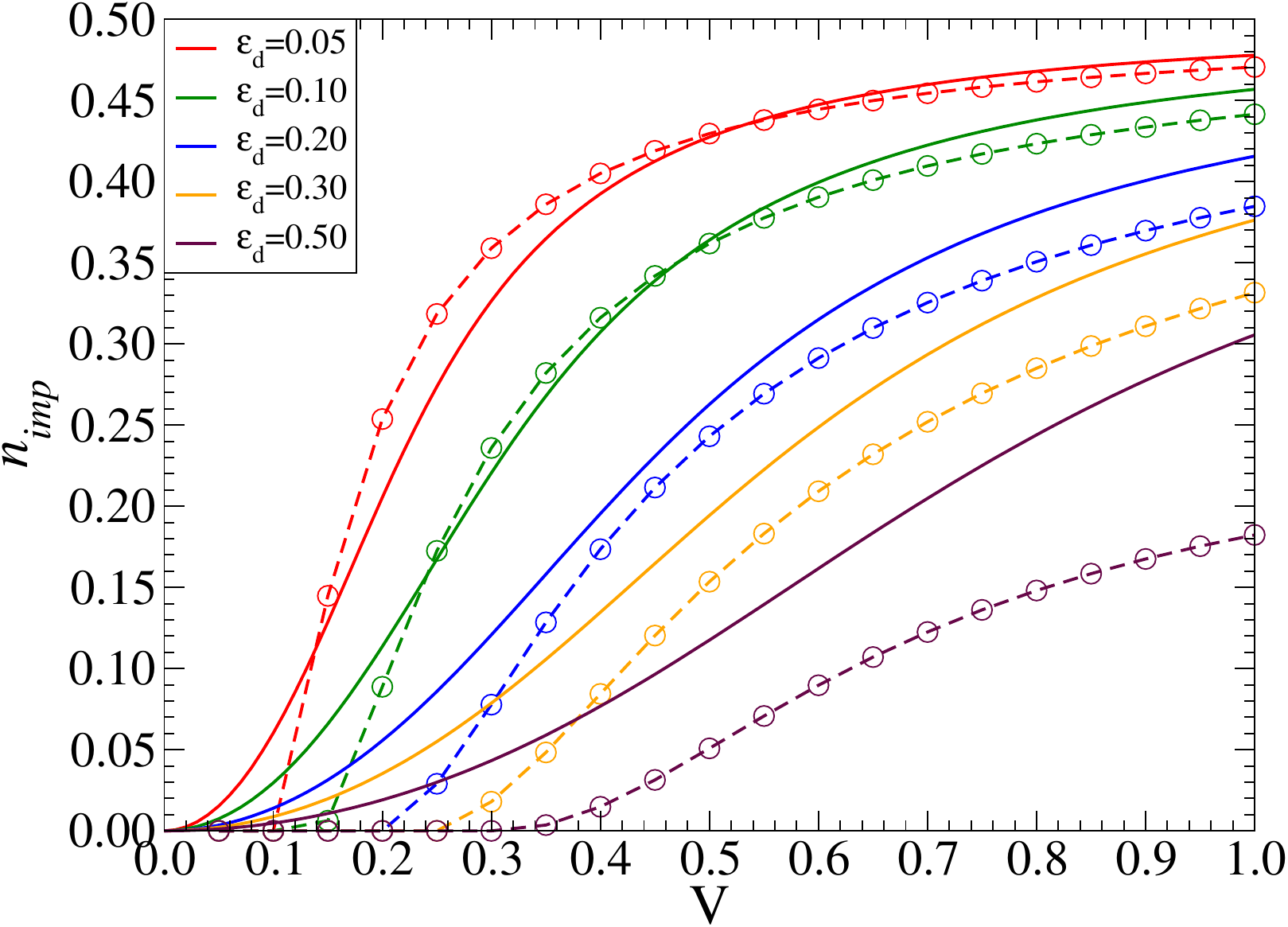}
    \caption{Occupation of the impurity site as a function of $V$ for different onsite energies for the resonance level model with $L=1000$ chain sites. The dashed lines correspond to the results obtained by our method, while the solid lines represent the exact results.
    }
\label{Fig:Resonance_level_model_Equilibrium}
\end{figure}

First, we perform calculations in equilibrium and compare the exact solution (solid lines) with the results obtained by our approach (dashed lines) in Fig.~\ref{Fig:Resonance_level_model_Equilibrium}. 
In particular, we investigate the dependence of the occupation of the impurity site $n_{\rm imp}$ on the hybridization $V$ for different onsite energies $\varepsilon_\mathrm{d}$. 
When $\varepsilon_\mathrm{d} = 0$, i.e.~for the half-filled system, both calculations coincide, resulting in $n_{\rm imp}=0.5$ for all hybridization strengths $V$ (not shown).

For finite onsite energies and small values of $V$, the mean field parameter $g_\sigma=f_\sigma=0$ and the impurity and the chain are decoupled from each other while for ${\varepsilon_\mathrm{d}>0}$ the impurity site is empty ($n_{\rm imp}=0)$. 
Upon increasing the hybridization strength $V$ beyond a critical value $V_{\mathrm{c}}$, we obtain finite mean field parameters $g_\sigma$ and $f_\sigma$. 
The impurity and the chain are coupled to each other, and the occupation of the impurity site $n_{\rm imp}$ increases with a further increase of the hybridization $V$.
The critical value $V_{\mathrm{c}}$ itself, after which the mean field parameters $g_\sigma$ and $f_\sigma$ are finite, increases with increasing $\varepsilon_\mathrm{d}$.

In general, we observe good qualitative agreement between our mean field (dashed lines) and the exact (solid lines) results for $n_{\rm imp}$ as a function of $V$. 
For small onsite energies $\varepsilon_\mathrm{d}$ even a reasonably good quantitative agreement is achieved. 
However, let us point out that in the exact solution, we find $n_{\rm imp}=0$ {\em only} at $V=0$ while for all finite $V$, $n_{\rm imp}>0$ in contrast to our mean field treatment where
$n_{\rm imp}>0$ only for $V>V_\mathrm{c}$. 
This is a typical mean field behavior where at finite $V$ a (spurious) phase transition is indicated by the emergence of the order parameter\cite{Mean-field-artifact} $n_{\rm imp}$.

\begin{figure}[t!]
\centering \includegraphics[width=0.45\textwidth]{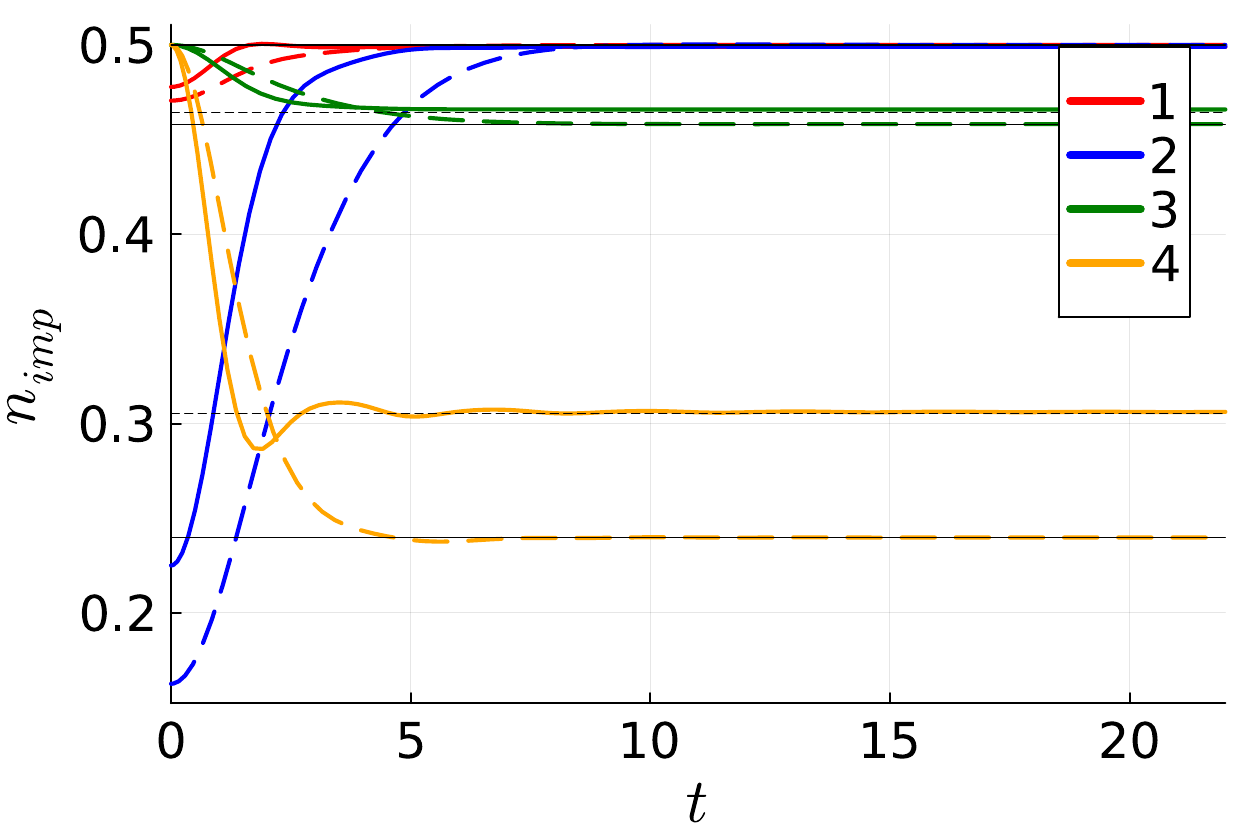}
    \caption{Occupation of the impurity site as a function of time for different quenches $(V_{\rm in},\varepsilon_{\mathrm{d},\rm in})\!\rightarrow\!(V_{\rm fi},\varepsilon_{\mathrm{d},\rm fi})$ for the resonance level model with $L\!=\!1000$ chain sites. The dashed lines correspond to the results obtained by our method, while the solid lines represent the exact results.
    We consider the following quenches: 1)~$(1.00,0.05) \rightarrow (1.00,0.00)$, 2)~$(0.75,0.50) \rightarrow (0.75,0.00)$,  3)~$(0.75,0.00) \rightarrow (0.75,0.05)$, 4)~$(1.00,0.00) \rightarrow (1.00,0.50)$.
    The horizontal black lines correspond to the equilibrium impurity filling for the final parameter set.
    }
\label{Fig:Resonance_level_model_Time_evolution}
\end{figure}

In the second step, we investigate the behavior of the system after quenching the onsite energy $\varepsilon_\mathrm{d}$  between the impurity and the chain. 
We present our results for the time evolution of the particle density at the impurity site $n_{\rm imp}(t)$ in Fig.~\ref{Fig:Resonance_level_model_Time_evolution} for four different quenches $(V_{\mathrm{in}},\varepsilon_{\mathrm{d}\mathrm{in}}) \rightarrow (V_\mathrm{fi},\varepsilon_{\mathrm{d},\mathrm{fi}})$ where the subscript ``in'' denotes the initial and the subscript ``fi'' the final value of the respective parameter after the quench. 
We observe that both our mean field results (dashed lines) and the exact solution (solid lines) quickly relax to the equilibrium result for $n_{\rm imp}$ defined by $(V_{\rm fi},\varepsilon_{\mathrm{d},\rm fi})$. 
As in the equilibrium case in Fig.~\ref{Fig:Resonance_level_model_Equilibrium}, the overall qualitative agreement between our method is good and for small values of $\varepsilon_{\mathrm{d},\rm fi}$ even the quantitative agreement is reasonable. 
Hence, we expect good results from our mean field calculations also for the SIAM, in particular for impurity occupations at or close to half-filling, which are investigated in this paper.

\section{Results}\label{Results}

In this section, we present our results for the SIAM at zero temperature ($T\!=\!0$) in which the impurity is coupled to the first site ($r\!=\!1$) of the chain with OBC ($J_{L,\sigma}\!=\!0$).
The system size is set to $L\!=\!1000$ as the results appear to be converged at this chain length (for the corresponding finite-size analysis, see Appendix~\ref{App:Finite-size-scaling}).
We consider a special case of the general model in Eq.~\eqref{Hamiltonian} which is characterized by the following choice of parameters:
\begin{itemize}
\item[(1)] The hybridization between impurity and chain is spin-independent: $V_{\sigma}\!=\!V$.
\item[(2)] The hopping amplitude is site and spin independent: $J_{i,\sigma}\!=\!J\!=\!1$, $i<L$, $J_{L,\sigma}=0$.
\item[(3)] The onsite energies on the chain are zero: $\varepsilon_{i,\sigma}\!=\!0$.
\item[(4)] The onsite energy at the impurity is set to\\ $\varepsilon_{\mathrm{d},\sigma}\!=\!-U/2$.
\item[(5)] The system is half-filled: $\mu_{\sigma}\!=\!0$.
\end{itemize}

First, we discuss the behavior of the system in equilibrium before the quench, i.e., when it is in its ground state.
We are particularly interested in the correlated site (described by the density matrix $\hat{\rho}$) and the mean field parameters $f_\sigma$ and $g_\sigma$, which characterize the coupling between the impurity and the chain. 
We will analyze these quantities as a function of the Hubbard interaction $U$ and/or the hybridization $V$. 

In the second step, we turn out attention to the dynamics of these (and some other) observables after a quench of the Hubbard interaction $U$ and/or hybridization $V$.

 \subsection{Equilibrium}\label{Equilibrium}

\subsubsection{Numerical results}

To obtain the equilibrium results for $f_\sigma\!\equiv\!f_\sigma(t\!=\!0)$, $g_\sigma\!\equiv\!g_\sigma(t\!=\!0)$ and $\hat{\rho}\!\equiv\!\hat{\rho}(t\!=\!0)$ in the SU(2) symmetric case, where $f_\uparrow\!=\!f_\downarrow\!\equiv\!f$ and $g_\uparrow\!=\!g_\downarrow\!\equiv\!g$, we numerically solve Eqs.~\eqref{Hamiltonian_Decoupled} and \eqref{Meanfield_parameters} self-consistently.
To this end, we start with an initial guess for the mean field parameters $f$ and $g$ in Eqs.~\eqref{Hamiltonian_Decoupled}.
We can then calculate the ground state of the impurity Hamiltonian in Eq.~\eqref{Hamiltonian_impurity_Decoupled} for the given $g$ by exact diagonalization since the corresponding Hilbert space spanned by the impurity $\hat{d}_\sigma$ and the Majorana $\hat{\gamma}_\sigma$ is only $16$ dimensional. 
Moreover, the Hilbert space consists of $4$ equivalent $4$ dimensional subspaces, and we can restrict ourselves to only one of those, reducing the problem to the determination of the eigenvalues and eigenvectors of a $4\times 4$ matrix (for details, see Appendix~\ref{App:Solving_the_equilibrium_problem_for_the_impurity}). 
The Hilbert space of the decoupled chain in Eq.~\eqref{Hamiltonian_chain_Decoupled_sigma}, on the other hand, is much larger ($2^{L+1}$) but the system is non-interacting. 
This allows us to transform the corresponding Hamiltonian for a given $f$ to a diagonal form by means of a unitary $2(L+1)$ dimensional matrix (for one spin species) where both the creation and the annihilation operators of Majoranas and chain electrons have to be taken into account explicitly within a Nambu spinor (for details, see Appendix~\ref{App:Solving_the_equilibrium_problem_for_the_chain}).
In the second step $g$ and $f$ can be updated via Eqs.~\eqref{Meanfield_parameters} where the corresponding expectation values are calculated from the respective ground states of the impurity and the chain obtained in the previous step. 
This self-consistent cycle is iterated until convergence of $f$ and $g$ is reached.

\begin{figure}[t!]
\centering \includegraphics[width=0.55\textwidth]{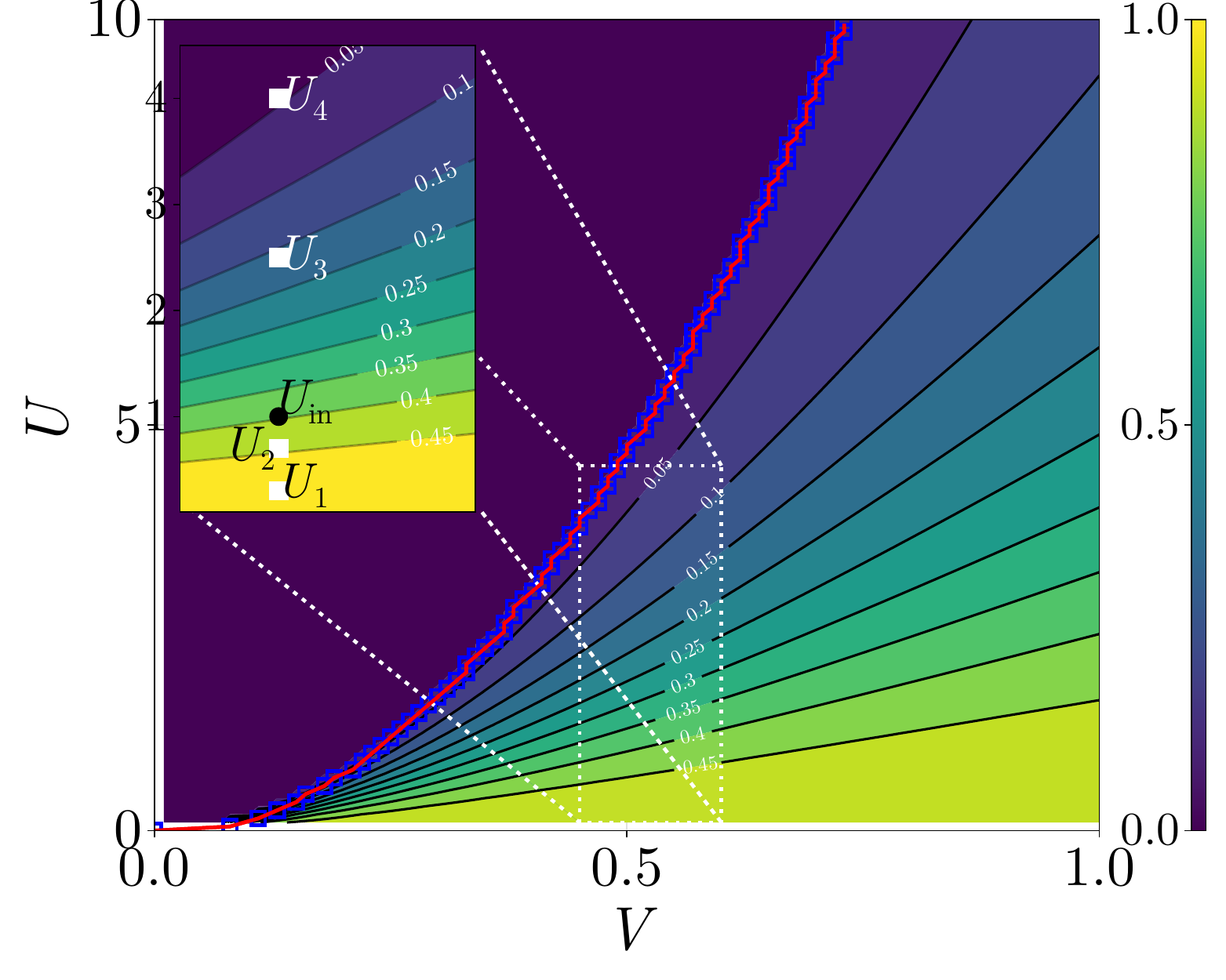} \\
\caption{$f$ as a function of $V$ and $U$. 
The calculations have been performed for 
$J\!=\!1$, $\varepsilon_{d,\sigma}\!=\!-U/2$, $\mu_\sigma\!=\!\varepsilon_{i,\sigma}\!=\!0$. 
$f\!=\!0$ (dark region) corresponds to the local moment regime where the impurity is decoupled from the chain and, hence, charge fluctuations are fully suppressed. The region where $f\!\ne\!0$ (colorful region) corresponds to the phase where impurity and chain are coupled to each other and, hence, charge fluctuations on the impurity site can be observed. The red line with symbols represents the phase transition separating the local moment regime from the valence bond regime. The blue line corresponds to $U=18.6022V^2$ (see the discussion in Sec. \ref{Analytical_analysis}).
In the inset, a specific $U_{\rm in}$ and four $U_{\rm fi}$, for which calculations have been performed, are marked by a black circle and white squares, respectively.}
\label{Fig:PD}
\end{figure}

Let us now start by investigating the mean field parameter $f$ as a function of the interaction strength $U$ between two electrons at the impurity site and the hybridization strength $V$ between the impurity and the non-interacting chain. 
The corresponding numerical results are presented by a color-coded plot in Fig.~\ref{Fig:PD} (for $g$ completely analogous results are found). 
The dark area in this phase diagram corresponds to $f\!=\!g\!=\!0$ representing a state in which the impurity and the chain are decoupled.
Consequently, a well-defined (spin-$1/2$) local moment is found at the impurity in this area of the phase diagram. 
This implies that no charge fluctuations at the impurity site emerge. 
The local moment regime is separated by a second order (quantum critical) phase transition (red line) from the region where $f\!\ne\!0$ (colorful area). 
In this mixed (or intermediate) valence regime of the phase diagram, the impurity and the chain are coupled and charge fluctuations between the impurity and the chain can be observed and become stronger for larger values of $V$ and lower interaction strength $U$ in accordance with the correspondingly higher values of $f$ in this region.
It is worth noting that for any finite value of the hybridization $V$, there exists a valence bond regime ($f \neq 0$). However, the corresponding critical interaction $U_c$ decreases as the hybridization $V$ decreases.

One should note that the obtained phase transition is the mean field feature as it is also indicated by our benchmark calculations for the resonant level model in Sec.~\ref{Resonance_Level_Model}.
In reality, we would rather expect a smooth crossover between these two regimes. 
To verify this and assess the accuracy of our approach, we compare our results with those obtained using the numerical renormalization group (NRG). The corresponding results and discussion are presented in the next subsection (Sec. \ref{Magnetic_susceptibility}).

In fact, it is rather common for crossovers observed in (numerically) exact calculations to turn into a phase transition when a mean field approach is used to solve the problem, since mean field theories generically overestimate ordered phases due to a neglect of fluctuations.\cite{Mean-field-artifact}
In Sec.~\ref{Analytical_analysis} and Appendix~\ref{App:Analytical_analysis} we provide a more detailed discussion of the phase transition including analytical calculations close to the transition line.

\begin{figure}[t!]
\centering \includegraphics[width=0.45\textwidth]{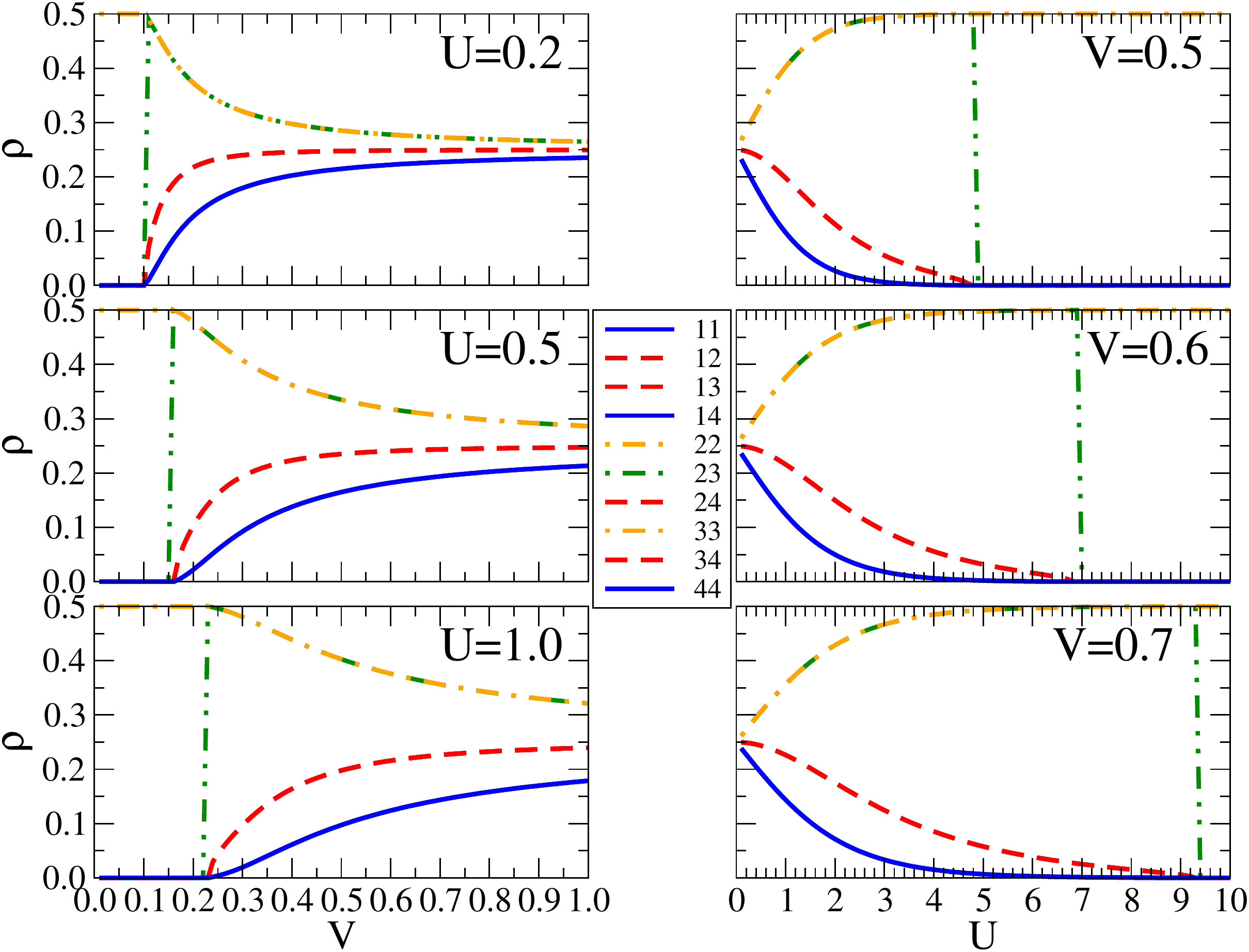} \\
\caption{Different matrix elements of the density matrix $\rho_{ij}$ as a function of $V$ (left) for different Hubbard interactions $U=0.2,\,0.5,\,1.0$ (from top to bottom) and as a function of $U$ (right) for different values of the hybridization $V=0.5,\,0.6,\,0.7$ (from top to bottom), respectively.}
\label{Fig:rho_L1000}
\end{figure}

Let us now turn our attention to the density matrix $\hat{\rho}$ of the impurity. 
In Fig.~\ref{Fig:rho_L1000} we plot various elements $\rho_{ij}$ of the $4\!\times\!4$ density matrix as a function of $V$ for different values of the Hubbard interaction $U$ (left row) as well as a function of $U$ for different hybridization values $V$ (right row), respectively.
For small values of the hybridization and/or strong interactions, where the mean field parameters $f\!=\!g\!=\!0$, all matrix elements are zero apart from $\rho_{22}\!=\!\rho_{33}\!=\!0.5$ (dashed-dotted orange lines). 
These two components of the density matrix correspond to the projectors $\lvert\uparrow\rangle\langle\uparrow\rvert$ and $\lvert\downarrow\rangle\langle\downarrow\rvert$ to the spin-$\uparrow$ and spin-$\downarrow$ states, respectively, which are occupied with the probability of $1/2$. 
This confirms that for $f\!=\!g\!=\!0$ the system is in the local moment regime.
Upon increasing the hybridization $V$ or decreasing the interaction $U$, we observe that for $V\!>\!V_\mathrm{c}$ (left panel) or $U\!<\!U_\mathrm{c}$ (right panel) all matrix elements become finite, so the impurity site can be in all $4$ possible states: $\lvert0\rangle$ (empty), $\lvert\uparrow\rangle$, $\lvert\downarrow\rangle$ and $\lvert\uparrow\downarrow\rangle$ [see also Eq.~\eqref{App:Basis_Impurity} in Appendix~\ref{App:Solving_the_equilibrium_problem_for_the_impurity}].

An interesting behavior can be observed for the matrix element $\rho_{23}\!=\!\lvert\uparrow\rangle\langle\downarrow\rvert$ across the phase transition. 
Within the local moment regime ($V\!<\!V_\mathrm{c}$ or $U\!>\!U_\mathrm{c}$) this component of $\rho$ is zero, indicating that spin flips do not occur at the impurity. 
In fact, in the SU$(2)$ symmetric case, such spin flips have to be mediated via the exchange with the non-interacting chain, which is decoupled from the impurity in this part of the phase diagram. 
Immediately after the phase transition ($V\!\gtrsim\!U_\mathrm{c}$ and $U\!\lesssim\!U_\mathrm{c}$) this matrix element features a finite jump and becomes equivalent to $\rho_{22}\!=\!\rho_{33}$. 
Let us point out that at the verge of the mixed valence regime (colorful area) still well-defined local moments are prevailing (indicated by $\rho_{22}\!=\!\rho_{33}\!\sim\!0.5$) which, however, can now feature fluctuations moderated by the coupling with the non-interacting chain. 
Upon further decrease of $V$ or increase of $U$, $\rho_{22}$, $\rho_{33}$ and $\rho_{23}$ decrease. 
At the same time, the matrix elements representing the doubly- and unoccupied states $\rho_{11}\!=\!\rho_{44}\!=\!\rho_{14}$ as well as matrix elements that mix these states with the singly-occupied ones ($\rho_{12}\!=\!\rho_{13}\!=\rho_{24}\!=\!\rho_{34}$) increase. 
In the limit of $U\!=\!0$ or $V\!\rightarrow\!\infty$ all elements of the density matrix become equal and acquire a value of $\rho_{ij}\!=\!0.25$.

\begin{figure}[h!]
\centering \includegraphics[width=0.475\textwidth]{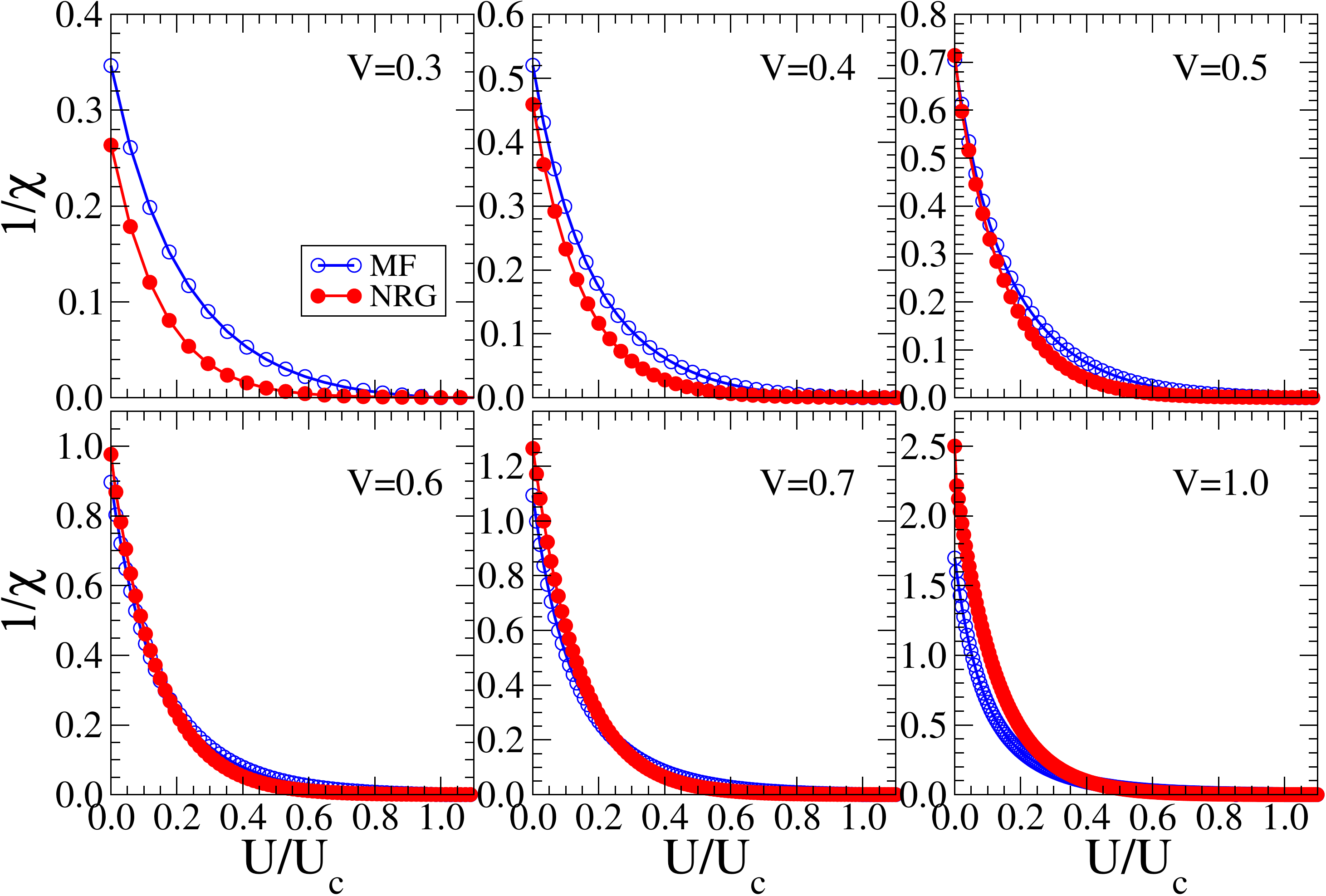}
\caption{The magnetic susceptibility $\chi=\mathrm{d}m/\mathrm{d}h$ as a function of Hubbard interaction $U$ for different values of hybridization $V$. Curves with filled symbols correspond to numerical renormalization group (NRG) results, while curves with open symbols indicate results obtained by our new mean field method.}
\label{Fig:chi_vs_U}
\end{figure}

\subsubsection{Magnetic susceptibility}
\label{Magnetic_susceptibility}

To gain more insight into the magnetic properties of the system we have studied the local impurity spin susceptibility $\chi\!=\!\chi_d(\omega=0,T=0)$ at zero frequency and temperature
\begin{equation}
\label{chi}
\chi=\frac{\mathrm{d}m_d}{\mathrm{d}h}=\frac{1}{2}\frac{\mathrm{d}}{\mathrm{d}h}\left(n_{d,\uparrow}-n_{d,\downarrow}\right)
\,
\end{equation}
where the derivative with respect to the external magnetic field $h$ has been taken numerically by considering results for $n_{d,\uparrow}$ and $n_{d,\downarrow}$ for a number of small (but finite) values of the external magnetic field $h$ applied to the impurity site.
We present our results in Fig.~\ref{Fig:chi_vs_U}, where we have plotted the inverse magnetic susceptibility $1/\chi$ as a function of the Hubbard interaction $U$ for different values of the hybridization $V$ as obtained from our mean field approach (empty blue circles). 
We observe that with increasing Hubbard interaction, the magnetic susceptibility increases (i.e., $1/\chi$ decreases).
This is indeed the expected behavior as for gradually larger values of $U$ a (screened) local moment develops at the impurity site which strongly reacts to an external magnetic field.
Eventually when the interaction reaches its critical value $U_c$, the impurity is decoupled from the chain giving rise to a free spin at the (half-filled) impurity site which is reflected in the divergence of the zero-temperature magnetic susceptibility (in accordance with the standard Curie-Weiss behavior $\chi\!\sim\!1/T$), i.e., $1/\chi=0$. 
Thus, the previously found mean field quantum phase transition can be also characterized by a diverging magnetic susceptibility at the impurity.

To estimate the accuracy of our new method in the different regimes defined by the parameters $V$ and $U$, we have also performed numerical renormalization group (NRG)\cite{bu.co.08, wils.75, hews.93, kr.wi.93, hofs.00} calculations for this thermodynamic observable. 
Our NRG results\cite{NRG} are presented alongside our mean field findings in Fig.~\ref{Fig:chi_vs_U} by red filled circles.
For impurity chain couplings $V\!\gtrsim\!0.5$ (corresponding to the half of the coupling $J$ inside the non-interacting chain) we observe that our mean field calculations predict a larger spin susceptibility (i.e., a smaller $1/\chi$) for the non-interacting case $U\!=\!0$ with respect to NRG (upper right and lower panels in Fig.~\ref{Fig:chi_vs_U}).
This can be understood in terms of the loss of quantum entanglement in our approach: In fact, this entanglement between the chain and the almost non-interacting impurity will lead to a screening of the magnetic moment which can only rudimentary be represented by the mean field $f$ in Eq.~\eqref{Hamiltonian_impurity_Decoupled}.
Upon increasing $U$ this screening effect becomes weaker also in the exact solution and, hence, the difference between the NRG and mean field results decrease until they become equivalent at a specific ($V$ dependent) value $U_\text{eq}$ (for $V\!=\!0.5$ this happens immediately after $U\!=\!0$). 
For even larger values of $U$ it appears that the averaging process within our mean field technique starts to underestimate the emerging local moment and, therefore, a larger spin susceptibility (i.e., a smaller $1/\chi$) is predicted by the exact NRG calculations.
Finally, at $U_c$ (corresponding to $U/U_c\!=\!1$ in the figure) the mean field spin susceptibility diverges (i.e., $1/\chi\!=\!0$) while the NRG one remains finite albeit very large (such that $1/\chi$ from NRG cannot be distinguished from zero at this interaction value in our figure).

For hybridizations $V\!<\!0.5$ (left and middle panel in the upper row of Fig.~\ref{Fig:chi_vs_U}) our mean field method underestimates the spin susceptibility already at $U\!=\!0$.
This indicates that at such low hybridization values the screening of the local moment already becomes ineffective in the exact NRG solution and, as discussed above, the averaging process of our mean field technique seems to suppress this observable.
As for larger values of $V$,  the spin susceptibility strongly increases upon increasing $U$ for both methods and eventually diverges in our mean field calculations while it remains finite (albeit very large) within NRG (apart from the limiting case $V\!=\!0$ where the spin susceptibility obviously diverges for all values of $U$ in both approaches).


The comparison of the impurity spin susceptibilities obtained from NRG and our mean field treatment of the problem demonstrates that our new method captures the essential physical of the model qualitatively correctly.
Only the observed phase transition is a typical mean field feature which predicts a diverging susceptibility in an interaction regime where, however, also the exact results for this observable are very large.
In some regions with sufficient impurity-chain coupling ($V\!\gtrsim\!0.5$) where the interaction $U$ is already large enough to induce the emergence of a local moment even good quantitative agreement can be achieved.
However, it should be noted that quality of such a quantitative agreement will depend on the observable under consideeration.
For instance, while the impurity double occupancy $d=\langle n_{d,\uparrow}n_{d,\downarrow} \rangle$ shows the correct qualitative behavior as a function of $V$ and $U$ within our mean field method its deviation from NRG results is larger than the corresponding differences observed for the spin susceptibility (not shown). 
This can be attributed to the fact the dynamics of the charge degrees of freedom might be stronger affected than the corresponding spin dynamics by the mean field decoupling of the impurity in our method.

It is finally worth noting that the Kondo temperature is inversely proportional to the magnetic susceptibility\cite{hews.93, wils.75}, i.e.
\begin{equation}
\label{Kondo_temperature}
T_K \sim \frac{1}{\chi} \,.
\end{equation}
Hence, our analysis of the spin susceptibility presented above also provides an estimate how well our new approach can estimate the Kondo temperature.
As NRG is one of the most reliable methods to capture Kondo physics the qualitatively (and for specific parameter regimes even quantitatively) good agreement of the spin susceptibility between our method and NRG indicates that our approach is able to well capture this thermodynamic observable.
Let us, however, points out that within our approach we cannot address other aspects of the Kondo problem, in particular, scaling and frequency dependence of impurity spectral functions.
In fact, after the mean field parameters $f$ and $g$ are self-consistently determined the Hilbert space of the decoupled impurity in Eq.~\eqref{Hamiltonian_impurity_Decoupled} is only four dimensional and, hence, the spectral function consists only of six (delta-like) peaks.

\subsubsection{Analytical analysis near the phase transition}
\label{Analytical_analysis}

We conclude our discussion of the equilibrium behavior with an analytical analysis of the system close to the phase transition (red line in Fig.~\ref{Fig:PD}).
Here we present only the main results while detailed calculations are provided in the Appendix~\ref{App:Analytical_analysis}.

\!\!In the vicinity of the phase transition both mean field parameters $f$ and $g$ are small and, hence, the terms proportional to these parameters in the mean field Hamiltonian in Eqs.~\eqref{Hamiltonian_chain_Decoupled_sigma} and \eqref{Hamiltonian_impurity_Decoupled} can be treated by perturbation theory:
\begin{subequations}\label{fgexpansion}
\begin{align}\label{fgexpansionf}
    & f = 
    \langle\hat{\gamma}_\sigma^{\phantom\dagger} \hat{d}^{\dagger}_\sigma\rangle (g)
    = F_0 + F_1g + F_2g^2 + \mathcal{O}(g^3)\\
\label{fgexpansiong}
    & g = \langle\hat{\gamma}_\sigma^{\phantom\dagger} \hat{c}^{\dagger}_{r,\sigma}\rangle(f)
    = G_0 + G_1 f + G_2 f^2 + \mathcal{O}(f^3),
\end{align}
\end{subequations}
%
The coefficients $F_{\rm i}$ and $G_{\rm i}$ can be determined by a perturbation theory calculation up to second order\cite{expectation_for_f} in the parameters $g$ and $f$, respectively (for details, see Appendix~\ref{App:Analytical_analysis}). 
As a result, we obtain
\begin{subequations}\label{fgexpansionexpressions}
\begin{align}
    &f = - \frac{4Vg}{U}
         + \mathcal{O}(g^3)\label{fgexpansionespressionsf}\\
    &g = - \frac{a V}{|J|}f 
         + \frac{b V^2}{J^2}f^2 
         + \mathcal{O}(f^3),\label{fgexpansionespressionsg}
\end{align}
\end{subequations}
where the constants $a$ and $b$ are given in Eqs.~\eqref{a} and \eqref{b} of Appendix~\ref{App:Analytical_analysis}.

Neglecting the terms of the order $\mathcal{O}(f^3)$ and $\mathcal{O}(g^3)$, the system of two equations~\eqref{fgexpansionexpressions} can be now solved for the two unknowns $f$ and $g$. 
Apart from the trivial solution $f\!=\!g\!=\!0$ we obtain
\begin{subequations}\label{fgsolution}
\begin{align}\label{fgsolutionf}
    & f = \frac{U J^2}{4b V^3} \left(
            \frac{4a V^2}{U |J|}-1
          \right)
\\
\label{fgsolutiong}
    & g = -\frac{U^2 J^2}{16b V^4}\left(
              \frac{4a V^2}{U |J|}-1
           \right) \,.
\end{align}
\end{subequations}
The phase transition to the local moment regime is indicated by $f\!=\!g\!=0$.
Applying this condition to Eqs.~\eqref{fgsolution} we can derive the critical hybridization strength $V_\mathrm{c}$ as a function of the interaction $U$ or, vice versa, the critical interaction $U_\mathrm{c}$ as a function of $V$
\begin{align}\label{VcUc}
    & V_\mathrm{c} = \sqrt{\frac{U |J|}{4a}} 
        & U_\mathrm{c} = \frac{4a V^2}{|J|}  \,.
\end{align}
This result can be easily understood by mapping the SIAM to the Kondo model by means of a Schrieffer-Wolff transform. 
The corresponding Kondo coupling between impurity and bath spin is then given by $J_{\rm ex}\!=\!8V^2/U$. 
Comparing this expression with Eq.~\eqref{VcUc}, the onset of the Kondo regime occurs when $J_{\rm ex}\!=\!2\lvert J \rvert/a$.

Let us also note, that in Fig.~\ref{Fig:PD} the analytically predicted transition curve $U_c(V)$ given in Eq.~\eqref{VcUc} [blue line] is indeed in excellent agreement with the numerically obtained transition (red line).
\begin{figure*}[t!]
\begin{center}
\centering \includegraphics[width=0.48\textwidth]{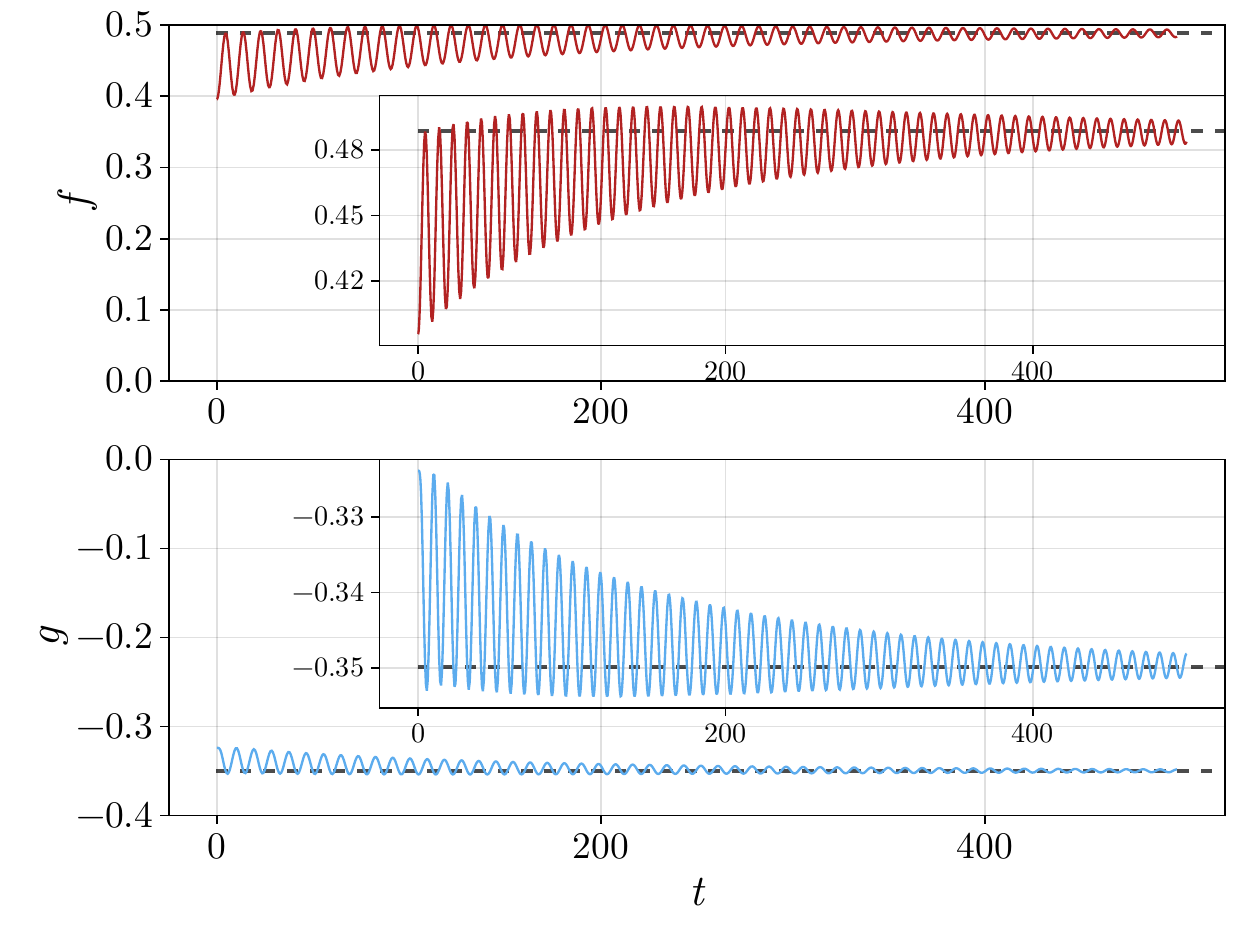}
\centering \includegraphics[width=0.48\textwidth]{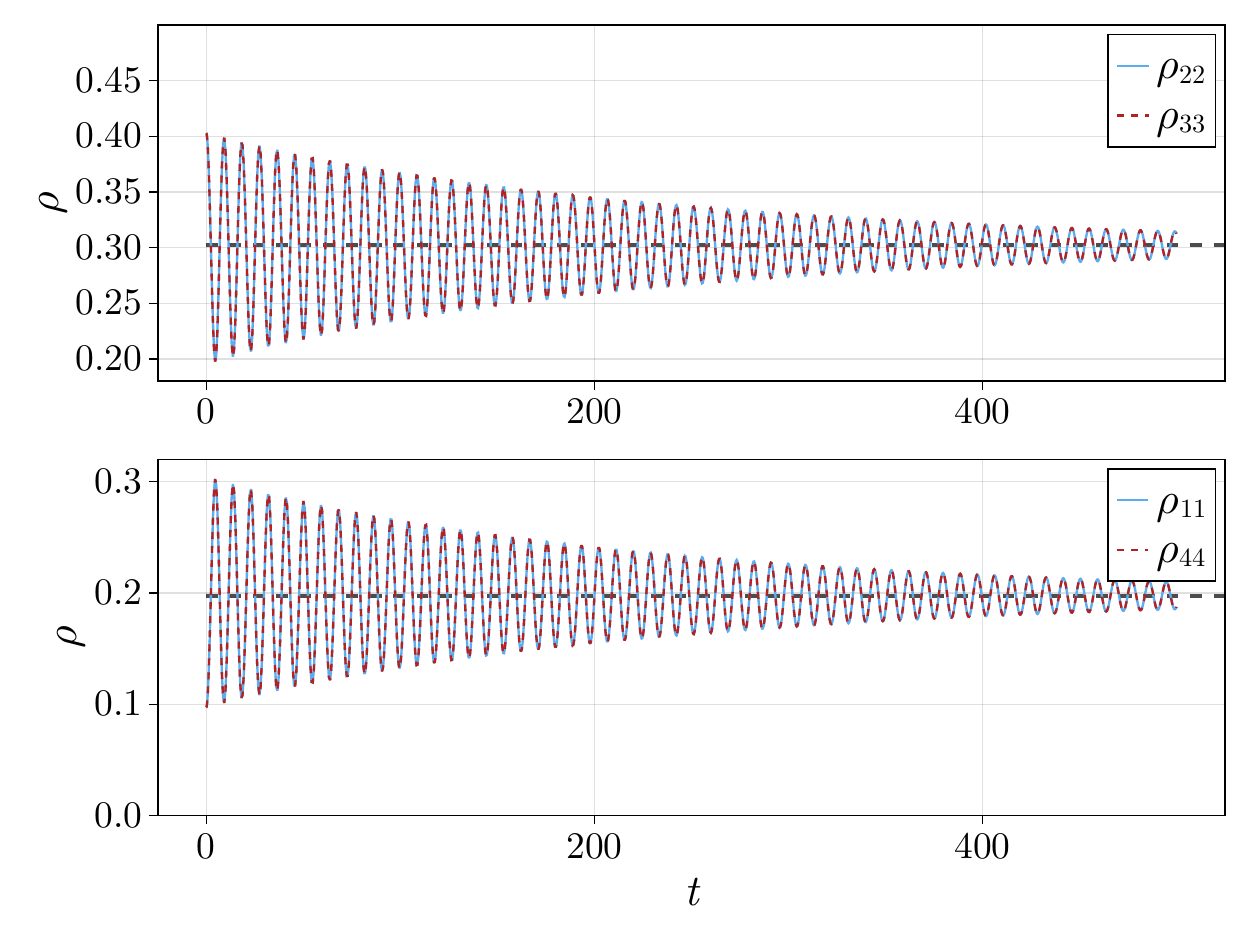}
\end{center}
\caption{Time evolution after the quench $(V_{\rm in}=0.5, U_{\rm in}=1.0)\rightarrow(V_{\rm fi}=0.5, U_{\rm fi}=0.3)$: Mean field parameters $f(t)$ and $g(t)$ (left) and diagonal elements of the density matrix $\rho_{ii}(t)$ (right). Case (i).}
\label{Fig:Case:i}
\end{figure*}

\begin{figure*}[t!]
\begin{center}
\centering \includegraphics[width=0.48\textwidth]{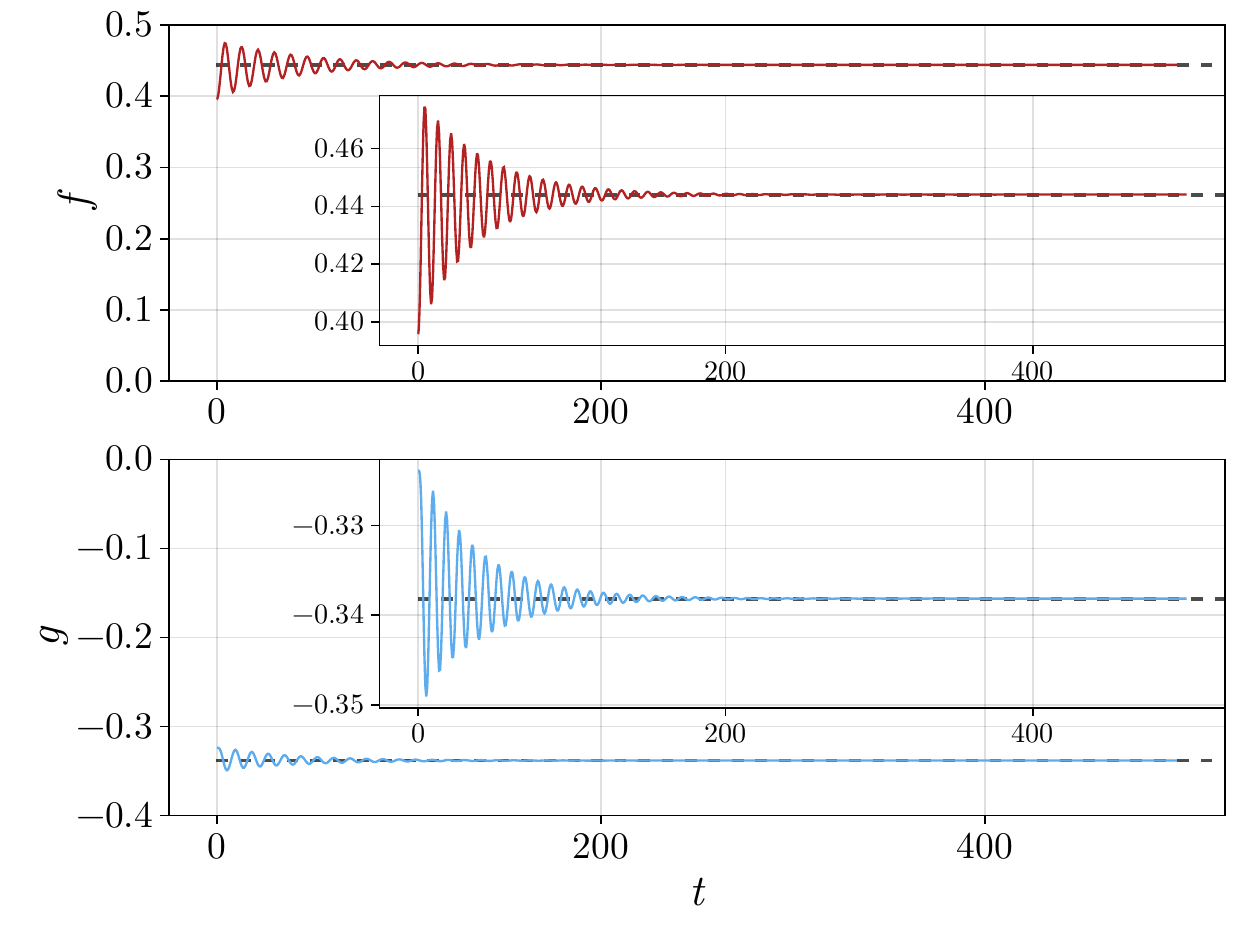}
\centering \includegraphics[width=0.48\textwidth]{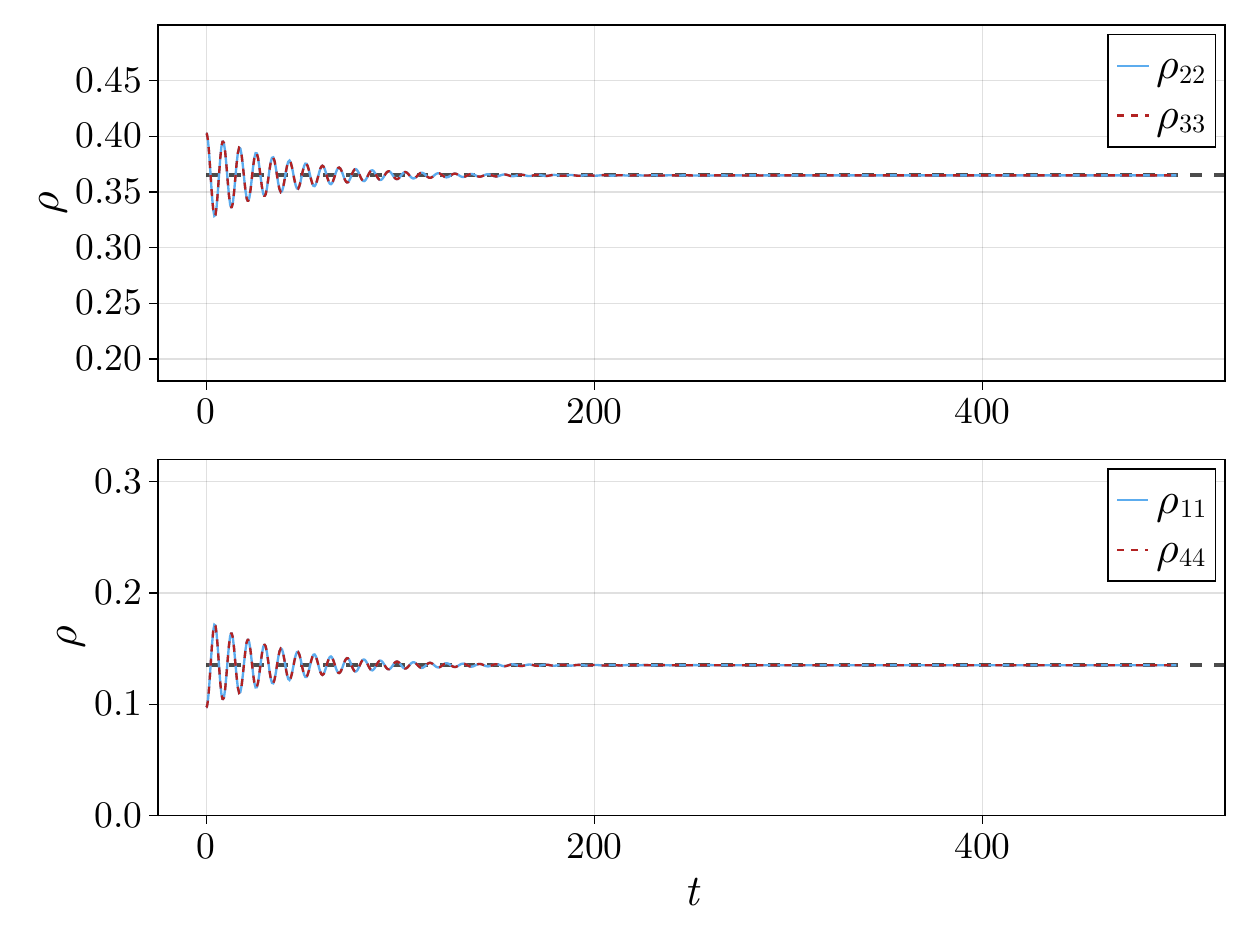}
\end{center}
\caption{Same as in Fig.~\ref{Fig:Case:i} but for the quench $(V_{\rm in}\!=\!0.5,U_{\rm in}\!=\!1.0)\!\rightarrow\!(V_{\rm fi}\!=\!0.5,U_{\rm fi}\!=\!0.7)$. Case (ii).}
\label{Fig:Case:ii}
\end{figure*}

\begin{figure*}[t!]
\begin{center}
\centering \includegraphics[width=0.48\textwidth]{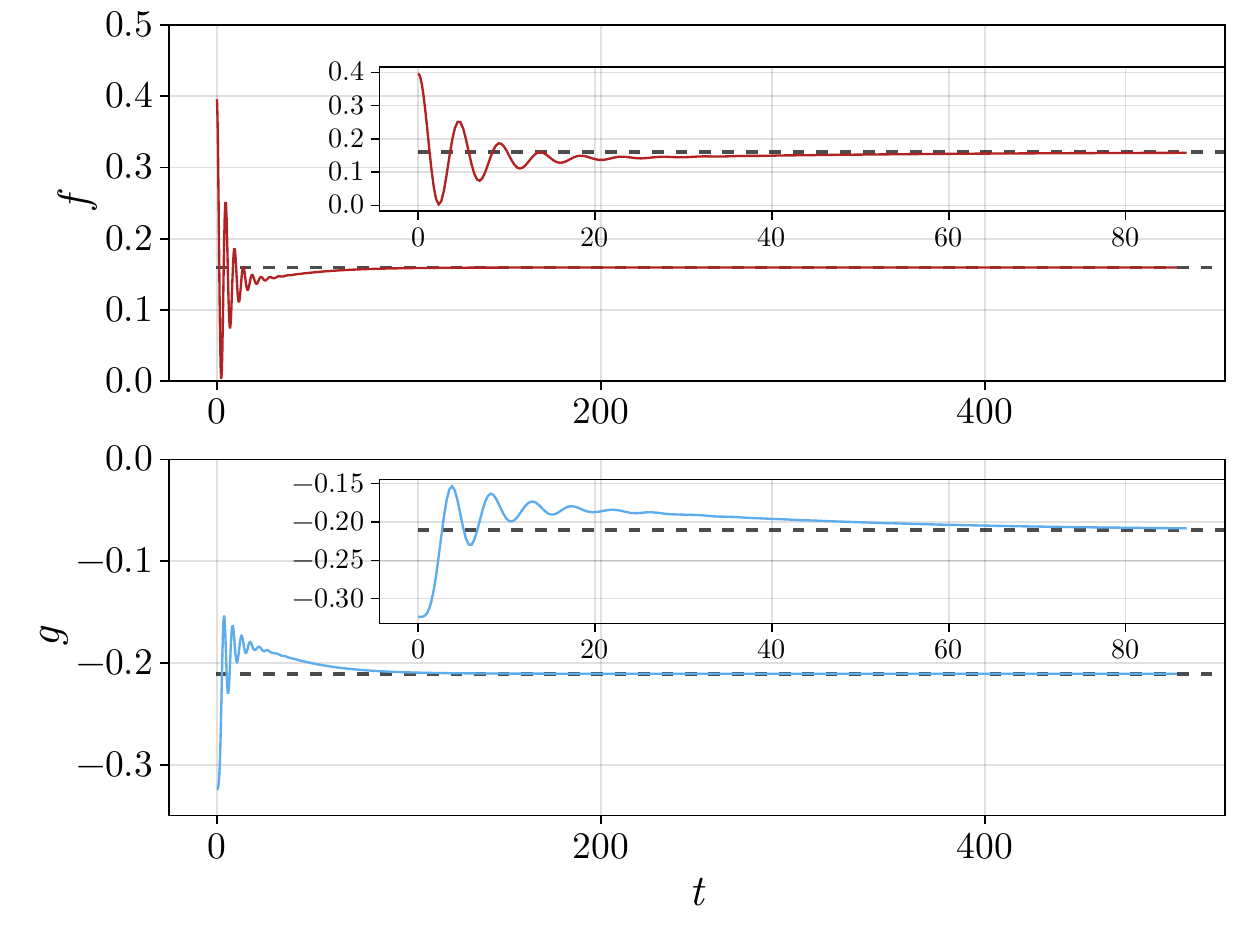}
\centering \includegraphics[width=0.48\textwidth]{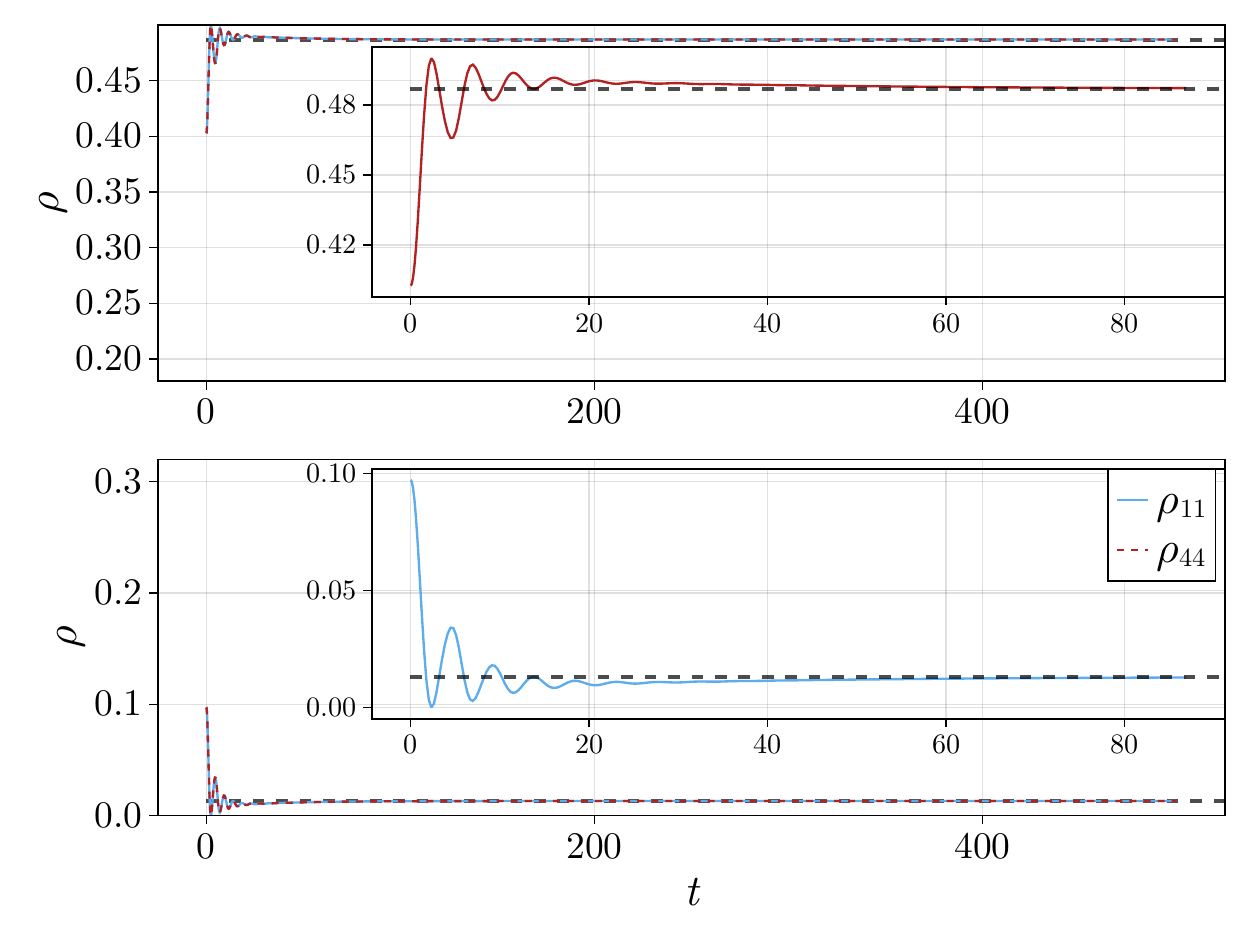}
\end{center}
\caption{Same as in Fig.~\ref{Fig:Case:i} but for the quench $(V_{\rm in}\!=\!0.5,U_{\rm in}\!=\!1.0)\!\rightarrow\!(V_{\rm fi}\!=\!0.5,U_{\rm fi}\!=\!2.5)$. Case (iii).}
\label{Fig:Case:iii}
\end{figure*}

\begin{figure*}[t!]
\begin{center}
\centering \includegraphics[width=0.48\textwidth]{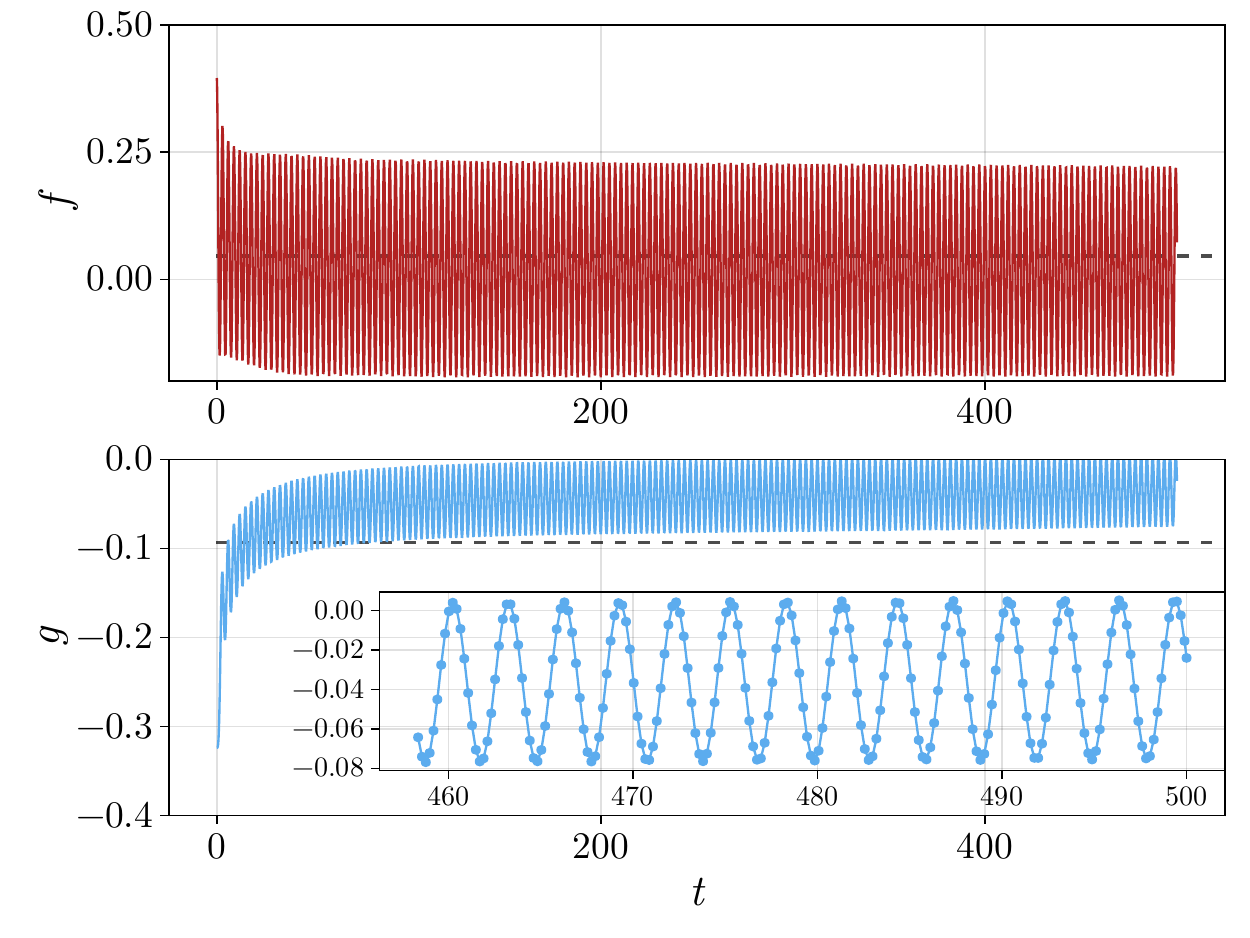}
\centering \includegraphics[width=0.48\textwidth]{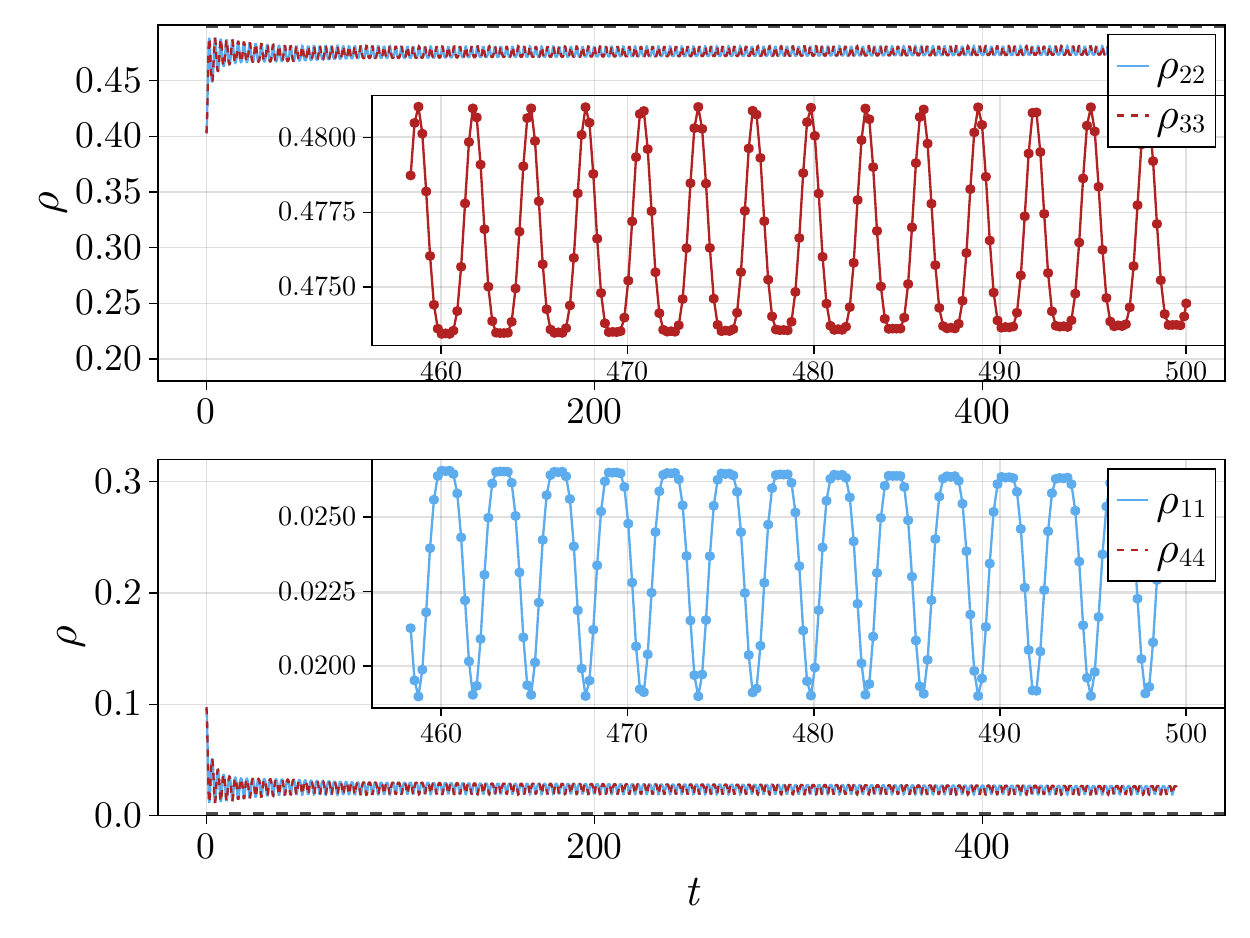}
\end{center}
\caption{Same as in Fig.~\ref{Fig:Case:i} but for the quench $(V_{\rm in}\!=\!0.5,U_{\rm in}\!=\!1.0)\!\rightarrow\!(V_{\rm fi}\!=\!0.5,U_{\rm fi}\!=\!4.0)$. The insets show the time dependence at large times in more detail. Case (iv).}
\label{Fig:Case:iv}
\end{figure*}

\begin{center}
\begin{figure*}[htb]
\includegraphics[width=\textwidth]{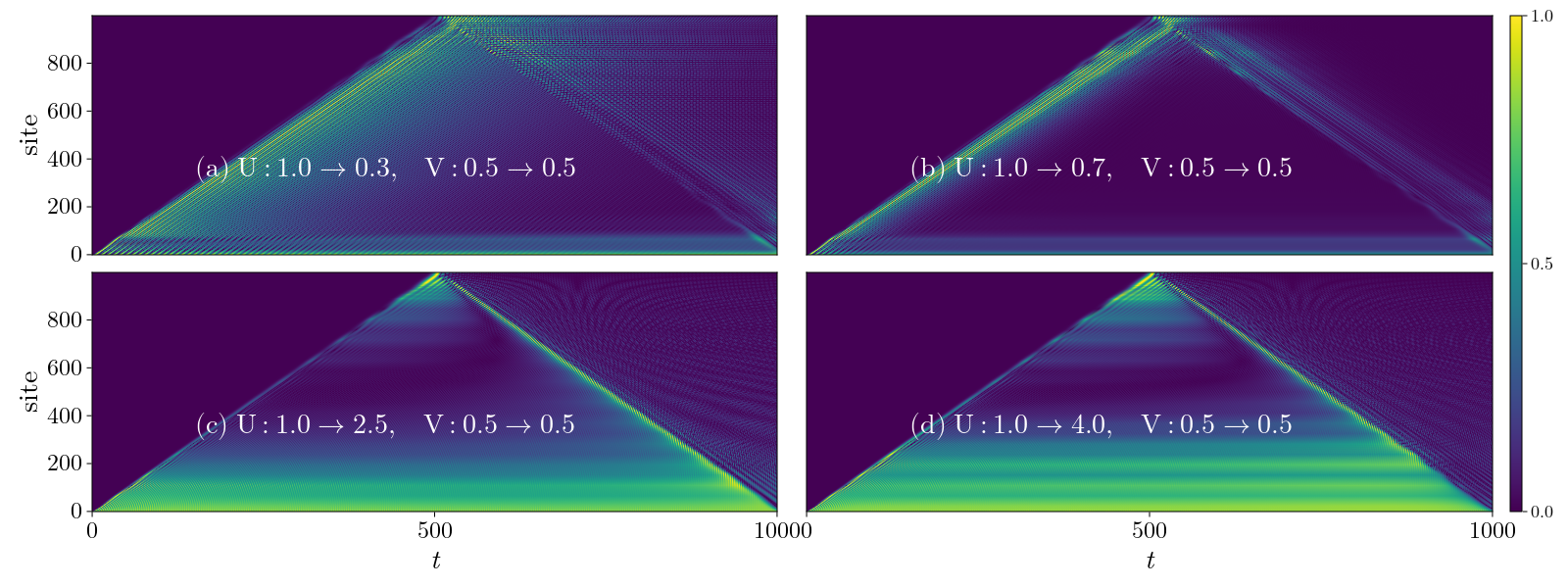}
\caption{Color coded plot of $C_{i(i+1)}(t)\!-\!C_{i(i+1)}(0)$ for the four different quenches (i)-(iv) $(V_{\rm in},U_{\rm in})\!\rightarrow\!(V_{\rm fi},U_{\rm fi})$ discussed above for $f(t)$, $g(t)$ and $\rho_{ii}(t)$ in Figs.~\ref{Fig:Case:i}-\ref{Fig:Case:iv}. The $x$- and $y$ axes refer to time $t$ and chain site $i$, respectively. For better contrast, the color coding of $C_{i(i+1)}(t)\!-\!C_{i(i+1)}(0)$ corresponds to the value $\frac{|C_{i(i+1)}(t) - C_{i(i+1)}(0)|}{\underset{t}{\rm{max}}|C_{i(i+1)}(t) - C_{i(i+1)}(0)|}$.}
\label{Fig:Cij_color_t}
\end{figure*}
\end{center}

\begin{figure*}[t!]
\begin{center}
\centering \includegraphics[width=0.90\textwidth]{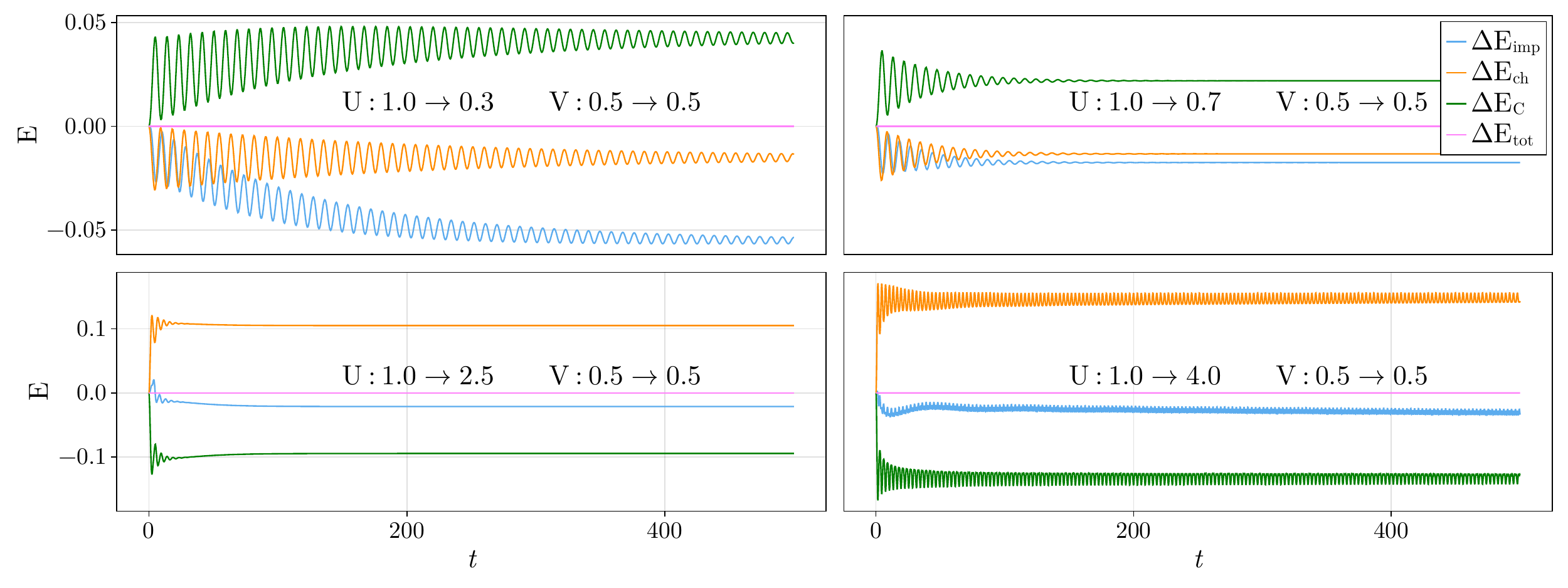}
\end{center}
\caption{Time evolution of the total energies of the impurity $E_{\rm imp}(t)$ (blue), the chain $E_{\rm ch}(t)$ (orange) and the coupling energy $E_C(t)\!=\!{\cal C}(t)$ defined in Eq.~\eqref{Ct} (green) relative to their values directly after the quench at $t=0^+$, i.e., $\Delta E_{\rm x}(t)\!=\!E_{\rm x}(t)-E_{\rm x}(0^+)$. 
Here $E_{\rm x}(0^+)$ represents the corresponding energy of the impurity (${\rm x}\!=\!{\rm imp})$, the chain (${\rm x}\!=\!{\rm ch}$) and the coupling energy (${\rm x}\!=\!C)$ for $(V_{\rm fi},U_{\rm fi})$ at $t=0^+$. The purple line ($\Delta E_{\rm tot}$) represents the sum of all three contributions.
}
\label{Fig:Energy_t}
\end{figure*}

\subsection{Time evolution} \label{Time_Evolution}

In this section, we investigate the time evolution of the system after a quench of the hybridization $V$ and/or the interaction $U$. 
More specifically, we start from the equilibrium values for $\rho_{ij}(t\!=\!0)\!=\!\rho_{ij}$, $C_{ij}(t\!=\!0)\!=\!C_{ij}$ and $K_{ij}(t\!=\!0)\!=\!K_{ij}$ for a given set of initial parameters $(V_{\rm in},U_{\rm in})$ at time $t\!=\!0$ (see Sec.~\ref{Equilibrium}) and analyze their time evolution under the equations of motion [Eqs.~\eqref{equation_of_motion_Density_Operator} and \eqref{equation_of_motion}] for a (final) set of parameters $(V_{\rm fi},U_{\rm fi})$. 
We will focus on the dynamics of the diagonal elements of the density matrix of the impurity $\rho_{ii}(t)$, the mean field parameter $f(t)$, which is related to some of the off-diagonal impurity density matrix elements via Eqs.~\eqref{Define_f_via_rho_up} and \eqref{Define_f_via_rho_down}, as well as the mean field parameter $g(t)$ which corresponds to the coupling $C_{01}(t)$ between the Majorana and the first site of the chain [see Eq.~\eqref{Define_g_via_C} for $r\!=\!1$]. 
Moreover, we will also address the nearest-neighbor correlators $C_{i(i+1)}(t)$ as a function of time and chain site $i$ (see also Appendix~\ref{App:Correlator}) as well as the time evolution of the total energies of the different parts of the system. 
Note that due to the SU($2$) symmetry of the system, all quantities are spin independent, and hence, we suppress all spin indices in our notation.

Let us point out that the diagonal elements $\rho_{ii}(t)$ of the density matrix, as well as its eigenvalues, remain non-negative real numbers throughout the time evolution. 
Additionally, for the half-filled impurity and chain considered here, the mean field parameters $f(t)$ and $g(t)$, as well as the nearest-neighbor correlators  $C_{i(i+1)}(t)$ are also real numbers at all times.
On the contrary, some of the off-diagonal elements of the density matrix $\rho_{i\ne j}(t)$ acquire an imaginary part after the quench, which, depending on the final set of parameters, might eventually vanish.
For the half-filled system, these off-diagonal elements are subject to a number of constraints. 
In particular, ${\rho_{12}(t)\!=\!\rho_{13}(t)\!=\!\rho_{24}^*(t)\!=\!\rho_{34}^*(t)}$. According to Eqs.~\eqref{Define_f_via_rho_up} and \eqref{Define_f_via_rho_down} the real parts of these matrix elements are then related to the mean field parameter $f(t)$ by $f(t)/2\!=\!\Re\rho_{12}(t)\!=\!\Re\rho_{13}(t)\!=\!\Re\rho_{24}^*(t)\!=\!\Re\rho_{34}^*(t)$. 
On the other hand, $\rho_{14}(t)\!=\!\rho_{11}(t)\!=\!\rho_{44}(t)$ and $\rho_{23}(t)\!=\!\rho_{22}(t)\!=\!\rho_{33}(t)$ remain real during the time evolution. 
Therefore, only the imaginary parts $\Im\rho_{12}(t)\!=\!\Im\rho_{13}(t)\!=\!\Im\rho_{24}^*(t)\!=\!\Im\rho_{34}^*(t)$ carry additional information which is not yet contained in $f(t)$ or $\rho_{ii}(t)$. 
The time evolution of these quantities will be briefly discussed in Appendix~\ref{App:Rho_offdiagonal}.

In the following, we will consider a number of different quenches $(V_{\rm in},U_{\rm in})\!\rightarrow\!(V_{\rm fi},U_{\rm fi})$ where the initial state is always located in the mixed valence regime (i.e., the colorful region) of the phase diagram in Fig.~\ref{Fig:PD}. 
In fact, for a starting point in the local moment regime (dark region in Fig.~\ref{Fig:PD}) the initial values $f\!=\!g\!=\!0$ correspond to a stationary point of our system of differential equations, and hence no dynamics can be observed. 
Let us, however, point out that our numerical calculations indicate that we are concerned with a meta-stable fixed point, and therefore arbitrarily small numerical deviations from the stationary starting values might lead to a (spurious) dynamics in our simulations.
(On the contrary, the situation where no quench is performed, that is $V_{\rm fi}\!=\!V_{\rm in}$ and $U_{\rm fi}\!=\!U_{\rm in}$, represents a stable fixed point of the system of differential equations.)
A more detailed analysis of this question is left for future research work.

For weak to intermediate interactions ($U \lesssim 4J$) the time dependence of the observables $X(t)\!=\!f(t)$, $ g(t)$ and $\rho_{ii}(t)$ after a settling time $t_{\rm S}$ can be well described by the following expression for the time dependence of a damped harmonic oscillator
\begin{align}\label{Fitting}
    X(t) = A + \left[B \cos(\omega t +\phi)-C\right] e^{-\frac{t}{t_{\rm rel}}}
\end{align}
Here, $\omega$ denotes the oscillation frequency, $B$ is the oscillation amplitude and $t_{\rm rel}$ corresponds to the relaxation time of the system. 
The parameter $A$ represents the asymptotic value to which the system converges for $t\!\rightarrow\!\infty$ and the parameter $C$ describes the non-oscillating contribution to $X(t)$. 
The phase shift $\phi$ does not carry physically relevant information, as it depends on the choice of the time interval for which the numerical data is fitted to Eq.~\eqref{Fitting} which, in turn, is associated with the settling time $t_{\rm S}$.

Note that while the settling time $t_{\rm S}$ itself and the time evolution for $t\!<\!t_{\rm S}$ obviously depend on the initial parameters $(V_{\rm in},U_{\rm in})$, the long time behavior is predominantly governed by the final values $(V_{\rm fi},U_{\rm fi})$ after the quench.
In particular, the fitting parameters $A$, $\omega$ and $t_{\rm rel}$ in Eq.~\eqref{Fitting} are independent of the equilibrium state from which we start our time evolution.
For moderate final interactions $U_{\rm fi}$ the parameter $A$ becomes equivalent to the equilibrium value $X_{\rm equ}$ for the set of final parameters $(V_{\rm fi},U_{\rm fi})$.
However, the oscillation amplitude $B$ and the parameter $C$ characterizing the non-oscillatory decay depend on $(V_{\rm in},U_{\rm in})$. 
This observation is also further addressed in Sec.~\ref{Discussion}.

The analytical expression for the time dependence of our observables $X(t)$ for times $t\!\gg\!t_{\rm S}$ allows us to identify different types of long-time behavior associated with final parameters $(V_{\rm fi},U_{\rm fi})$ located in different regimes of the phase diagram of Fig.~\ref{Fig:PD}. 
More specifically, we can group the long-time evolution of $X(t)$ into four different classes depending on the final interaction $U_{\rm fi}$ where we fix the value of the hybridization between impurity and chain to $V\!=\!0.5$ and the initial interaction to $U_{\rm in}\!=\!1.0$. 
The locations of the four final interactions $U_1$ to $U_4$ which correspond to these four classes are marked by white squares in the inset of Fig.~\ref{Fig:PD} while the initial interaction $U_{\rm in}\!=\!1.0$ is indicated by a black dot.
The dependence on $V$ as well as on the initial parameters $(V_{\rm in},U_{\rm in})$ will be briefly addressed in Sec.~\ref{Discussion}. 

\begin{figure*}[th!]
\includegraphics[width=0.48\textwidth]{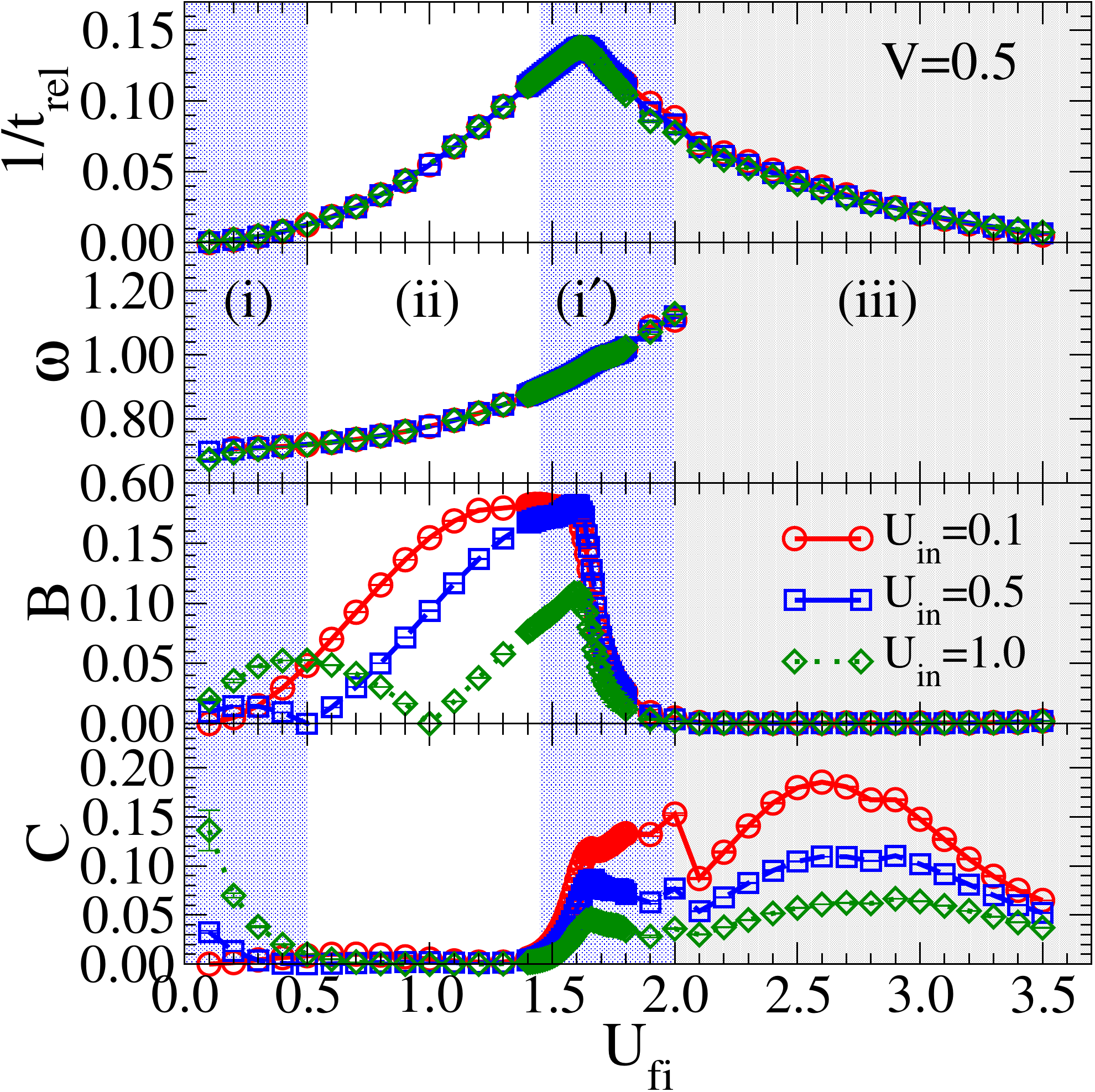}
\hspace{0.005\textwidth}
\includegraphics[width=0.48\textwidth]{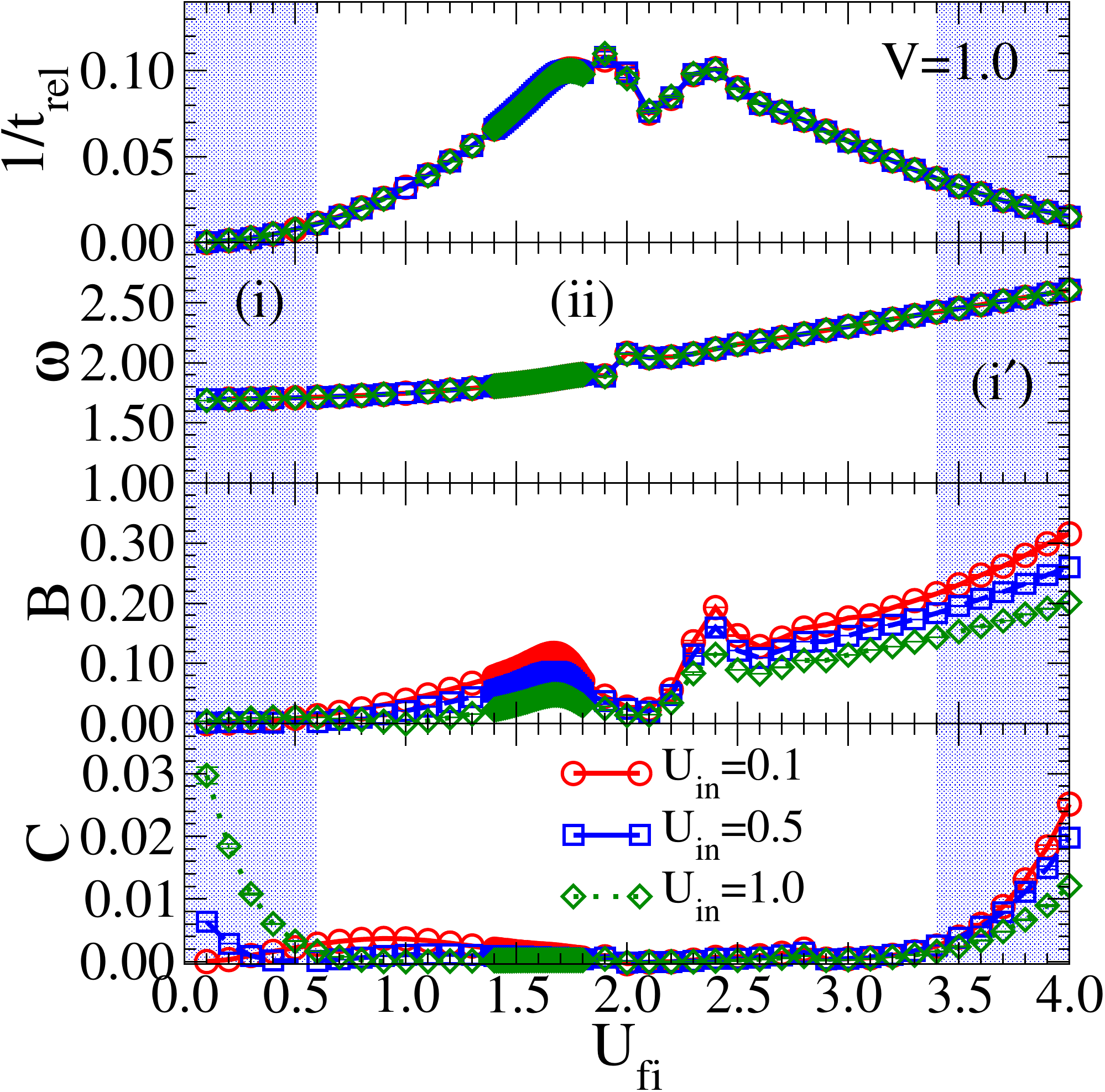}
\caption{Inverse relaxation time $1/t_{\rm rel}$, oscillation frequency $\omega$, oscillation amplitude $B$ and non-oscillating decay parameter $C$ as a function of the final interaction $U_{\rm fi}$ for different initial interactions $U_{\rm in}$ and fixed hybridization $V_{\rm in}=V_{\rm fi}=0.5$ (left) and $V_{\rm in}=V_{\rm fi}=1.0$ (right), respectively. 
All parameters are obtained by fitting the numerical data for the mean field parameter $f(t)$ to Eq.~\eqref{Fitting}. 
Different shadings indicate the different types (i), (ii), (i') and (iii) of long-time behavior of the observables.}
\label{Fig:Fitting_f}
\end{figure*}

    \noindent\underline{(i) $U_{\rm fi}\!\lll\! 4$ (see Fig.~\ref{Fig:Case:i}):} After a (very short) settling time $t_{\rm S}$ we observe that $X(t)$ features two contributions. 
    One corresponds to an exponentially decaying oscillation with the decay rate $1/t_{\rm rel}$ and the other is a non-oscillatory term that decreases exponentially with the same decay rate as the amplitude of the oscillation. 
    For very weak final interactions $U_{\rm fi}$, the relaxation time is large ($t_{\rm rel}\! > \!100$) and decreases with increasing interaction. 
    For $t\!\rightarrow\!\infty$ our observable $X(t)$ converges to the (static) equilibrium value $X_{\rm equ}$ defined by the final set of parameters $(V_{\rm fi},U_{\rm fi})$ after the quench. 
    In summary, this regime is characterized by the following set of fitting variables:
    \begin{equation*}
        B\!\ne\!0, 
        100\!<\!t_{\rm rel}\!<\!\infty, 
        \lim\limits_{t\rightarrow\infty}X(t)\!=\!A\!=\!X_{\rm equ}, 
        C\!\ne \!0.
    \end{equation*}
    An example for this type of quench is shown in Fig.~\ref{Fig:Case:i} where on the left-hand side the time evolution of the mean field parameters $f(t)$ and $g(t)$ and on the right-hand side the diagonal density matrix elements $\rho_{ii}(t)$ are depicted. 
    Note that the constant $C$, which describes the non-oscillating part, emerges in the time evolution of $f(t)$ and $g(t)$ while it is (almost) absent in $\rho_{ii}(t)$. 
    Hence, the value of this fitting parameter depends on the specific observable under consideration.
    We also note that with the increase of $U_{\rm fi}$ a regime similar to (i) reappears after regime (ii), which we will denote as (i'). 
    It shares the emergence of a finite value of $C$ with regime (i); however, the related retardation time $t_{\rm ret}$ is much shorter compared to (i).
    Regime (ii) is then sandwiched between (i) and (i') [see also Fig.~\ref{Fig:Fitting_f}].

    \noindent\underline{(ii) $U_{\rm fi}\!\ll\! 4$ (see Fig. \ref{Fig:Case:ii}):} Similarly to the previous quench, we observe a damped oscillatory time evolution for the observable $X(t)$. 
    However, in contrast to case (i), the non-oscillatory part is missing and $X(t)$ oscillates around the equilibrium value $X_{\rm equ}$ defined by the parameters $(V_{\rm fi},U_{\rm fi})$. 
    Moreover, the relaxation time is considerably shorter than for the first type of quenches.
    Therefore, regime (ii) is characterized by the following fitting parameters:
    \begin{equation*} 
        B\!\ne\!0, 
        10\!<\!t_{\rm rel}\!<\!100, 
        \lim\limits_{t\rightarrow\infty}X(t)\!=\!A\!=\!X_{\rm equ},
        C\!\simeq\! 0.
    \end{equation*}
    An example for this type of quench is depicted in Fig.~\ref{Fig:Case:ii}. 
    We indeed observe that the relaxation time $t_{\rm rel}$ is much shorter compared to quenches to lower values of $U_{\rm fi}$ as in Fig.~\ref{Fig:Case:i}, i.e., the observable $X(t)$ converges more rapidly to its equilibrium value. A completely equivalent relaxation behavior for the same model has been observed in Ref.~\cite{li.de.15}. There a configuration interaction method has been exploited for the calculations which was, however, restricted to much smaller chain lengths due to its higher numerical complexity.
        
    \noindent\underline{(iii) $U_{\rm fi}\!<\! 4$, $V \lesssim 0.7$ (see Fig. \ref{Fig:Case:iii}):} Increasing the final interaction further, we observe a qualitative change in the long-time behavior. 
    More specifically, after the settling time $t_{\rm S}$ the oscillations vanish ($B\!\simeq\!0$) and the observable $X(t)$ decays exponentially to the equilibrium value $X_{\rm equ}$ for $(V_{\rm fi},U_{\rm fi})$.  
    Consequently, no oscillation frequency can be defined. 
    Let us point out that the interval for $U_{\rm fi}$, for which this type of long-time behavior emerges, decreases with increasing $V$ and eventually vanishes for hybridization strengths $V\!\gtrsim\!0.7J$. 
    Regime (iii) is thus characterized by the following set of fitting parameters:
    \begin{equation*}
        B\!\simeq\!0, 
        t_{\rm rel}\!<\!\infty, 
        \lim\limits_{t\rightarrow\infty}X(t)\!=\!A\!=\!X_{\rm equ}, 
        C\!\ne\! 0.
    \end{equation*}
    An example for this type of long-time behavior is provided in Fig.~\ref{Fig:Case:iii}. Although at small times ($t\!<\!t_{\rm S}$), we still observe a limited number of oscillations.
    This oscillatory behavior vanishes for larger times, where only an exponential decay prevails.

    \noindent\underline{(iv) $U_{\rm fi}\!\simeq\!4$ and larger (see Fig. \ref{Fig:Case:iv}):}
    For interactions on the order of the bandwidth of the chain ($U_{\rm fi}\!\sim\!4$), we observe the reappearance of oscillatory behavior for the observable $X(t)$. 
    However, these oscillations appear to be persistent, i.e., $X(t)$ does not decay to its equilibrium value $X_{\rm equ}$ for $t\!\rightarrow\!\infty$. 
    Therefore, the relaxation time approaches infinity ($t_{\rm rel}\!\rightarrow\!\infty$).
    However, in practice due to finite size effects, it is impossible to distinguish between infinitely large and very large but finite $t_{\rm rel}$ in our numerical calculations.
    This makes it difficult to find stable values for $A$ and $C$ in this situation, leading to a less reliable (or even unfeasible) fitting procedure compared to cases (i)-(iii). 
    Moreover, we point out that for strong interactions ($U_{\rm fi} \gtrsim 4J$) the time evolution of the diagonal elements of the density matrix feature a double oscillatory behavior and therefore cannot be described by Eq.~\eqref{Fitting}.
    Since the fitting procedure causes these additional difficulties, we will discuss regime (iv) in more detail and from different perspectives in Sec.~\ref{Discussion}. 
    Keeping the above-mentioned limitations in mind, we can nevertheless characterize regime (iv) by the following set of parameters:
    \begin{equation*}
        B\!\ne\!0, t_{\rm rel}\!=\!\infty.
    \end{equation*}
    An example of this type of long-time behavior is provided in Fig.~\ref{Fig:Case:iv}.
    In the left panel, we indeed observe persistent oscillations for $f(t)$ and $g(t)$ with a single frequency $\omega$ while in the right panel, the diagonal elements of the density matrix feature an alternating oscillation period giving rise to a second oscillation frequency (see insets).
    Let us also mention that such persistent oscillations were found in Ref.~\cite{li.de.15} exploiting a configuration interaction approach for this problem.

We point out that the four regimes described above do not correspond to well-defined phases. 
They have to be understood as a rough discrimination of different regions in the parameter space which are connected by smooth crossovers rather than sharp transitions. 
Furthermore, the exact values of the fitting parameters in Eq.~\eqref{Fitting} can (weakly) depend on the fitting range, i.e., on the value of the settling time $t_{\rm S}$ after which the fit is performed. 
Only region (iv) seems to be more clearly separated from the other ones, although here a finite decay rate $1/t_{\rm rel}$ might not be observed due to the limitations of a finite time and chain length cutoff. 
Therefore, in Sec.~\ref{Discussion} we will analyze our numerical findings [with a particular focus on regime (iv)] in more detail.

Next, we briefly discuss the time evolution of the nearest-neighbor chain correlators $C_{i(i+1)}(t)$ in Fig.~\ref{Fig:Cij_color_t}.
The data are presented as a color-coded two-dimensional heat map where the $x$- and $y$-axis correspond to time $t$ and chain site $i$, respectively.
First, we observe for all four quenches that the changes of nearest-neighbor correlators are zero in the region $t<i/2$.
This indicates a universal propagation velocity $v\!=\!i/t\!=\!2$ of the dynamics induced by the quench at $t=i=0$ which is independent of the final (and also initial) quench parameters.
This observation is explained by the fact that the propagation velocity is exclusively determined by the hopping amplitude $J$ between neighboring chain sites, which is set constant ($J\!=\!1$) in our calculations.

The four different types (i)-(iv) of long-time behavior, which have been discussed above for $f(t)$, $g(t)$ and $\rho_{ii}(t)$ (see Figs.~\ref{Fig:Case:i}-\ref{Fig:Case:iv}), can be also identified for the nearest-neighbor correlators in Fig.~\ref{Fig:Cij_color_t}.
In the left upper panel we show our data for the quench to $(V_{\rm fi}\!=\!0.5,U_{\rm fi}\!=\!0.3)$ corresponding to case~(i) discussed above. 
We can see that the wave front reaches a given chain site $i$ at time $t\!=\!i/2$. 
For later times, the absolute amplitude of the nearest-neighbor correlator decays, but remains sizable.
This indicates a very long relaxation time, which is consistent with the long relaxation times for $f(t)$, $g(t)$ and $\rho_{ii}(t)$ in regime (i) [see Fig.~\ref{Fig:Case:i}].
On the contrary, for the quench to $(V_{\rm fi}\!=\!0.5,U_{\rm fi}\!=\!0.7)$ [Fig.~\ref{Fig:Cij_color_t}, upper right panel] we observe that $C_{i(i+1)}(t)$ rapidly decreases for $t\!>\!i/2$ indicated by the dark (black) color that sets in soon after the wave front has reached the lattice site at $t\!=\!i/2$.
This is indeed the expected behavior for a quench in class (ii) where, also for $f(t)$, $g(t)$ and $\rho_{ii}(t)$, a rapid equilibration is observed in Fig.~\ref{Fig:Case:ii}.
A similar behavior arises for the quench to $(V_{\rm fi}\!=\!0.5,U_{\rm fi}\!=\!2.5)$ in Fig.~\ref{Fig:Cij_color_t} (lower left panel), albeit with slightly longer relaxation time, consistent with the regime (iii) discussed in Fig.~\ref{Fig:Case:iii} for $f(t)$, $g(t)$ and $\rho_{ii}(t)$.
Although the relaxation time for the other two cases in Fig.~\ref{Fig:Cij_color_t} differs only slightly for different lattice sites, we can clearly see in case (iii) that the relaxation time increases significantly with increasing site index $i$.
Finally, for the quench to $(V_{\rm fi}\!=\!0.5,U_{\rm fi}\!=\!4.0)$ in Fig.~\ref{Fig:Cij_color_t} we observe that the maxima and minima of the oscillations are almost constant in time, which is indicated by the uniform color along horizontal lines corresponding to the specific lattices site $i$.
This demonstrates that persistent oscillations also emerge in $C_{i(i+1)}(t)$ consistent with the analogous observations for $f(t)$, $g(t)$ and $\rho_{ii}(t)$.
The minimal relative amplitude of these persistent oscillations is found at $i\!\sim\!500$, which corresponds to the middle of the chain.
This can be indeed expected for a finite system where the amplitude reaches its largest value at the borders (see Fig.~\ref{Fig:Cij_equilibrium} in the Appendix~\ref{App:Correlator}).

Let us also mention that the signal will reach the end of the chain of length $L$ after a time $t\!=\!L/2$. 
There it will be reflected and travel in the opposite direction. 
In this paper, however, we are not interested in these finite-size effects and therefore consider large chains and times below half of the chain length.
For completeness, we demonstrate this reflection in Fig.~\ref{Fig:Cij_color_t}, but will not make any other references to it.

Let us finally briefly address the time evolution of the total energies of the impurity and the chain presented in Fig.~\ref{Fig:Energy_t}. 
There, we depict the time dependence of the respective energy relative to the corresponding energy directly after the quench at $t=0^+$ for the final set of parameters $(V_{\rm fi},U_{\rm fi})$. 
Let us stress that we also include the energy $E_C(t)\!=\!{\cal C}(t)$ defined in Eq.~\eqref{Ct} which corresponds to the constant term in equilibrium, arising in the mean field decoupling of the impurity and the chain [see last term in Eqs.~\eqref{Decoupling}].
Although such a constant contribution represents just an overall energy shift and can be neglected in equilibrium calculations\cite{Neglecting_Constant} it has to be taken into account to preserve energy conservation in the non-equilibrium situation.
From a physical perspective, it corresponds to an explicit coupling energy between the impurity and the chain.
Indeed, all three energy contributions (relative to their value at $t=0^+$), including the term $E_C(t)\!=\!{\cal C}(t)$, add up to zero (purple line in Fig.~\ref{Fig:Energy_t}) at all times, guaranteeing the conservation of energy at each moment of the time evolution.
This is expected as after the quench the Hamiltonian is time-independent, and therefore the total energy must be conserved.
Apart from this, the long-time behavior of the energies follows the corresponding long-term behavior of the mean field parameters $f(t)$ and $g(t)$ as well as the diagonal part of the density matrix $\rho_{ii}(t)$ and correlators $C_{i(i+1)}$.
In particular, for quenches to large values $U_{\rm fi}$ [bottom right panel of Fig.~\ref{Fig:Energy_t}] we observe persistent oscillations in the partial energies as for all other observables.
These oscillations correspond to a (periodic) redistribution of the total energy between the impurity, the chain, and the impurity-chain coupling.

\begin{figure*}[t!]
\includegraphics[width=\textwidth]{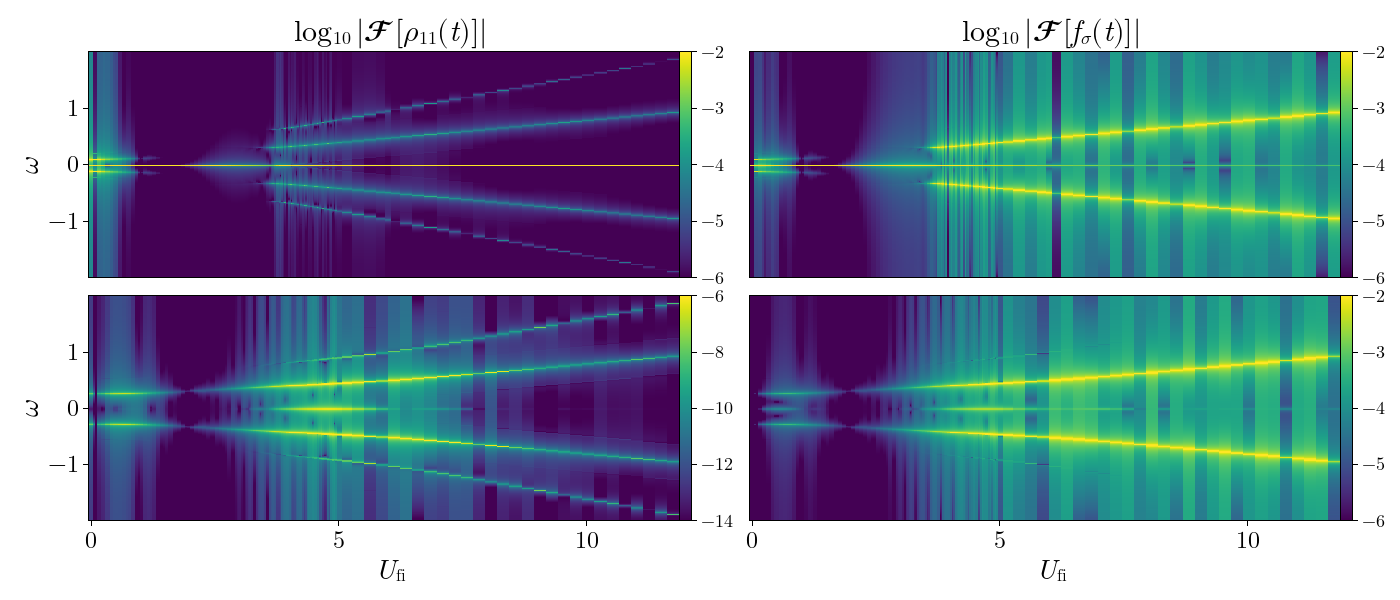}
\caption{Discrete Fourier transform of $\rho_{11}(t)$ [left] and $f(t)$ [right] for $(V_{\rm in}\!=\!V_{\rm fi},U_{\rm in})\!=\!(0.5,1.0)$ [top] and $(V_{\rm in}\!=\!V_{\rm fi},U_{\rm in})\!=\!(1.0,0.1)$ [bottom], respectively, as a function of $U_{\rm fi}$ (horizontal axis) and $\omega$ (vertical axis). 
Data are shown as a two-dimensional color map on a logarithmic scale.}
\label{Fig:Fourier}
\end{figure*}

\section{Discussion}
\label{Discussion}

Let us now investigate the four different regimes of long-time behavior (i)-(iv) defined in the previous section in more detail.
To this end, we have extended our calculations to a wider range of initial parameters $(V_{\rm in},U_{\rm in})$ and analyzed the time evolution of the observables after a quench as a function of the final interaction $U_{\rm fi}$ for $V_{\rm fi}\!=\!V_{\rm in}$.
We have then fitted our numerical data for $f(t)$ and $\rho_{11}(t)$ to Eq.~\eqref{Fitting} and extracted the (inverse) relaxation time $1/t_{\rm rel}$, the oscillation frequency $\omega$, the oscillation amplitude $B$ and the non-oscillatory contribution $C$.
In Fig.~\ref{Fig:Fitting_f} we show these fitting parameters for the mean field $f(t)$ as a function of $U_{\rm fi}$ for the two different initial hybridization strengths $V_{\rm in}\!=\!0.5$ (left panel) and $V_{\rm in}\!=\!1.0$ (right panel), respectively, for three different initial interactions $U_{\rm in}\!=\!0.1$, $0.5$ and $1.0$.
The corresponding analysis in which the parameters are obtained from a fit of $\rho_{11}(t)$ is presented in Appendix~\ref{App:Fitting}.
We observe that the relaxation time $t_{\rm rel}$ and the oscillation frequency $\omega$ (top panels) are independent of the initial interaction strength $U_{\rm in}$ (within the range under investigation). 
The oscillation amplitude $B$ and the magnitude of the non-oscillatory contribution $C$ on the other hand vary with $U_{\rm in}$.
This is a direct consequence of the fact that the time evolution is performed with the Hamiltonian defined by the final parameter values, while the initial values appear only in initial conditions\cite{Schroedinger_equation}.
We can now identify in Fig.~\ref{Fig:Fitting_f} (left panel) the first three regimes (i)-(iii) as well as the intermediate regime (i') of long-term behavior discussed in the previous section.
We indicate the different regimes with different shadings in the figure.
At very low $U_{\rm fi}\!\lesssim\!0.5$ we observe a finite amplitude $C$ of the non-oscillatory contribution and a very large relaxation time $t_{\rm rel}\!>\!100$ ($1/t_{\rm rel}\!<\!0.01$) indicative of regime (i) of final parameters (see Fig.~\ref{Fig:Case:i}). 
Note that this regime cannot be identified from the fit for $\rho_{ii}(t)$ in Appendix~\ref{App:Fitting} (see Fig.~\ref{Fig:Fitting_rho11}) as $C\!\simeq\!0$ at low $U_{\rm fi}$ in this case.
For $0.5\!\lesssim\!U_{\rm fi}\!\lesssim\!1.5$ the parameter $C$ vanishes, while the relaxation time decreases, which corresponds to the regime (ii) of final parameters.
Finally, for $U_{\rm fi}\!\gtrsim \!2$ the amplitude of the oscillation vanishes, and hence the oscillation frequency can no longer be defined consistently in regime (iii), as discussed above.
Let us point out that this regime is absent for $V_{\rm in}\!=\!1$ (right panel in Fig.~\ref{Fig:Fitting_f}) as already mentioned in Sec.~\ref{Results}.
For $V_{\rm in}\!=\!0.5$ regimes (ii) and (iii) are connected by a transition region $1.5\!<\!U_{\rm fi}\!<\!2$ where $C$ rapidly increases from zero to a finite value, while $B$ rapidly drops to zero.
The finite values of $B$ and $C$ are characteristic for regime (i) while the correlation time is much shorter than in (i).
Therefore we have labeled this region between (ii) and (iii) as (i') in Sec.~\ref{Results} as it shares some properties with regime (i).
On the contrary, for $V_{\rm in}\!=\!1$ (right panel of Fig.~\ref{Fig:Fitting_f}) we observe an increase of $C$ at large $U$ with a finite value of $B$ indicating a reappearance of long-time behavior similar to type (i) at stronger coupling. 
However, as for $V_{\rm in}\!=\!0.5$ the relaxation time is larger than in regime (i) which is why we denote this region again as (i').
The results for region (iv) corresponding to $U_{\rm fi}\!>\!3.5$ are not shown in Fig.~\ref{Fig:Fitting_f} (and Fig.~\ref{Fig:Fitting_rho11}) as the fitting procedure becomes unstable for such large final interactions.
We will address this regime in more detail below.
Let us now briefly discuss the $U_{\rm fi}$ dependence of the fitting parameters.
The oscillation amplitude $\omega$ monotonically increases with increasing $U_{\rm fi}$ (as long as it can be defined). 
This is the expected behavior, as oscillation frequencies are usually related to the eigen energies in a quantum mechanical system. 
Therefore, an overall increase in the energy scales induced by the increase in $U_{\rm fi}$ will lead to faster oscillations.
On the other hand, the inverse of the relaxation time $1/t_{\rm rel}$ increases for small values of $U_{\rm fi}$, reaches a maximum at $U_{\rm fi}\!\sim\!1.5$ and then decreases monotonously to zero.
This observation can be explained by considering the limiting cases of $U_{\rm fi}\!=\!0$ and $U_{\rm fi}\!\rightarrow\!\infty$. 
For the non-interacting situation, the impurity site becomes just an additional non-interacting lattice site. 
In this case, the particles propagate freely through the chain and cannot redistribute the energy they had before the quench due to the interaction, corresponding to an infinite relaxation time and consequently $1/t_{\rm rel}$ becomes zero. 
For $U_{\rm fi}\!\rightarrow\!\infty$ on the other hand, the energy added at the impurity becomes too large and cannot be completely dissipated to the chain. 
Therefore, the system is trapped in a non-equilibrium state which does not decay. 
In this situation, we also expect an infinite relaxation time, i.e.~$1/t_{\rm rel}\!=\!0$. 
Since in both limits $1/t_{\rm rel}\!=\!0$ we expect a maximum of this quantity (that is, a minimal relaxation time) to emerge for some finite $U_{\rm fi}$. 
The value $U_{\rm fi}\!\sim\!1.5$ where this maximum appears is close to the half bandwidth $W/2\!=\!2$ of the non-interacting chain which marks the transition from a weakly coupled to a moderately correlated regime.
Let us also briefly mention the tiny dip right after the maximum of $1/t_{\rm rel}$ for $V_{\rm in}\!=\!1$ in the right panel of Fig.~\ref{Fig:Fitting_f}.
This feature seems to be a consequence of the absence of the non-oscillatory region (iii) for this specific set of initial parameters.
It emerges at $U_{\rm fi}\!\sim\!2.1$ where the system unsuccessfully tries to turn to regime (iii) indicated by a minimum in the oscillation amplitude $B$.
The amplitude $C$ of the non-oscillatory drift is finite for small $U_{\rm fi}$ and decays to zero until it reappears again for larger values of the final interaction. 
However, since this quantity depends on the initial set of parameters, it is difficult to find a clear-cut physical interpretation.
Moreover, for small final interactions, a finite value of $C$ is observed only for fits of the mean field parameters $f(t)$ but is absent for $\rho_{11}(t)$, as is apparent in Fig.~\ref{Fig:Fitting_rho11} in Appendix~\ref{App:Fitting}.
The scattering amplitude $B$ depends even more strongly on $(V_{\rm in},U_{\rm in})$.
It has the peculiar feature that it vanishes continuously at $(V_{\rm in},U_{\rm in})\!=\!(V_{\rm fi},U_{\rm fi})$ when no quench is performed (while for all other parameters, this corresponds just to a singular point at which the fit is not well defined).
The most striking feature is the disappearance of $B$ at roughly $U_{\rm fi}\! \sim \!2$ for $V_{\rm in}\! \sim \!0.5$ where the oscillations cease (left panel in Fig.~\ref{Fig:Fitting_f}).
It first seems that the same happens for $V_{\rm in}\!=\!1$ (right panel in Fig.~\ref{Fig:Fitting_f}), however, after a rapid drop of $B$ around $U_{\rm fi}\!=\!2.0$ it starts to grow again, and hence oscillations prevail in the entire region of final interactions.

To support the validity of our fitting procedure and the classification of the long time behavior into four different classes we have also performed (discrete) Fourier transforms\cite{is.mu.11B} of our observables.
In Fig.~\ref{Fig:Fourier} we show the Fourier transforms of $\rho_{11}(t)$ [left panels] and $f(t)$ [right panels] for two different sets of initial parameters (top and bottom panels) specified in the caption of the figure. 
The data are presented in a color-coded form on a logarithmic scale as a function of the final interaction $U_{\rm fi}$ and the frequency $\omega$.
Bright features in the figure indicate the presence of well-defined oscillations, and their locations on the $y$-axis correspond to the oscillation frequency.
For small values of $U_{\rm fi}$ ($0\!<\!U_{\rm fi}\!\lesssim\! 2.0$) we can indeed identify one bright line at a small (positive) frequency $\omega$ (and, of course, its mirror image in the negative frequency region) corresponding to finite oscillations compatible with regimes (i) and (ii).
Let us stress that the Fourier transform does not allow us to distinguish between these two regimes.
Upon increasing $U_{\rm fi}$ to $2.0\!\lesssim\!U_{\rm fi}\!\lesssim\! 4$ these bright lines disappear for $V_{\rm in}\!=\!0.5$ (top panels) which signals the onset of regime (iii) where no oscillations can be observed.
Consistent with our previous discussion, this is not the case for $V_{\rm in}\!=\!1.0$ (bottom panel) where the bright lines at a single frequency prevail within this parameter region.
We also want to mention that the Fourier transform also captures the increase of the oscillation frequency with the final interaction that we have already noticed in Fig.~\ref{Fig:Fitting_f}.

Let us now turn our attention to the strong coupling regime $U_{\rm fi}\!\gtrsim\!4$ which we could not analyze previously due to instabilities in the fitting procedure. 
In the Fourier transform we observe for this parameter region a (re)appearance of strong oscillations indicated by the return of bright lines.
Interestingly, for $\rho_{11}(t)$ two lines emerge at positive frequencies corresponding to the double oscillations of this quantity at large final interaction discussed in Sec.~\ref{Results}.
Moreover, the oscillation frequency increases further with increasing $U_{\rm fi}$ as expected for systems with higher energies.

\begin{figure}[t!]
\centering \includegraphics[width=0.45\textwidth]{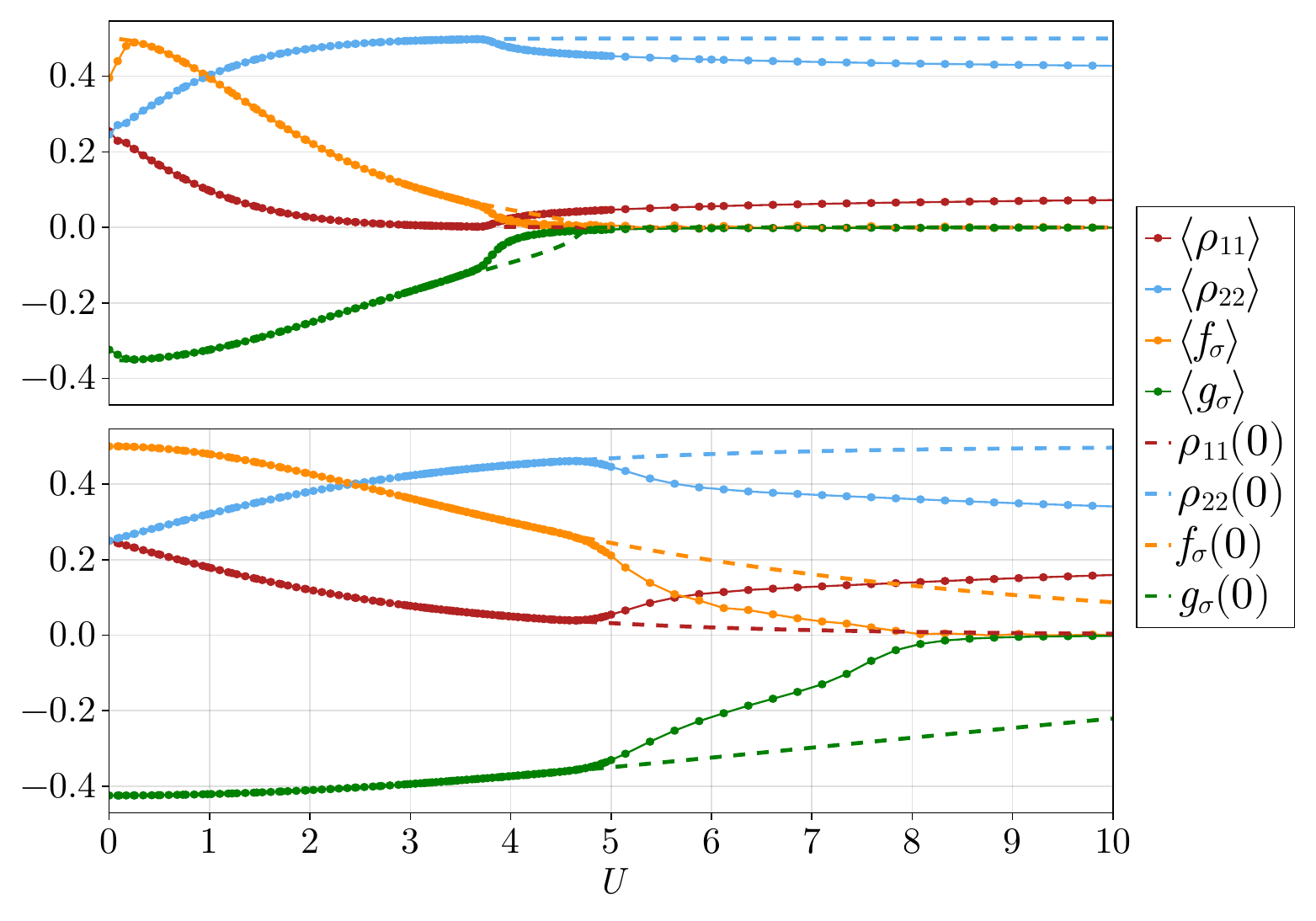}
\caption{$\langle f_\sigma^{\phantom*} \rangle$, $\langle g_\sigma^{\phantom*} \rangle$, $\langle \rho_{11}^{\phantom*} \rangle$, and $\langle \rho_{22}^{\phantom*} \rangle$ as a function of $U_{\rm fi}$ for fixed value of $(V_{\rm in}=V_{\rm fi}=0.5,U_{\rm in}=1.0)$ (upper panel) and $(V_{\rm in}=V_{\rm fi}=1.0,U_{\rm in}=0.1)$ (lower panel).
Dashed lines indicate the corresponding equilibrium values for the final set of parameters.}
\label{Fig:average_X}
\end{figure}

\subsection{Non-equilibrium phase transition}
\label{PDNonequ}

Our numerical data in Fig.\ref{Fig:Case:iv} suggest that the oscillations observed at strong coupling ($U_{\rm fi}\!\gtrsim\!4.0$) are persistent, i.e., they do not decay to the equilibrium value corresponding to the final set of parameters.
This might indicate the emergence a non-equilibrium dynamic phase transition between a region of final parameters where the system thermalizes after a quench at low $U_{\rm fi}$ and a region where the system is trapped in a non-thermal state at large $U_{\rm fi}$.
To address this question in more detail we introduce the following time average of an observable $X(t)$ over $m$ oscillation periods $T\!=\!2\pi/\omega$ at very large times~$t\!\gg\!t_{\rm S}$:
\begin{equation}\label{average_X_def}
    \langle X \rangle =\lim_{t \rightarrow \infty} \bar{X}(t) =\lim_{t \rightarrow \infty} \frac{1}{Tm}\int_{t-\frac{1}{2}Tm}^{t+\frac{1}{2}Tm} \dd \tau X(\tau) \,.
\end{equation}
When $X(t)$ converges to its equilibrium value as in regions (i)-(iii) then, trivially, $\langle X \rangle\!=\!X_{\rm equ}$. 
Moreover, $\bar{X}(t)$ converges faster to $X_{\rm equ}$ than $X(t)$, for example in region (ii) taking almost immediately the limiting value.
This motivates us to use in region (iv) $\langle X \rangle$ as a proxy for the fully converged values of $\lim\limits_{t \to \infty} X(t)$ in regimes (i)-(iii).
In Fig.~\ref{Fig:average_X} we show the large time averages for $f(t)$, $g(t)$, $\rho_{11}(t)$ and $\rho_{22}(t)$ as well as their corresponding equilibrium values for $(V_{\rm fi},U_{\rm fi})$ as a function of $U_{\rm fi}$ for two different sets of initial parameters detailed in the caption of the figure.
At the onset of regime (iv) where persistent oscillations emerge [at $U_{\rm fi}\!\sim\!3.75$ for $(V_{\rm in}=V_{\rm fi}=0.5,U_{\rm in}=1.0)$ in the upper panel and at $U_{\rm fi}\!\sim\!4.8$ for $(V_{\rm in}=V_{\rm fi}=1.0,U_{\rm in}=0.1)$ in the lower panel, respectively] we can see that $\langle X\rangle$ starts to deviate from the associated equilibrium value $X_{\rm equ}$ in Fig.~\ref{Fig:average_X}.
The value of $U_{\rm fi}$ where this happens is smaller than the corresponding interaction strength of the equilibrium phase transition indicated by $f\!=\!g=\!0$ for the given value of $V_{\rm fi}$.
Hence, this might be interpreted as a dynamical phase transition that separates the regions where equilibration to a thermal state can be observed or is absent.
For even larger values of $U_{\rm fi}$ $\langle f(t)\rangle$ and $\langle g(t)\rangle$ eventually vanish.
Let us, however, stress that this does not mean that the system thermalizes since $f(t)$ and $g(t)$ still oscillate around zero although their averages vanish.
This is different for $\rho_{11}(t)$ and $\rho_{22}(t)$ whose time averages do not converge to their corresponding equilibrium values for large $U_{\rm fi}$ given by $\rho_{11}\!=\!0$ and $\rho_{22}\!=\!0.5$.
This is a further indication that the system is indeed trapped in a non-thermal state. 

{However, it is worth mentioning} that the above results depend more or less strongly on the initial set of parameters and the chain length.
That makes it indeed difficult to verify the emergence of a dynamic phase transition numerically. In fact, it is indeed possible that oscillations that appear to be persistent for a given chain length $L$ decay for longer chains.
To this end we have performed a finite size scaling of the $\langle\rho_{11}\rangle$ in Fig.~\ref{Fig:scaling_rho11} in Appendix~\ref{App:Finite-size-scaling:Time_evolution}.
These results indeed indicate that the dynamic phase transition is not just an artifact of the finite chain length but should also prevail when approaching the thermodynamic limit. 

Let us finally discuss whether the observed non-equilibrium phase phase transition is a physical property of the system or a mere feature of the mean field treatment which we use for our calculations.
In this respect let us first stress that a similar dynamical phase transition has also been observed in the SIAM using configuration interaction method\cite{li.de.15} which goes beyond a mere mean field ansatz.
However, the times which can be reached by this approach are limited (as it is the case for many other more involved or exact techniques) which prevents us from drawing definite conclusions for the long-time behavior of the system on long chains.
In fact, persistent oscillations and the associated sharp phase transitions are rarely found in exact solutions for correlated many-body systems and are typically features of mean field theories. 
Such transitions have been indeed found for a number of models exploiting different flavors of mean field approaches, for instance, in the large-component-number limit of the generalized Bose-Anderson model\cite{ch.ri.16,ch.ru.17} or in the fermionic Hubbard model treated with a Gutzwiller wave function ansatz\cite{sc.fa.10, sc.fa.11, sa.sc.12, sa.fa.13}
The, in principle, spurious mean field phase transitions obtained by such approaches can indicate a (more or less sharp) crossover between different relaxation behaviors in solutions provided by more sophisticated methods. 
A good example is the time-dependent Gutzwiller approach for the Hubbard model in Refs. \cite{sc.fa.10, sc.fa.11}, where a sharp dynamical phase transition is observed between quenches to interaction values $U_{\rm fi}$ below and above a critical value of interaction. 
Within this approximation, the two regimes of interaction values are characterized by qualitatively different persistent oscillations, which are characteristic of the mean field-like approach. 
In fact, in the more advanced dynamical mean field theory (DMFT) treatment of the problem in Ref. \cite{ec.ko.09}, no such persistent oscillations are found. 
However, the phase transition predicted by the mean field-like Gutzwiller approach is reflected in a sharp crossover between two different relaxation regimes. 
This demonstrates that mean field methods can capture specific features of exact solutions, although these features may manifest themselves differently compared to numerically exact results.
Therefore, we believe that the dynamical phase transition observed in our calculations is a genuine property of the system.

{Additionally, we observe that before the transition, the diagonal elements of the density matrix $\rho_{ii}$ oscillate with a single frequency, whereas after the transition, their pattern features two distinct frequencies.}
Moreover, before the dynamical phase transition, the average values of different observables ($\langle \rho_{ii} \rangle$, $\langle f \rangle$, and $\langle g \rangle$) taken over several oscillation periods for large times match their equilibrium values for the final interaction strength after the quench. 
On the contrary, after the transition, these averages differ from the corresponding equilibrium values for $U_{\rm fi}$ (see Fig.~\ref{Fig:average_X}).  
An analogous behavior has also been reported in the aforementioned DMFT calculations for the 2D Hubbard model \cite{ec.ko.09}. 
Thus, our paper may serve as a prompt for researchers studying the time evolution of the system with more accurate methods like NRG and DMRG to capture this dynamical phase transition or a related crossover.

\section{Conclusions  and Outlook}\label{Conclusions}

We have presented a new method to study the SIAM at equilibrium as well as its time evolution after a quench of the impurity interaction and/or the hybridization of the impurity with the non-interacting chain.
The idea of our approach is the mean field decoupling of the impurity and the chain which is achieved by introducing a pair of auxiliary Majorana fermions between theses two parts of the model. 
This leads to a set of (algebraic or differential) equations for the impurity and the chain which are coupled by specific mean field parameters.

First, we solved the problem in equilibrium where we are concerned with an algebraic system of equations for the mean field parameters.
For small interactions (w.r.t. the hybridization) we find finite values for the mean fields indicative of the mixed-valence regime of the SIAM.
On the contrary, for large interactions (w.r.t. the hybridization) the mean field parameters vanish and, hence, a well-defined local moment emerges at the impurity.
In this phase, no charge fluctuations appear and the impurity is occupied either by a fermion with spin up or by a fermion with spin down.
Let us however mention that the sharp transition between these two phases is a mean field feature while the exact solution should exhibit a smooth crossover between the two regimes.
In particular, while the impurity spin susceptibility will diverge in the local moment regime obtained from the mean field approach while it remains finite (albeit very large) within the numerically exact NRG treatment.

In the next step, we studied the time evolution of the system after a quench of the hybridization between the impurity and the chain and/or the Hubbard interaction at the impurity. 
This requires the solution of a set of coupled differential equations for the mean field parameters $f$ and $g$, density matrix elements $\rho_{ij}$, and correlators $C_{ij}$ and $K_{ij}$.
We have identified four different types of long time behavior for selected observables which correspond to certain regimes for the final interaction strength and the final hybridization after the quench while the dependence on the initial parameters is mostly insignificant.
For weak to intermediate interactions, i.e., in regions (i)-(iii), all quantities converge to their equilibrium results for the final set of parameters. 
On the contrary, for large final interactions, i.e.~in region (iv), persistent oscillations emerge in the system and the observables do not relax to their equilibrium values.
We speculate that the border between the region where the system equilibrates and the regions where persistent oscillations prevail corresponds to a dynamic phase transition.
To gain more evidence for the existence of such a dynamic phase transition, we have studied the long-term averages of the mean field parameters ($\langle f \rangle$ and $\langle g \rangle$) and the diagonal elements of the density matrix ($\langle \rho_{11} \rangle$ and $\langle \rho_{22} \rangle$). 
We observe that these quantities deviate from their equilibrium values for a final set of parameters after some critical interaction,  which coincides exactly with the interaction strength where persistent oscillations emerge.
This further supports our assertion of a dynamical phase transition between a region where the systems thermalizes and a regime where persistent oscillations emerge and the system is trapped in a non-thermal state.
Finally, it is worth noting that while persistent oscillations and related sharp phase transitions are likely artifacts of the mean field treatment of the problem, they may indicate an approximate integral of motion in exact solutions. This phenomenon has been observed in several other models and mean field techniques in the literature.
Hence, while our description of the weak to moderate correlated regimes is correct at all time scales, for large final interaction our approximation is fully valid only at moderate time scales
whereas persistent oscillations and related sharp phase transitions are likely artifacts of the mean field treatment of the problem.
 
In a next step we plan to investigate impurity models out of half filling and also consider quenches of the impurity onsite energy.
Moreover, our method can be straightforwardly generalized to the two impurity Anderson model (TIAM)\cite{al.an.64, ti.sc.12} and the dilute periodic Anderson model (DPAM)\cite{co.mu.88, sc.ti.13} to study local and nonlocal susceptibilities out of equilibrium or the crossover between the RKKY and the inverse indirect magnetic exchange regime \cite{sc.ti.13}, respectively. 

\section*{Acknowledgments}
We thank M.~Eckstein, M.~Potthoff, and H.~Strand for useful discussions. We acknowledge financial support from the Deutsche Forschungsgemeinschaft (DFG) through Projects No.~407372336 and No.~449872909. The work was
supported by the North-German Supercomputing Alliance (HLRN).


\begin{appendix}

\section{Mapping to the Spin Hamiltonian}
\label{App:Mapping_to_Spin}

In this appendix, we discuss the equivalence of our decoupled fermionic model in Eqs.~\eqref{Hamiltonian_Decoupled} and a corresponding decoupled spin chain under a Jordan Wigner transform.
This justifies our decoupling procedure based on the introduction of auxiliary Majorana fermions between the impurity and the non-interacting chain.
We consider the case where the impurity is coupled to the first site of the chain with OBC. We use the Jordan-Wigner transformation
\begin{subequations}
\label{App:Jordan-Wigner}
\begin{align}
\label{App:Jordan-Wigner_s-}
&\hat s_{i,\sigma}^{-} =\left\{
\begin{array}{lll}
\hat d_{\sigma}^{\phantom\dagger} &\quad&i=0\\
\hat c_{i,\sigma}^{\phantom\dagger} \exp \left[{ i\pi \left(\sum\limits_{j=1}^{i-1} \hat c_{j,\sigma}^{\dagger}\hat c_{j,\sigma}^{\phantom\dagger} + \hat d_{\sigma}^{\dagger}\hat d_{\sigma}^{\phantom\dagger}
\right)}\right]
&&i>0
\end{array}
\right.
\\
\label{App:Jordan-Wigner_sz}
&\hat s_{i,\sigma}^{z} =\left\{
\begin{array}{lll}
\hat d_{\sigma}^{\dagger}\hat d_{\sigma}^{\phantom\dagger} - \frac{1}{2} &\quad&i=0\\
\hat c_{i,\sigma}^{\dagger}\hat c_{i,\sigma}^{\phantom\dagger} - \frac{1}{2} &     &i>0
\end{array}
\right.
\\
\label{App:Jordan-Wigner_d}
&\hat d_{\sigma}^{\phantom\dagger} = \hat s_{0,\sigma}^{-}
\\
\label{App:Jordan-Wigner_c}
&\hat c_{i,\sigma}^{\phantom\dagger} =\hat s_{i,\sigma}^{-}e^{-i\pi \sum\limits_{j=0}^{i-1}\left(\hat s_{j,\sigma}^{z}+\frac{1}{2}\right)}
\end{align}
\end{subequations}
to rewrite our original fermionic Hamiltonian in Eq.~\eqref{Hamiltonian} in terms of the spin Hamiltonian:
%
\begin{subequations}
\label{spinhamiltonian}
\begin{align}
&{\cal H}=\sum_\sigma {\cal H}_{0,\sigma} +\sum_\sigma\left[{\cal H}_{1,\sigma} + {\cal H}_{01,\sigma} \right]
+{\cal C}_S
\\
&{\cal H}_0=U \hat s_{0,\uparrow}^{z} \hat s_{0,\downarrow}^{z} -\sum_\sigma h_{0,\sigma}s_{0,\sigma}^z
\\
&{\cal H}_{1,\sigma}=\sum_{i=1}^L \left[J_{i,\sigma}\left(\hat s_{i,\sigma}^{+}\hat s_{i+1,\sigma}^{-} + \hat s_{i+1,\sigma}^{+}\hat s_{i,\sigma}^{-}\right) - h_{i,\sigma}s_{i,\sigma}^z\right]
\\
&{\cal H}_{01,\sigma}=V_\sigma \left(\hat s_{0,\sigma}^{+}\hat s_{1,\sigma}^{-} + \hat s_{1,\sigma}^{+}\hat s_{0,\sigma}^{-}\right)\label{spinhamiltoniancoupled}
\end{align}
\end{subequations}
Here
\begin{align}
h_{i,\sigma} =\left\{
\begin{array}{lll}
\mu_\sigma -\varepsilon_{d,\sigma} - \frac{1}{2}U &\quad&i=0\\
\mu_\sigma -\varepsilon_{i,\sigma} &     &i>0
\end{array}
\right.
\end{align}
and
\begin{align}
&{\cal C}_{S}=\frac{1}{4}U +
\sum_\sigma\left[ \frac{L+1}{2}\mu_\sigma -\frac{1}{2}\varepsilon_{d,\sigma}
-\frac{1}{2}\sum_{i=1}^L \varepsilon_{i,\sigma}\right]
\end{align}
Note that we have here two types of spins $\hat{s}_{i,\sigma}$ which interact only at the impurity site $i\!=\!0$. Moreover, after the transformation, the chain does not correspond to a non-interaction system since neighboring spins of the same type interact via an exchange coupling $J$.

The next step is the mean field decoupling of ${\cal H}_{01,\sigma}$.
For this purpose, we express the spin ladder operators in terms of their average (expectation) values plus a fluctuation term, i.e.
$
\hat s_{i,\sigma}^{\pm}=\langle \hat s_{i,\sigma}^{\pm} \rangle + \delta \hat s_{i,\sigma}^{\pm}
$.
Inserting this expression into ${\cal H}_{01,\sigma}$ in Eq.~\eqref{spinhamiltoniancoupled} and omitting the terms corresponding to the square of the fluctuation, we obtain
\begin{align}
{\cal H}_{01,\sigma}& \simeq V_\sigma \Bigl(\langle \hat s_{0,\sigma}^{+} \rangle \delta\hat s_{1,\sigma}^{-}
+ \delta \hat s_{0,\sigma}^{+} \langle \hat s_{1,\sigma}^{-} \rangle
+\langle \hat s_{0,\sigma}^{+} \rangle \langle \hat s_{1,\sigma}^{-} \rangle
+ \langle \hat s_{1,\sigma}^{+}\rangle \delta\hat s_{0,\sigma}^{-}
+ \delta\hat s_{1,\sigma}^{+}\langle \hat s_{0,\sigma}^{-} \rangle
+\langle \hat s_{1,\sigma}^{+}\rangle \langle \hat s_{0,\sigma}^{-} \rangle
\Bigl)
\nonumber\\
&=V_\sigma \Bigl(\langle \hat s_{0,\sigma}^{+} \rangle \hat s_{1,\sigma}^{-}
+ \hat s_{0,\sigma}^{+} \langle \hat s_{1,\sigma}^{-} \rangle
+ \langle \hat s_{1,\sigma}^{+}\rangle \hat s_{0,\sigma}^{-}
+ \hat s_{1,\sigma}^{+}\langle \hat s_{0,\sigma}^{-} \rangle
-\langle \hat s_{0,\sigma}^{+} \rangle \langle \hat s_{1,\sigma}^{-} \rangle
- \langle \hat s_{1,\sigma}^{+}\rangle \langle \hat s_{0,\sigma}^{-} \rangle \Bigl) \,.
\end{align}

After this mean field decoupling, we obtain the following mean field Hamiltonian
\begin{subequations}
\label{meanfieldspin}
\begin{align}
&{\cal H}={\cal H}_\mathrm{imp} + \sum_\sigma {\cal H}_{\mathrm{chain},\sigma} + {\cal C}_{S}
\\
\label{App:Himp_S}
&{\cal H}_\mathrm{imp}={\cal H}_0 + V_\sigma \Bigl( \langle \hat s_{1,\sigma}^{-} \rangle \hat s_{0,\sigma}^{+}
+ \langle \hat s_{1,\sigma}^{+}\rangle \hat s_{0,\sigma}^{-}\Bigl) \\
\label{App:Hchain_S}
&{\cal H}_{\mathrm{chain},\sigma}={\cal H}_1 + V_\sigma \Bigl( \langle \hat s_{0,\sigma}^{-} \rangle \hat s_{1,\sigma}^{+}
+ \langle \hat s_{0,\sigma}^{+}\rangle \hat s_{1,\sigma}^{-}\Bigl)
\end{align}
\end{subequations}
where the expectation values $\langle\hat{s}_{0,\sigma}^+\rangle$ and $\langle\hat{s}_{1,\sigma}^+\rangle$ have to be determined self-consistently.

As the next step, we demonstrate that the mean field Hamiltonian for the spin system defined in Eqs.\eqref{meanfieldspin} is equivalent to the decoupled fermionic Hamiltonian in Eqs.\eqref{Hamiltonian_Decoupled}.
To achieve this, we represent both Hamiltonians in their corresponding spin and occupation number bases, respectively.

First, we discuss the bases for the Hamiltonians describing the impurity, i.e., for Eqs.~\eqref{App:Himp_S}
and \eqref{Hamiltonian_impurity_Decoupled}. The Hilbert space of the spin Hamiltonian \eqref{App:Himp_S} is of dimension $4$, and its basis is given by:
\begin{align}
\label{App:Spin_Basis_Impurity}
|1\rangle =|\uparrow;\uparrow\rangle, 
|2\rangle =|\uparrow;\downarrow\rangle, 
|3\rangle =|\downarrow;\uparrow\rangle, 
|4\rangle =|\downarrow;\downarrow\rangle.
\end{align}
In contrast, the Hilbert space of the fermionic Hamiltonian \eqref{Hamiltonian_impurity_Decoupled} has a dimension of $16$. However, as will be discussed in detail in the next section (Sec.~\ref{App:Solving_the_equilibrium_problem_for_the_impurity}), it has a block diagonal structure. Due to the behavior of the Majorana fermion $\hat \gamma_\sigma^{\dagger} =\hat \gamma_\sigma^{\phantom\dagger}$, all of these blocks are equivalent. The basis for one of these sectors is given by Eq.~\eqref{App:Basis_Impurity}, where the first two (last two) numbers describe spin up (spin down) states for Majorana and Dirac fermions, respectively.
By performing simple calculations in their corresponding bases mentioned above, we find that both Hamiltonians yield the same Hamiltonian matrix. Consequently, $\langle \hat\gamma_\sigma \hat d_\sigma \rangle$ and $\langle \hat s_{0,\sigma}^{-} \rangle$ will give the same results. This indicates that, after mean field decoupling, our fermionic Hamiltonian properly describes the behavior of the impurity.

Now we discuss the bases for the Hamiltonians describing the chain, i.e., for Eqs.~\eqref{App:Hchain_S} and \eqref{Hamiltonian_chain_Decoupled_sigma}. The basis of the spin Hamiltonian \eqref{App:Hchain_S} is given by:
\begin{equation}
\label{App:Basis_Spin_Chain}
|\Psi_{\sigma,i}^{\rm spin}\rangle = |s_{1,\sigma}, s_{2,\sigma}, \ldots, s_{L,\sigma}\rangle \,.
\end{equation}
Here, $s_{i,\sigma}=\uparrow, \downarrow$ describes the spin state at the $i$-th site of the chain for the $\sigma$ branch.

For the fermionic Hamiltonian, we consider the following basis:

\begin{equation}
\label{App:Basis_Chain}
|\Psi_{\sigma,i,\pm}\rangle = \frac{1}{2}\left(|0\rangle|\psi_{\sigma,i}\rangle \pm |1\rangle|\psi_{\sigma,i}\rangle \right) = |\pm\rangle |\psi_{\sigma,i}\rangle ,.
\end{equation}

Here, $|0\rangle$,  $|1\rangle$, and $|\pm\rangle$ describe Majorana fermions, while
\begin{equation}
\label{App:Basis_Chain_dirac}
|\psi_{\sigma,i}\rangle = |n_{1,\sigma}, n_{2,\sigma}, \ldots, n_{L,\sigma}\rangle ,
\end{equation}
describes Dirac fermions on the chain. Here $n_{i,\sigma}=0,1$ describes the number of spin $\sigma$ fermions at site $i$.

We note that $|\pm\rangle$ states are eigenvectors for the Majorana fermion with eigenvalues $\pm 1$. Therefore, the Hilbert space of the Hamiltonian \eqref{Hamiltonian_chain_Decoupled_sigma} has a block diagonal structure corresponding to the $|+\rangle$ and $|-\rangle$ states of Majorana fermions, and these two blocks are equivalent to each other. Due to the fact that there is only hopping between neighboring lattice sites and open boundary conditions, statistical differences between spin and fermionic systems will not arise. Direct calculations show that the Hamiltonian matrix corresponding to the spin Hamiltonian \eqref{App:Hchain_S} in the basis given by Eq.~\eqref{App:Basis_Spin_Chain} is equivalent to one of the blocks of the Hamiltonian matrix for the fermionic Hamiltonian \eqref{Hamiltonian_chain_Decoupled_sigma} in the basis given by Eq.\eqref{App:Basis_Chain}. Consequently, $\langle \hat\gamma_\sigma \hat c_{1,\sigma} \rangle$ and $\langle \hat s_{1,\sigma}^{-} \rangle$ will yield the same results. This indicates that, after mean field decoupling, our fermionic Hamiltonian also properly describes the behavior of the chain.

To conclude, the results obtained from the mean field Hamiltonian for fermions, given by Eqs.\eqref{Hamiltonian_Decoupled}, are consistent with those from the corresponding mean field spin Hamiltonian, given by Eqs.\eqref{meanfieldspin}. Since the mean field decoupling is well justified for spins it should be also a reasonable approximation in our fermionic system.



\section{Solving the equilibrium problem}
\label{App:Solving_the_equilibrium_problem}

\subsection{Solving the equilibrium problem for the impurity}
\label{App:Solving_the_equilibrium_problem_for_the_impurity}

The basis vectors of the impurity Hamiltonian, Eq.~\eqref{Hamiltonian_impurity_Decoupled}, are given by
\begin{equation}
\label{App:Impurity_vectors}
|m_{\uparrow}, n_{d\uparrow}; m_{\downarrow}, n_{d\downarrow} \rangle = (\hat a_\uparrow)^{m_{\uparrow}}
(\hat d_{\uparrow}^{\dagger})^{n_{{d}\uparrow}}
(\hat a_\downarrow)^{m_{\downarrow}}
(\hat d_{\downarrow}^{\dagger})^{n_{{d}\downarrow}}
|0\rangle
\end{equation}
and the size of its Hilbert space is $16$.
Here $m_\sigma$ and $n_{{d}\sigma}$ are eigenvalues of the operators $\hat a_{\sigma}^{\dagger} \hat a_{\sigma}^{\phantom\dagger}$ and $\hat d_{\sigma}^{\dagger} \hat d_{\sigma}^{\phantom\dagger}$ and describe the number of Majorana and Dirac fermions with spin $\sigma$, respectively. Based on the Pauli principle, they can take the values $0$ and $1$.
Due to the following commutation relation for the Hamiltonian Eq.~\eqref{Hamiltonian_impurity_Decoupled}
\begin{equation}
\label{App:Conservation_Impurity}
\left[e^{i\pi\left(\hat d_\sigma^{\dagger}\hat d_\sigma^{\phantom\dagger} + \hat a_\sigma^{\dagger}\hat a_\sigma^{\phantom\dagger}\right)} , \hat {\cal H}_\mathrm{imp} \right]=0
\end{equation}
its Hilbert space has a block diagonal structure. 
Each block is characterized by the parity of the total number of particles ($m_{\sigma} + n_{{d}\sigma}$)
for up and down spin fermions. 
Accordingly, we have four different blocks: 
($O\,,O$), ($O\,,E$), ($E\,,O$), ($E\,,\,E$). 
Here, $O$ and $E$ correspond to the odd and even total number of particles.
The first letter stands for up spin fermions and the second letter for down spin fermions.
Due to the behavior of the Majorana fermion, that $\hat\gamma_\sigma^{\dagger}=\hat\gamma_\sigma^{\phantom\dagger}$, all of these blocks perform equivalently.
Therefore, without loss of generality, we consider the block ($O\,,O$).
So we consider the following basis
\begin{align}
\label{App:Basis_Impurity}
\begin{array}{cc}
|1\rangle =|0,1;0,1\rangle\,,\,
&|2\rangle =|0,1;1,0\rangle\,,\,\\
|3\rangle =|1,0;0,1\rangle\,,\,
&|4\rangle =|1,0;1,0\rangle\,.
\end{array}
\end{align}
In this basis, the Hamiltonian Eq.~\eqref{Hamiltonian_impurity_Decoupled} in the matrix form reads
\begin{equation}
\label{App:Impurity_Hamiltonian_Matrix}
\hat {\cal H}_\mathrm{imp} =
\left(
\begin{array}{cccc}
 U+\sum\limits_\sigma(\varepsilon_{d,\sigma}-\mu_{\sigma}) & V_\downarrow g_\downarrow^{\phantom*}(t) & V_\uparrow g_\uparrow^{\phantom*}(t) & 0 \\
 V_\downarrow g_\downarrow^{*}(t)& \varepsilon_{d,\uparrow}-\mu_{\uparrow} & 0 & V_\uparrow g_\uparrow^{\phantom*}(t)  \\
 V_\uparrow g_\uparrow^{*}(t)& 0 & \varepsilon_{d,\downarrow}-\mu_{\downarrow} & V_\downarrow g_\downarrow^{\phantom*}(t)  \\
 0 & V_\uparrow g_\uparrow^{*}(t) & V_\downarrow g_\downarrow^{*}(t) & 0
\end{array}
\right)
\end{equation}

The Hamiltonian matrix Eq.~\eqref{App:Impurity_Hamiltonian_Matrix} is $4 \times 4$, so it is very straightforward to find its eigenvectors $|\Psi_l \rangle$ and eigenvalues ${\cal E}_{l,\sigma}$. In particular, we are interested in the ground state
\begin{align}
|\mathrm{GS}\rangle = \alpha_1|1\rangle + \alpha_2|2\rangle + \alpha_3|3\rangle + \alpha_4|4\rangle \,.
\end{align}

To study the ground state properties of the impurity as well as its time evolution, we use the density matrix
\begin{equation}
\label{App:Impurity_Density_Operator}
\hat \rho (t)=\sum_{i,j}\rho_{ij}|i\rangle\langle j|=
\left(
\begin{array}{cccc}
\rho_{11}(t) & \rho_{12}(t) & \rho_{13}(t) & \rho_{14}(t) \\
 \rho_{21}(t) & \rho_{22}(t) & \rho_{23}(t) & \rho_{24}(t) \\
 \rho_{31}(t) & \rho_{32}(t) & \rho_{33}(t) & \rho_{34}(t) \\
 \rho_{41}(t) & \rho_{42}(t) & \rho_{43}(t) & \rho_{44}(t)
\end{array}
\right) \,.
\end{equation}
Here
\begin{equation}
\label{App:Matrix_operatoprs_for_Impurity_T0}
\rho_{ij}(0)= \langle \mathrm{GS}|\hat \rho |\mathrm{GS} \rangle = \alpha_{i}^{\phantom*}\alpha_{j}^{*} \,.
\end{equation}
We are not interested in time dependence in this subsection. 
Here, and in the remainder of this subsection, we introduce time dependence for the sake of completeness for our notation so that we can use it in the future when we discuss time evolution. In this subsection, we describe the behavior of the system for $t=0$.

For the self-consistency loop, we need to calculate
\begin{align}
f_\sigma(t)&=\langle \hat\gamma_\sigma^{\phantom\dagger}  \hat d_\sigma^{\phantom\dagger} \rangle
= \rm{Tr}\left[\hat \rho(t) \hat\gamma_\sigma^{\phantom\dagger}  \hat d_\sigma^{\phantom\dagger} \right]
\end{align}
We obtain that
\begin{subequations}
\label{App:Define_f_via_rho}
\begin{align}
\label{App:Define_f_via_rho_up}
&f_{\uparrow}(t)=\rho_{13}(t)+\rho_{24}(t)
\\
\label{App:Define_f_via_rho_down}
&f_{\downarrow}(t)=\rho_{12}(t)+\rho_{34}(t)
\end{align}
\end{subequations}

\subsection{Solving the equilibrium problem for the chain}
\label{App:Solving_the_equilibrium_problem_for_the_chain}

To close the self-consistency loop, we need to calculate $g_\sigma(t)$ from the chain Hamiltonian Eq.~\eqref{Hamiltonian_chain_Decoupled_sigma}. For this purpose, we rewrite it in the matrix form
\begin{align}
\label{App:H_chain_sigma_matrix}
\hat {\cal H}_{\mathrm{chain},\sigma}=\frac{1}{2} \hat\Psi_{C}^{\dagger}
\left(
\begin{array}{cccc}
\hat {\cal H}_{c,\sigma} & \hat{0}                    &  \hat F_\sigma &  \hat F_\sigma \\
\hat{0}                  & - \hat {\cal H}_{c,\sigma} & -\hat F_\sigma & -\hat F_\sigma \\
\hat F_\sigma^T                 & -\hat F_\sigma^T                  & 0       & 0       \\
\hat F_\sigma^T                 & -\hat F_\sigma^T                  & 0       & 0       \\
\end{array}
\right)
\hat\Psi_{C}^{\phantom\dagger}
\end{align}
Here
\begin{align}
\hat\Psi_{C}^{\phantom\dagger}=\left(
\begin{array}{c}
c_{1,\sigma}^{\phantom\dagger}\\
c_{2,\sigma}^{\phantom\dagger}\\
\vdots \\
c_{L,\sigma}^{\phantom\dagger}\\
c_{1,\sigma}^{\dagger}\\
c_{2,\sigma}^{\dagger}\\
\vdots \\
c_{L,\sigma}^{\dagger}\\
a_{\sigma}^{\phantom\dagger}\\
a_{\sigma}^{\dagger}
\end{array}
\right)
\end{align}
is $2L+2$ dimensional vector.
\begin{align}
[\hat{\cal H}_\sigma]_{ij}&=(\delta_{i+1,j}+\delta_{i,j+1})J_{i,\sigma}
+(\delta_{i,1}\delta_{j,L}+\delta_{i,L}\delta_{j,1})J_{L,\sigma}
-\delta_{ij}\left(\mu_\sigma -\varepsilon_{i,\sigma}\right)
\end{align}
is $L\times L$ matrix and describes the behavior of the $\sigma$-spin fermion on the chain.
$\hat 0$ is also $L\times L$ matrix, where all components are zero. Finally
\begin{align}
[\hat F_\sigma]_{i1} =\delta_{ir} V_{\sigma} f_{\sigma}^{\phantom*}(0) =\delta_{ir}V_{\sigma} \langle \hat\gamma_\sigma \hat d_\sigma \rangle \,.
\end{align}
describes the coupling of the chain with the impurity site and is an $L\times 1$  matrix where all components are zero except for $r$-th component, to which impurity is coupled.

To investigate the ground state properties of the chain as well as its time evolution, we calculate the following correlation functions:
\begin{subequations}
\begin{align}
C_{ij,\sigma}(t)&=\left\{
\begin{array}{ccc}
\langle \hat c_{i,\sigma}^{\dagger} \hat c_{j,\sigma}^{\phantom\dagger} \rangle(t)
&\quad& 1 \leq i,j \leq L
\\
\langle \hat\gamma_{\sigma}^{\phantom\dagger} c_{j,\sigma}^{\phantom\dagger} \rangle(t)
&\quad& i=0,\,1 \leq j \leq L
\\
\langle \hat c_{i,\sigma}^{\dagger} \hat\gamma_{\sigma}^{\phantom\dagger} \rangle(t)
&\quad&1 \leq i \leq L,\,  j=0
\\
1
&\quad& i=j=0
\end{array}
\right.
\\
K_{ij,\sigma}&=\left\{
\begin{array}{ccc}
\langle \hat c_{i,\sigma}^{\phantom\dagger} \hat c_{j,\sigma}^{\phantom\dagger} \rangle(t)
&\quad& 1 \leq i \neq j \leq L
\\
0
&\quad& i=j
\end{array}
\right.
\end{align}
\end{subequations}
Here $C_{ij,\sigma}$ is $(L+1) \times (L+1)$ matrix, while $K_{ij,\sigma}$ is $L \times L$ matrix. For these correlators the following relations hold
\begin{subequations}
\begin{align}
C_{ji,\sigma}^{\phantom*}(t)&=C_{ij,\sigma}^{*}(t)
\\
K_{ji,\sigma}(t)&=-K_{ij,\sigma}^{\phantom*}(t)
\end{align}
\end{subequations}
Thus, we need only perform calculations for $i \leq j$ and reproduce the results for $j > i$ based on the aforementioned symmetry.

To obtain, the ground state properties of the chain we diagonalize the Hamiltonian Eq.~\eqref{App:H_chain_sigma_matrix} using Unitary transformation $\hat U_\sigma$. We obtain that
for $ 1 \leq i,j \leq L$ we have
\begin{subequations}
\begin{align}
C_{ij,\sigma}(0)&=\sum_{l=1}^{2L+2} U_{il,\sigma}^* U_{jl,\sigma}^{\phantom*} F(E_{l,\sigma})
\\
K_{ij,\sigma}(0)&=\sum_{l=1}^{2L+2} U_{(i+L)l,\sigma}^{*} U_{jl,\sigma}^{\phantom*}  F(E_{l,\sigma})
\end{align}
while when one of the indices is equal to zero we have
\begin{align}
C_{0j,\sigma}(0)=\sum\limits_{l=1}^{2L+2} \left(U_{(L+1)l,\sigma}^*+U_{(L+2)l,\sigma}^*\right) U_{jl,\sigma}^{\phantom*} F(E_{l,\sigma})
\end{align}
\end{subequations}
Here $E_{l,\sigma}$ is $l$-th eigenvalue and $F(E_{l,\sigma})$ is the Fermi function. At half-filling and zero temperature
\begin{align}
F(E_l)
=\left\{
\begin{array}{ccc}
1&\quad&1 \leq l \leq L \\
\frac{1}{2}&\quad&l=L+1, L+2 \\
0&\quad&l>L+2
\end{array}
\right.
\end{align}

One can easily notice that
\begin{align}
\label{App:Define_g_via_C}
g_\sigma^{\phantom*}(t)
=\langle \hat\gamma_{\sigma}^{\phantom\dagger} \hat c_{r,\sigma}^{\phantom\dagger} \rangle (t)= C_{0r,\sigma}(t) \,.
\end{align}

\begin{figure}
\centering \includegraphics[width=0.45\textwidth]{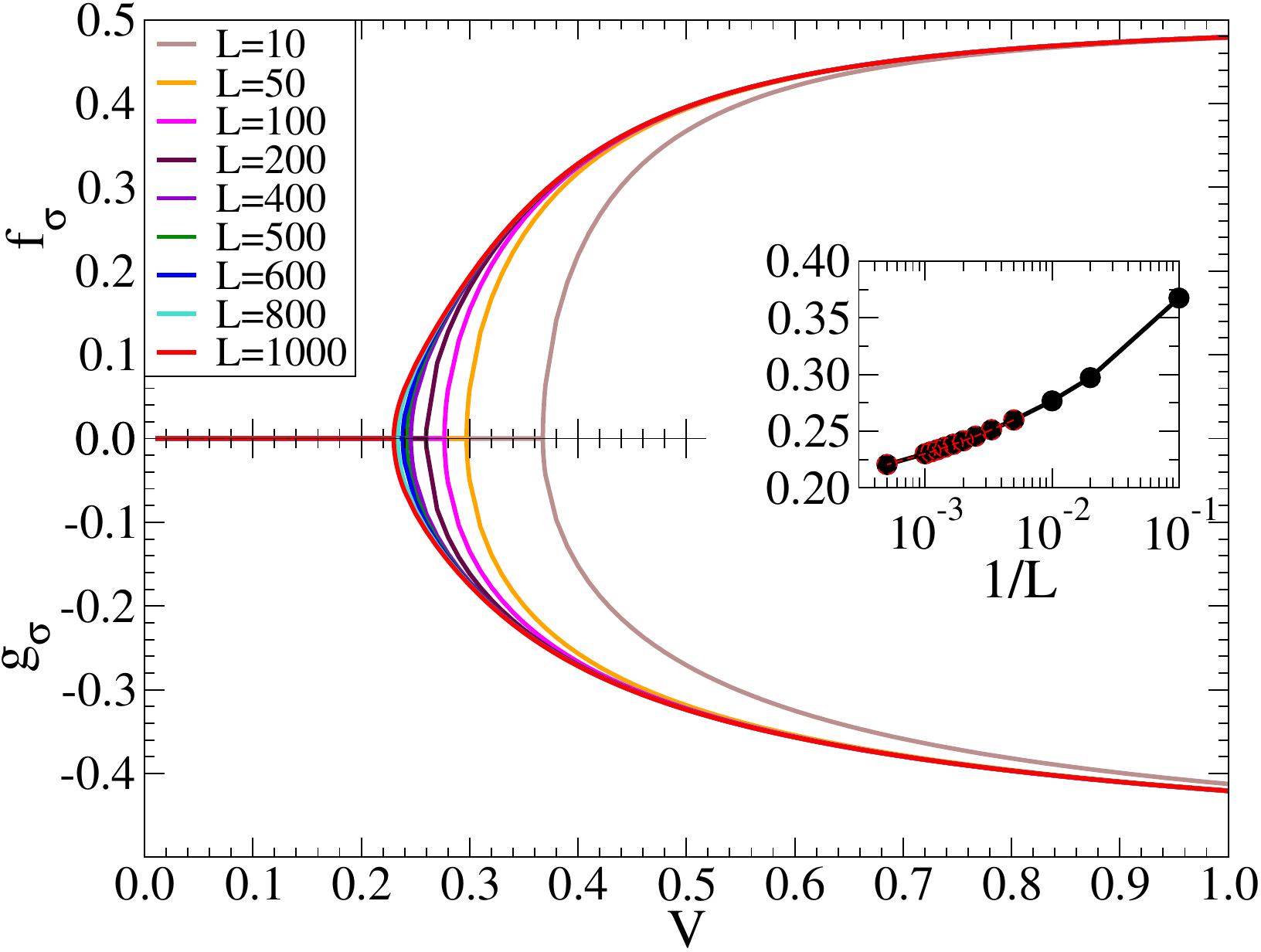}

\vspace{0.1cm}
\centering \includegraphics[width=0.45\textwidth]{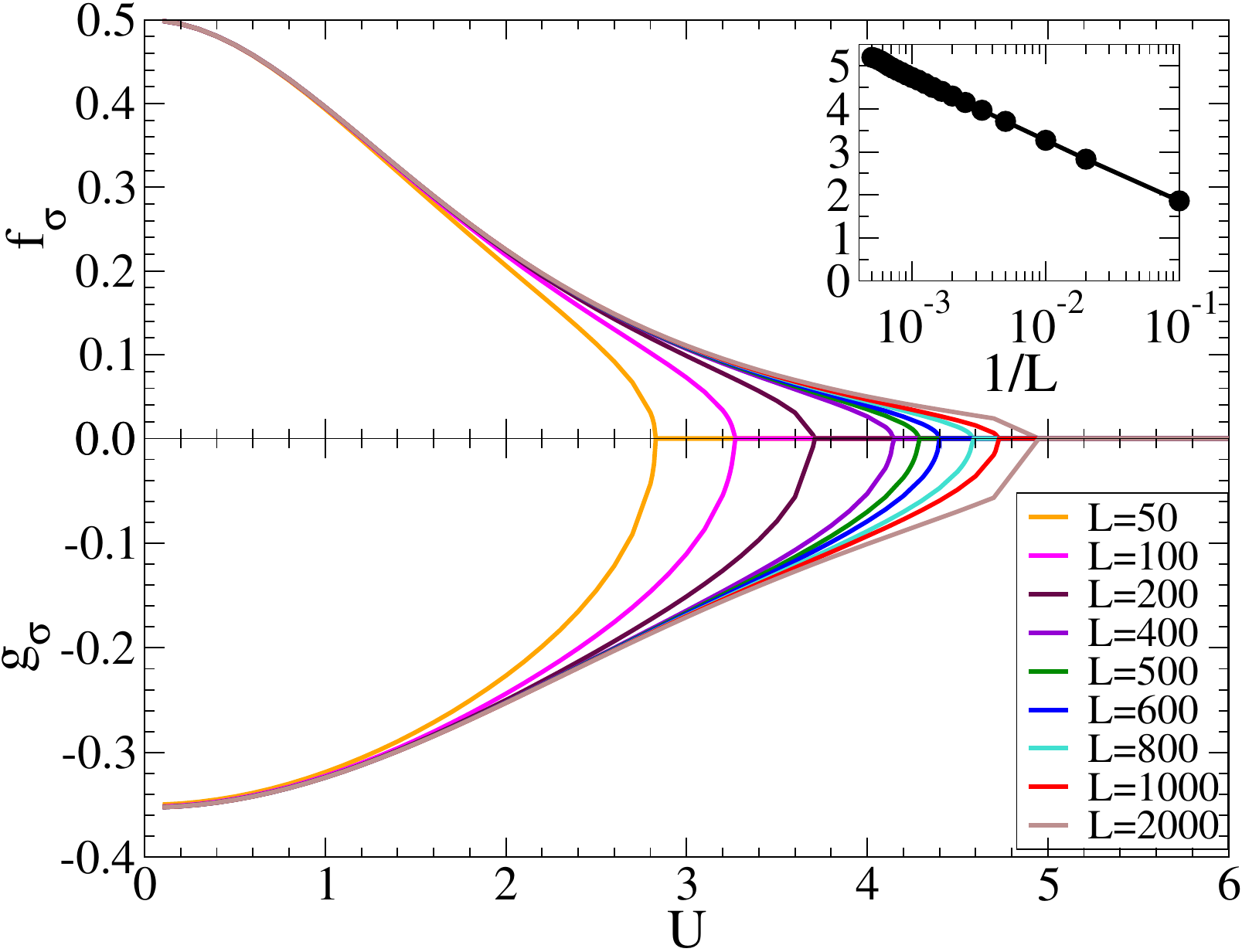}
    \caption{$f_\sigma^{\phantom*}\!=\!\langle \hat\gamma_\sigma^{\phantom\dagger} \hat d_{\sigma}^{\phantom\dagger}\rangle$ and $g_\sigma^{\phantom*}\!=\!\langle \hat\gamma_\sigma^{\phantom\dagger} \hat c_{r,\sigma}^{\phantom\dagger} \rangle$ for different system sizes $L$, as a function of hybridization $V$ (upper panel) at $U\!=\!1$ and as a function of interaction $U$ at $V\!=\!0.5$ (lower panel). 
    Inset (upper panel): critical value of $V_\mathrm{c}$ as a function of the inverse chain size. The results obtained are well fitted by the following function
    $
    V_\mathrm{c}(L)=A_0+A_1 \left(\frac{1}{L}\right)^{1/4}
    $, 
    where $A_0=0.1700$ and $A_1=0.3379$.
    Inset (lower panel): critical value of $U_\mathrm{c}$ as a function of the inverse chain size. 
}
\label{Fig:f_and_g_vs_V_and_U_for_L}
\end{figure}

\begin{figure}
\centering \includegraphics[width=0.45\textwidth]{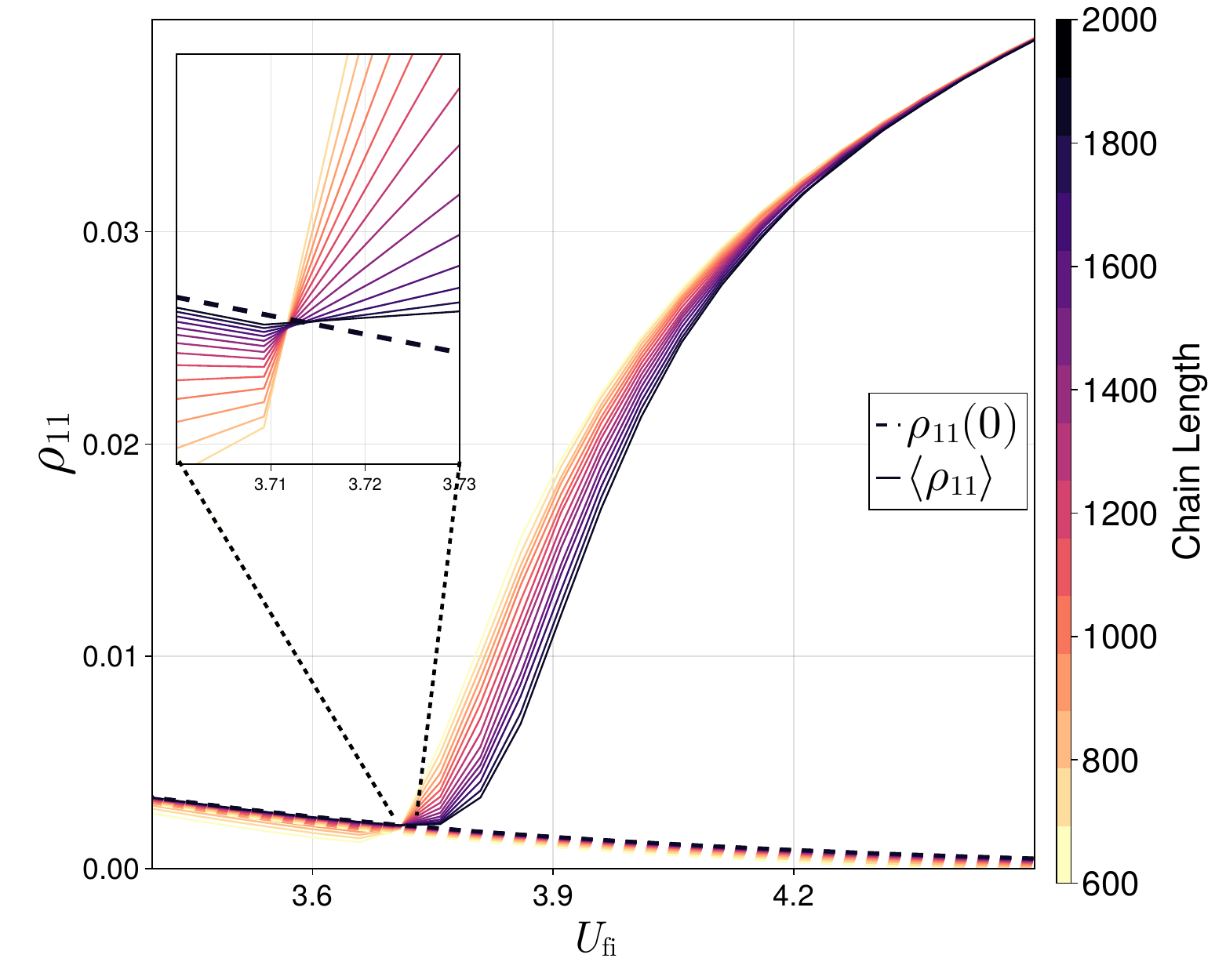}
    \caption{
    Scaling of the non-equilibrium average $\langle\rho_{11}\rangle$ defined in Eq.~\eqref{average_X_def} with the chain length $L$ for $U_\mathrm{in}\!=\!1.0$ and $V_\mathrm{in}\!=\!V_\mathrm{fi}\!=\!0.5$ as a function of the final interaction $U_{\rm fi}$. The equilibrium values are indicated with dashed lines.
}
\label{Fig:scaling_rho11}
\end{figure}

\subsection{Beyond the half-filling}
\label{App:Solving_the_equilibrium_problem_beyond_the_half_filling}

The half-filling can be ensured by taking $\mu_\sigma=0$ and $\varepsilon_{d}=-U/2$. For other fillings, one has to adjust the chemical potential during each self-consistent iteration to obtain the desired filling. For this reason, one should additionally ensure that the following relation holds
\begin{equation}
\label{App:Fixing_filling}
\left\{
\begin{array}{c}
\rho_{11}(0)+ \rho_{22}(0)+\sum\limits_{i=1}^{L}C_{ii,\uparrow}(0)=N_{\uparrow}\\
\rho_{11}(0)+ \rho_{33}(0)+\sum\limits_{i=1}^{L}C_{ii,\downarrow}(0)=N_{\downarrow}
\end{array}
\right.
\end{equation}
Here $N_\sigma$ is the total filling for spin $\sigma$ fermions.
At half-filling, $N_{\sigma}=L/2$. 
One should keep in mind that $\rho_{ij}$ and $C_{ij,\sigma}$ depend on $\mu_\sigma$. If $N_{\uparrow}=N_{\downarrow}$, then solving only one of these equations is sufficient since they will be identical.

\section{Finite size scaling}
\label{App:Finite-size-scaling}

\subsection{Equilibrium}
\label{App:Finite-size-scaling:Equilibrium}

In this section, we investigate the mean field parameters $f_\sigma^{\phantom*}$ and $g_\sigma^{\phantom*}$ as a function of the hybridization $V$ at $U\!=\!1$ and as a function of the interaction $U$ at $V=0.5$ for different system sizes (i.e., chain lengths) $L$ (see Fig.~\ref{Fig:f_and_g_vs_V_and_U_for_L}). 

For large values of the hybridization (for weak interactions), $f_\sigma$ and $g_\sigma$ depend only weakly on $L$.
Upon decreasing $V$ (increasing $U$) the dependence on $L$ increases, eventually leading to different critical values $V_\mathrm{c}(L)$ ($U_\mathrm{c}(L)$) where $f_\sigma$ and $g_\sigma$ vanish, indicating the phase transition to the local moment regime (see Sec.~\ref{Equilibrium}).
In the insets of Fig.~\ref{Fig:f_and_g_vs_V_and_U_for_L} we plot (upper panel) $V_\mathrm{c}(L)$ (for $U\!=\!1$) and (lower panel) $U_\mathrm{c}(L)$ (for $V\!=\!0.5$) as a function of the inverse system size $1/L$. The asymptotic behavior of the obtained results for the hybridization is well described by the function
\begin{equation}
V_\mathrm{c}(L)=A_0+A_1 \left(\frac{1}{L}\right)^{1/4}\,,
\end{equation}
where $A_0=0.1700$ and $A_1=0.3379$. 
Hence, at $U\!=\!1$ the critical hybridization strength approaches $V_\mathrm{c}(L)\!\rightarrow\!0.17$ in the limit of infinite system size $L\!\rightarrow\!\infty$.

We observe that $V_\mathrm{c}(L)$ and $U_\mathrm{c}(L)$ converge relatively quickly with the size of the system. 
In particular, our results indicate that a value of ${L\!=\!1000}$ is sufficient to capture features of the system in the thermodynamic limit.
Hence, in all our calculations we set the system size to $L\!=\!1000$.

\subsection{Time evolution}
\label{App:Finite-size-scaling:Time_evolution}

We have also performed a finite-size scaling analysis for the average $\langle \rho_{11} \rangle$ as defined in Eq.~\eqref{average_X_def} with respect to the equilibrium value $\rho^{\rm equ}_{11}$ as a function of the final interactions $U_{\rm fi}$ for a fixed initial interaction and a fixed hybridization strength. 
In Fig.~\ref{Fig:scaling_rho11} we observe that at a critical final interaction strength $U^{\rm ne}_c\!\sim\!3.75$ the difference $\langle\rho_{11}\rangle\!-\!\rho^{\rm equ}_{11}$ becomes positive.
Remarkably, the interaction value where this happens has negligible dependence on the chain length. 
Furthermore, $\langle \rho_{11} \rangle$ converges (in $L$) toward the equilibrium value of $\rho^{\rm equ}_{11}$ before but not after the intersection points.
Also, this feature is stable with respect to the chain length $L$.
Therefore, we conclude that $U^{\rm ne}_c\!\sim\!3.75$ indeed marks a dynamic phase transition which is considerably lower than the equilibrium phase transition at $U_\mathrm{fi}\!\simeq\!4.73$ for the given $V_{\rm fi}$ (for $L=1000$).

\begin{figure}[t!]
\centering \includegraphics[width=0.45\textwidth]{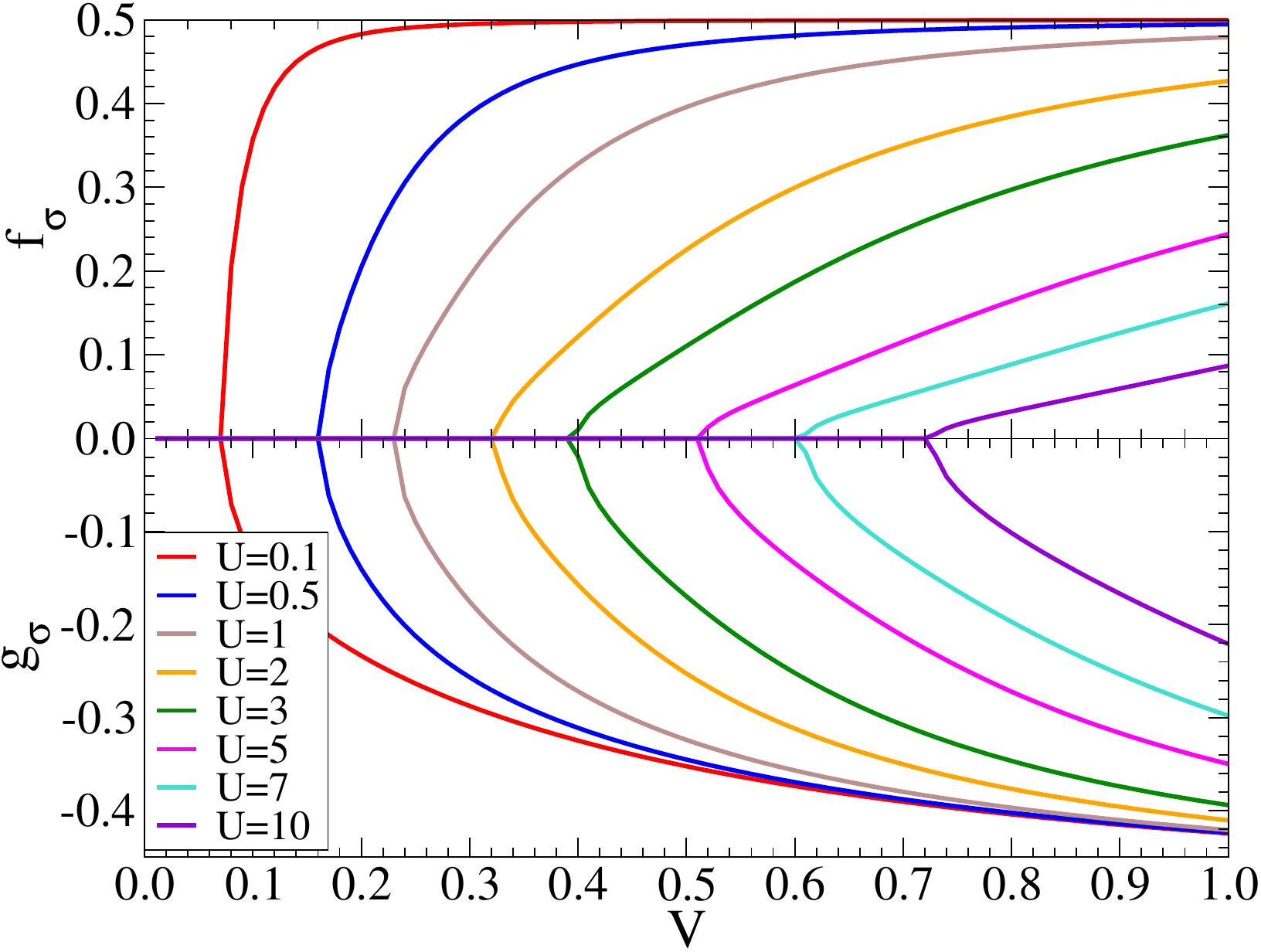} \\
\caption{$f^{\phantom*}$ and $g^{\phantom*}$ as a function of $V$ for various $U$.
}
\label{Fig:f_and_g_vs_V_L1000}
\end{figure}

\begin{figure}[t!]
\centering \includegraphics[width=0.45\textwidth]{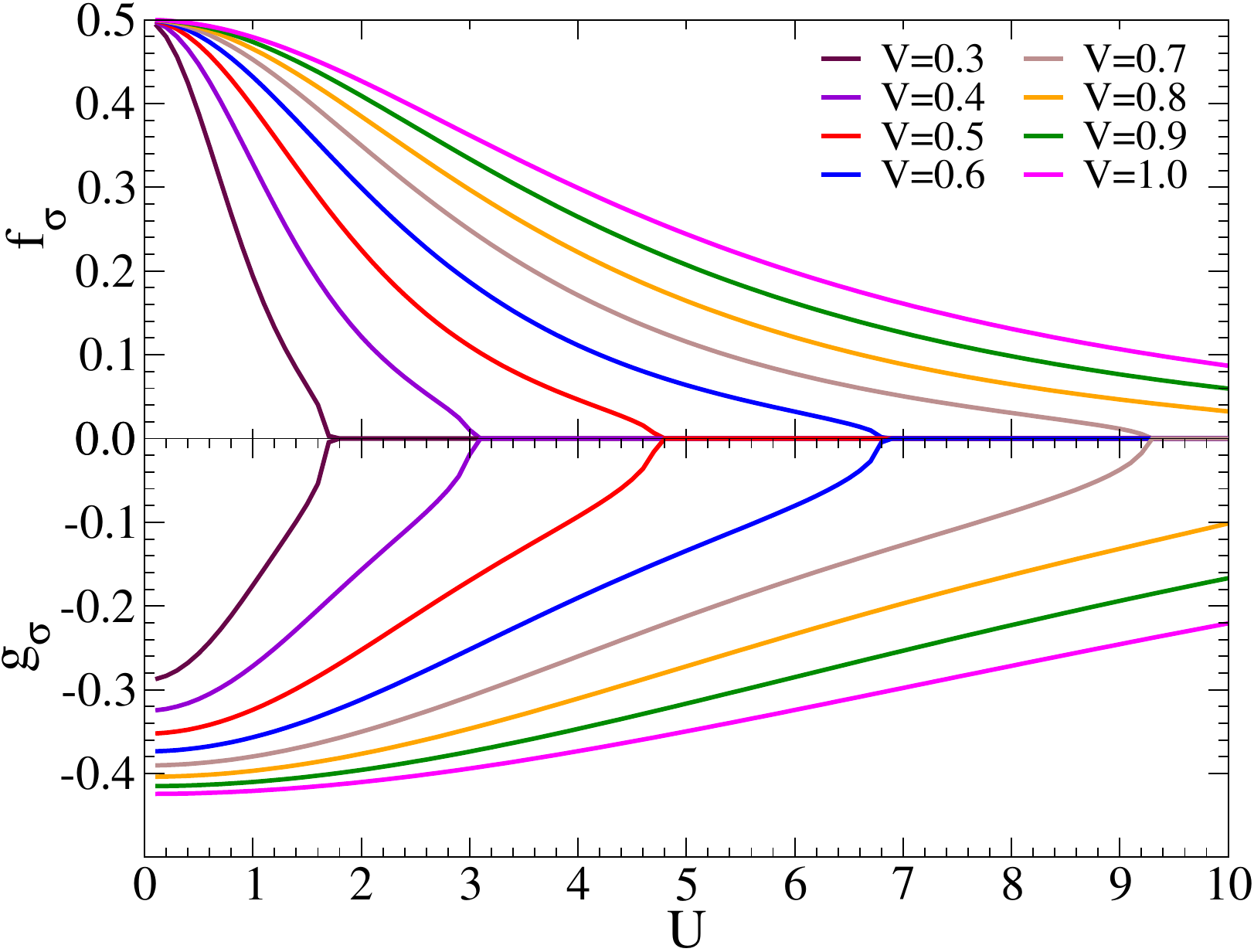} \\
\caption{$f^{\phantom*}$ and $g^{\phantom*}$ as a function of $U$ for various $V$.
}
\label{Fig:f_and_g_vs_U_L1000}
\end{figure}

\section{Analytical analysis near the phase transition}
\label{App:Analytical_analysis}

In this section, we derive analytical expressions for the mean field parameters $f_\sigma$ and $g_\sigma$ close to the phase transition between the mixed valence and the local moment regime (colorful and dark regions in Fig.~\ref{Fig:PD}). 
The condition $f_\sigma\!=\!g_\sigma\!=\!0$ will then allow us to derive the critical hybridization $V_\mathrm{c}$ as a function of the interaction strength $U$ as well as the critical interaction $U_\mathrm{c}$ as a function of the hybridization $V$.
Unlike our numerical calculations, we do not restrict our impurity to be coupled to the first site of the chain. Moreover, we have $g_\uparrow\!=\!g_\downarrow\!=\!g$ and $f_\uparrow\!=\!f_\downarrow\!=\!f$ due to SU($2$) symmetry.

\subsection{Impurity site}
\label{App_Analytical_Impurity}

The Hilbert space of the decoupled impurity Hamiltonian in  Eq.~\eqref{Hamiltonian_impurity_Decoupled} is spanned by the electrons at the impurity $\hat{d}_\sigma$ and the Majorana $\hat{\gamma}_\sigma$. Therefore it is 16 dimensional. However, as discussed in Sec.~\ref{App:Solving_the_equilibrium_problem_for_the_impurity} [see Eqs.~\eqref{App:Conservation_Impurity} and \eqref{App:Basis_Impurity}), it consists of four equivalent and independent 4 dimensional subspaces to one of which we can restrict ourselves reducing the problem to the diagonalization of a $4\!\times\!4$ matrix.
The ground state of the system at half-filling is then given by
\begin{equation}
\label{App:ground_state}
|\Psi_{0,\rm imp}\rangle =\frac{2Vg|1\rangle + E_{0,\rm imp}|2\rangle + E_{0,\rm imp}|3\rangle + 2Vg|4\rangle}
{\sqrt{2E_{0,\rm imp}^2+8V^2g^2}},
\end{equation}
where the states $\lvert n\rangle$, $n\!=1\!\ldots\!4$, are defined Eq.~\eqref{App:Basis_Impurity} and
\begin{equation}
\label{App:ground_state_Energy}
E_{0,\rm imp}=-\frac{1}{4}\left(U+\sqrt{U^2+64V^2g^2}\right)
\end{equation}
is the corresponding ground state energy.

From this, we can evaluate the mean field parameter $f$ according to Eq.~\eqref{Meanfield_f} as
\begin{equation}
\label{App:f_exact_Analytical}
f=\langle\Psi_{0,\rm imp}\lvert \hat\gamma_\sigma \hat{d}_\sigma\rvert\Psi_{0,\rm imp}\rangle = \frac{2E_{0,\rm imp} V g }{E_{0,\rm imp}^2+4V^2g^2}
\end{equation}
$f$ can be expressed as a function of ${4Vg}/{U}$. 
Since we are interested in the system's behavior close to the phase transition, i.e. in the vicinity of  $g \!=\! 0$ for any finite $U$, the condition ${|g| \ll U/4V}$ can be fulfilled. Thus, we can extend Eq.~\eqref{App:f_exact_Analytical} into a power series of ${4Vg}/{U}$ yielding
\begin{equation}
\label{App:f_approximate}
f \simeq -\frac{4Vg}{U}+2\left(\frac{4Vg}{U}\right)^3  \,.
\end{equation}
For the density matrix $\rho_{ij}$, for finite values of $g$ according to Eq.~\eqref{App:Matrix_operatoprs_for_Impurity_T0} the following relationships apply
\begin{subequations}
\label{App:rho_Analytical}
\begin{align}
    & \rho_{11} 
        =      \rho_{44}
        =      \rho_{14}
        =      \frac{4V^2 g^2}{2E_{0,\rm imp}^2+8V^2g^2} 
        \simeq \frac{1}{2}\left(\frac{4V g}{U}\right)^2
\\
&\rho_{22}=\rho_{33}=\rho_{23}=\frac{E_0^2}{2E_{0,\rm imp}^2+8V^2g^2}   \simeq \frac{1}{2}\left[1-\left(\frac{4V g}{U}\right)^2\right]  
\\
&\rho_{12}=\rho_{13}=\rho_{24}=\rho_{34}=\frac{2V g E_0}{2E_{0,\rm imp}^2+8^2g^2} \simeq -\frac{2Vg}{U} \,.
\end{align}    
\end{subequations}
For $g=0$ the ground state is degenerate and direct calculations show that all matrix elements are zero, apart from $\rho_{22}=\rho_{33}=0.5$. Immediately after the phase transition, $\rho_{23}$ therefore makes a finite jump and becomes equal to $\rho_{22}=\rho_{33}$.

$U=0$, corresponding to the resonance level model with $\varepsilon_d=0$, is a special case.
In this limiting case
\begin{equation}
\label{App:f_non-int}
f = \left\{
\begin{array}{ccc}
\frac{1}{2}\mbox{sign}\left(- V g\right)     &\quad&  V g \neq 0\\
0                                            &\quad&  V g =0
\end{array}
\right.
\end{equation}
So $f$ shows a step-like behavior. 
This fully agrees with our numerical calculations, which showed that for $V=0$ $f=0$ and any finite $V$ $f=0.5$ (see Sec. \ref{Resonance_Level_Model}.)

We also observe a similar step-like behavior for density matrix $\rho_{ij}$. For any finite $V$ we obtain that all matrix elements are equal to each other and $\rho_{ij}=0.25$, while for $V=0$ all matrix elements are zero apart from $\rho_{22}=\rho_{33}=0.5$.

\subsection{Chain}
\label{App:Analytical_Chain}

The next step is to calculate $g$ from Eq.~\eqref{Meanfield_g} using second order perturbation theory.
To this end, we consider the second term in Eq.~\eqref{Hamiltonian_chain_Decoupled_sigma}, which includes the Majorana $\hat{\gamma}_\sigma$ and is proportional to $f$, as a small perturbation of the non-interacting chain.
Hence, our calculations apply to the region close to the phase transition in Fig.~\ref{Fig:PD} where $f\!\ll\!1$.
Moreover, let us point out that the perturbative expansion is only applicable for finite values of $U$ (i.e., away from the lower left corner of the phase diagram), because, as mentioned above, at $U\!=\!0$ the mean field parameter $f$ features a step-like behavior when changing from a finite value to zero.
Let us finally note, that since the two spin projections are completely independent from each other in the chain Hamiltonian in Eq.~\eqref{Hamiltonian_chain_Decoupled_sigma}, we can treat them separately. Hence, we will suppress the spin index in the following calculations.

The unperturbed Hamiltonian is non-interacting and, therefore, its eigenvectors can be expressed as Slater determinants
\begin{equation}
\label{App:Eigenvectors_Non-perturbed}
|\Phi_{l}^{(0)}\rangle =\prod_{k \in {\cal M}_l} \hat{\tilde{c}}_{k}^{\dagger}|0\rangle,
\end{equation}
where for the case of equivalent hoping amplitudes between the lattice sites ($J_i\!=\!J$) and open boundary conditions the Fourier transform between $\hat{c}_i$ and $\hat{\tilde{c_k}}$ is given by
\begin{equation}
\label{cl_vs_ci}
\hat{\tilde c}_{k}^{\phantom\dagger} =\frac{\sqrt{2}}{\sqrt{L+1}}\sum_{i=1}^{L}\sin(ki)\;\hat{c}_{i}^{\phantom\dagger},
\end{equation}
where $k\!=\!\pi m/(L\!+\!1), m\!=\!1,\!\ldots\!,L$.
In Eq.~\eqref{App:Eigenvectors_Non-perturbed}, ${\cal M}_l$ is the set of filled single-particle levels in the $l$-th eigenvector associated with the eigenvalue $E_{l}^{(0)}$. The number of particles for each eigenvector is conserved and can take values between $0$ (all states are unoccupied) to $L$ (all states are occupied). For the ground state ($l=0$), the lowest $N$ states are filled (for half-filled system $N=L/2$).

The unperturbed part of the Hamiltonian in Eq.~\eqref{Hamiltonian_chain_Decoupled_sigma} is decoupled from the Majorana fermion. Therefore, each energy level is doubly degenerate (in addition to the spin degeneracy), corresponding to the two different states for the Majorana fermion, i.e., either the Majorana fermion is present or absent. 
The corresponding degenerate eigenvectors $|\Psi_{l,0}^{(0)}\rangle$ and $|\Psi_{l,1}^{(0)}\rangle$ are given by
\begin{subequations}
\label{App:0States_perturbative}
\begin{align}
\label{App:0States_0}
|\Psi_{l,0}^{(0)}\rangle&=|\Phi_{l}^{(0)}\rangle|0\rangle_M
=\prod_{k \in {\cal M}_l}c_{k}^{\dagger}|0\rangle
\\
\label{App:0States_1}
|\Psi_{l,1}^{(0)}\rangle&=|\Phi_{l}^{(0)}\rangle|1\rangle_M
=\prod_{k \in {\cal M}_l}c_{k}^{\dagger}a^{\dagger}|0\rangle,
\end{align}
where $|1\rangle_M$ and $|0\rangle_M$ denotes the state with and without Majorana, respectively.
Since these two configurations are coupled to each other by the perturbation [i.e., the second part of the chain Hamiltonian in Eq.~\eqref{Hamiltonian_chain_Decoupled_sigma} which contains the Majorana operators $\hat{\gamma}$], it is convenient to decouple them by the following linear combinations
\begin{align}
|\Psi_{l,\pm}^{(0)} \rangle=|\Phi_{l}^{(0)}\rangle|\pm\rangle_M
=\frac{1}{\sqrt{2}}\prod_{k \in {\cal M}_l} c_{k}^{\dagger}\left(1 \pm a^{\dagger}\right)|0\rangle
\end{align}
\end{subequations}
The states $|\Psi_{l,+}^{(0)} \rangle$ and $|\Psi_{l,-}^{(0)} \rangle$ form two equivalent invariant subspaces with respect to the chain Hamiltonian in Eq.~\eqref{Hamiltonian_chain_Decoupled_sigma}. Similarly, as for the impurity, we can, hence, restrict ourselves to one of these two subspaces although we will present the following equations for both of them for the sake of generality.

In the next step, we can now construct the first and second order corrections (in $f$) to the ground state $|\Psi_{l,\pm}^{(0)} \rangle$ which are generated by the perturbation term
\begin{align}
\hat{\cal H}_f =V \left(\hat c_{r}^{\dagger}\hat\gamma^{\phantom\dagger}
+ \hat\gamma\hat c_{r}^{\phantom\dagger}\right),
\end{align}
which is the second part of the chain Hamiltonian in Eq.~\eqref{Hamiltonian_chain_Decoupled_sigma}. The result reads
\begin{align}
\label{App:Perturbed_State}
|\Psi_{\pm}\rangle &=|\Psi_{0,\pm}^{(0)} \rangle
- f \sum_{l \neq 0} |\Psi_{l,\pm}^{(0)}\rangle
\frac{ \langle \Psi_{l,\pm}^{(0)}| \hat{\cal H}_f |\Psi_{0,\pm}^{(0)}\rangle}{E_l^{(0)}-E_0^{(0)}} 
+ f^2 \Biggl[\sum_{l, l' \neq 0} |\Psi_{l,\pm}^{(0)}\rangle \frac{
    \langle \Psi_{l,\pm}^{(0)}| \hat{\cal H}_f |\Psi_{l',\pm}^{(0)}\rangle
    \langle \Psi_{l',\pm}^{(0)}| \hat{\cal H}_f |\Psi_{0,\pm}^{(0)}\rangle}
    { (E_l^{(0)}-E_0^{(0)})(E_{l'}^{(0)}-E_0^{(0)}) }
-\frac{1}{2}|\Psi_{0,\pm}^{(0)}\rangle \sum_{l \neq 0} 
    \frac{|\langle \Psi_{l,\pm}^{(0)}| \hat{\cal H}_f |\Psi_{0,\pm}^{(0)}\rangle|^2}{(E_l^{(0)}-E_0^{(0)})^2} \Biggl] \,.
\end{align}

Using one of the two equivalent ground states $|\Psi_{\pm}\rangle$ we can calculate $g$ up to second order in $f$ yielding
\begin{equation}
\label{App:g_approximate}
g=\langle \Psi_{\pm} \lvert \hat{\mathcal{H}}_f \rvert \Psi_{\pm} \rangle \simeq - \frac{a V}{|J|}f + \frac{b V^2}{J^2}f^2 \,.
\end{equation}
Here
\begin{subequations}
\begin{align}
a & = \sum_{l \neq 0}\frac{A_{0l}^2+A_{l0}^2}{\Delta \tilde E_{l0}}\label{a}
\\
b & = 3\sum_{l_1,l_2 \neq 0}
    \frac{ \left(A_{l_1 0}^{\phantom*} + A_{0l_1}^{\phantom*}\right) \left(A_{l_2 0}^{\phantom*} + A_{0l_2}^{\phantom*}\right)}
         {\Delta \tilde E_{l_10} \Delta \tilde E_{l_20}}
    A_{l_1l_2}^{\phantom*}\label{b}
\end{align}
where
\begin{align}
A_{l_1l_2}^{\phantom*} = \langle \Psi_{l_1,\pm}^{(0)}|\gamma c_r |\Psi_{l_2,\pm}^{(0)}\rangle
\end{align}
and
\begin{align}
\Delta \tilde E_{l0} =\frac{1}{|J|} \left(E_l^{(0)}-E_0^{(0)}\right)\,.
\end{align}
\end{subequations}
Let us note that $A_{l_1l_2}^{\phantom*}$ are real numbers by construction.

Detailed derivation of coefficients $a$ and $b$ see in the supplemental material (see Sec. S3 in Ref.\cite{Supplemental_Material}).

\subsection{Behavior close to the phase transition}
\label{App:sec:fg_vs_V_and_U}
Now we try to determine the behavior of $g$ and $f$ close to the phase transition.
For this purpose, we combine the mean field expansions for $f$ and $g$ in Eqs.~\eqref{App:f_approximate} and~\eqref{App:g_approximate} keeping only contributions up to $g^2$ and $f^2$. More specifically, by inserting the expression for $f$ from Eq.~\eqref{App:f_approximate} into Eq.~\eqref{App:g_approximate} for $g$ we obtain
\begin{align}
g &= \frac{4a V^2}{U |J|}g +\frac{16b V^4}{U^2 J^2}g^2
\end{align}
Solving this equation for $g$ we obtain (apart from the trivial solution $g\!=\!0$)
\begin{subequations}
\label{App:fg_vs_V_and_U}
\begin{align}
\label{App:g_vs_V_and_U}
g=-\frac{U^2 J^2}{16b V^4}\left(\frac{4a V^2}{U |J|}-1\right).
\end{align}
Reinserting this result for $g$ into Eq.~\eqref{App:f_approximate} we obtain (neglecting contributions beyond second order in $g$)
\begin{align}
\label{App:f_vs_V_and_U}
f= \frac{U J^2}{4b V^3}\left(\frac{4a V^2}{U |J|}-1\right)
\end{align}
\end{subequations}
Substituting Eq. ~\eqref{App:g_vs_V_and_U} into Eq. ~\eqref{App:rho_Analytical} yields the following expressions for the density matrix:
\begin{subequations}
\label{App:rho_Analytical_U_and_V}
\begin{align}
    & \rho_{11} 
        =      \rho_{44}
        =      \rho_{14}
        \simeq \frac{U^2 J^4}{32 b^2 V^6}\left(\frac{4a V^2}{U |J|}-1\right)^2
\\
&\rho_{22} =
 \rho_{33} =
 \rho_{23} 
  \simeq \frac{1}{2} - \frac{U^2 J^4}{32 b^2 V^6}\left(\frac{4a V^2}{U |J|}-1\right)^2
\\
&\rho_{12} =
 \rho_{13} =
 \rho_{24} =
 \rho_{34} \simeq  \frac{U J^2}{8 b V^3}\left(\frac{4 a V^2}{U |J|}-1\right) \,.
\end{align}    
\end{subequations}

The phase transition to the local moment regime is indicated by $f\!=\!g\!=0$.
Applying this condition to Eqs.~\eqref{App:fg_vs_V_and_U} we can derive the critical hybridization strength $V_\mathrm{c}$ as a function of the interaction $U$ or, vice versa, the critical interaction $U_\mathrm{c}$ as a function of $V$
\begin{align}
\label{App:VcUc}
&V_\mathrm{c}=\sqrt{\frac{U |J|}{4a}}
&U_\mathrm{c}=\frac{4a V^2}{|J|}
\end{align}

Based on these critical values we can rewrite $f$ and $g$ solely as a function of the hybridization $V$
\begin{subequations}
\label{App:g_and_f_with_critical}
\begin{align}
g&=-\frac{a^2}{b}\left(\frac{V_\mathrm{c}}{V}\right)^4\frac{V+V_\mathrm{c}}{V_\mathrm{c}}\frac{V-V_\mathrm{c}}{V_\mathrm{c}}
\\
f&=\frac{a |J|}{b V_\mathrm{c}}\left(\frac{V_\mathrm{c}}{V}\right)^3\frac{V+V_\mathrm{c}}{V_\mathrm{c}} \frac{V-V_\mathrm{c}}{V_\mathrm{c}}
\end{align}
or exclusively as a function of the Hubbard interaction $U$
\begin{align}
g&=-\frac{a^2}{b}\frac{U}{U_\mathrm{c}} \frac{U_\mathrm{c}-U}{U_\mathrm{c}}
\\
f&=\sqrt{\frac{4 a^{3} |J|}{b^2 U_\mathrm{c}}} \frac{U_\mathrm{c}-U}{U_\mathrm{c}} \,.
\end{align}
\end{subequations}

\subsection{Comparison with numerical results}

Let us close this section by presenting data for $f$ and $g$ along specific slices in the phase diagram of Fig.~\ref{Fig:PD} to confirm the analytical results for these quantities close to the phase transition. 
In particular, Fig.~\ref{Fig:f_and_g_vs_V_L1000} shows the mean field parameters as a function of $V$ for various values of $U$ while Fig.~\ref{Fig:f_and_g_vs_U_L1000} provides the results as a function of $U$ for various values of $V$. We indeed observe a linear vanishing of $f$ and $g$ close to the respective transition points in accordance with the linear behavior predicted in Eqs.~\eqref{App:g_and_f_with_critical}. Moreover, the slope of the transition lines is very steep for small values of $V_\mathrm{c}$ and $U_\mathrm{c}$. This can be also understood from our analytical calculations, where this slope is defined by the inverse of $V_\mathrm{c}$ and $U_\mathrm{c}$, respectively.

\begin{center}
\begin{figure*}[t!]
\subfigure[$(0.5,1.0) \rightarrow (0.5,0.3)$]{
\label{Fig:imrho_ij_case:i}
\begin{minipage}[b]{0.45\textwidth}
\centering \includegraphics[width=1\textwidth]{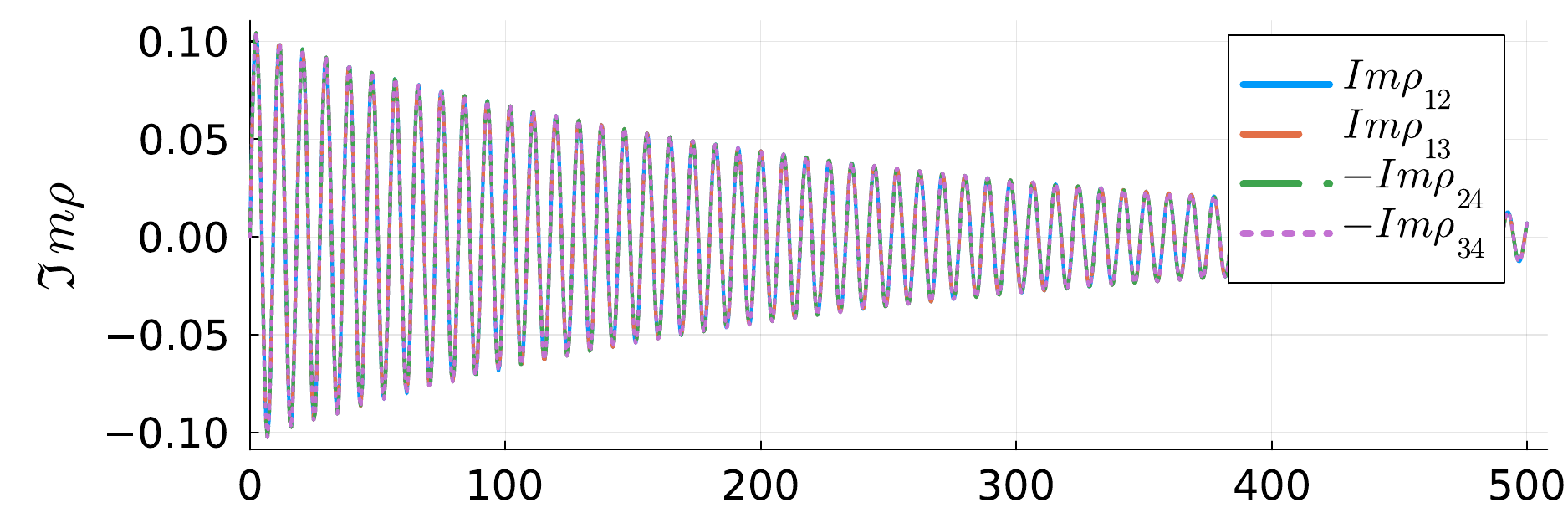}
\end{minipage}}
\hspace{0.025\textwidth}
\subfigure[$(0.5,1.0) \rightarrow (0.5,0.7) $]{
\label{Fig:imrho_ij_case:ii}
\begin{minipage}[b]{0.45\textwidth}
\centering \includegraphics[width=1\textwidth]{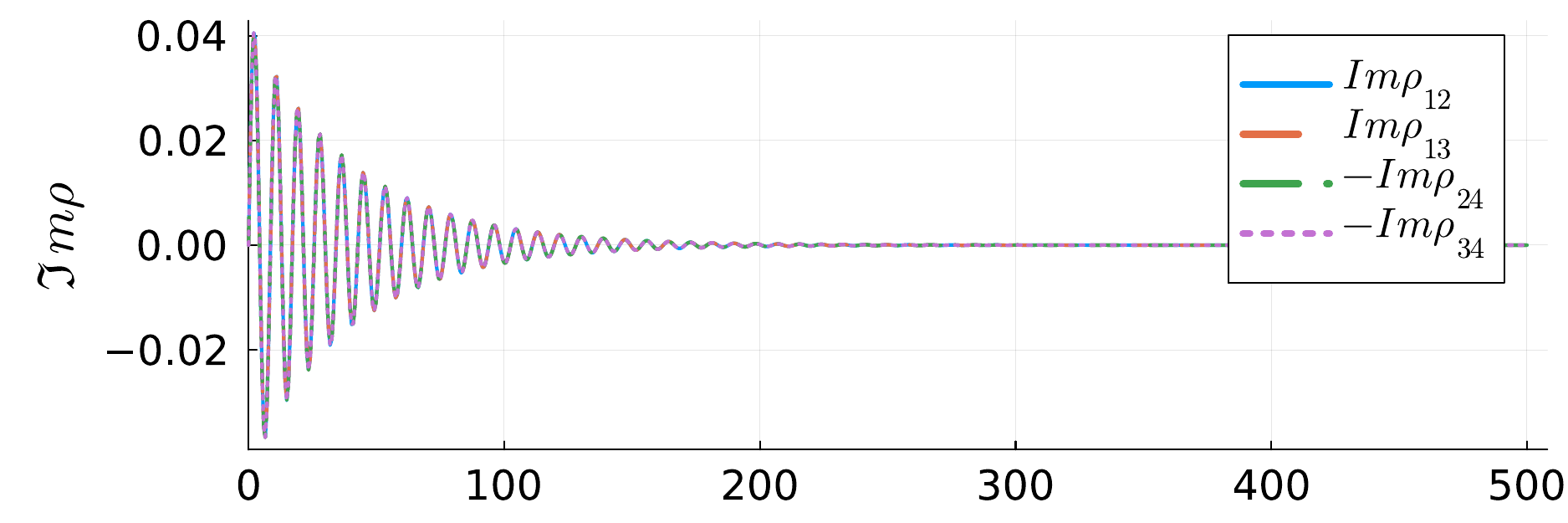}
\end{minipage}}
\\
\subfigure[$(0.5,1.0) \rightarrow (0.5,2.5)$]{
\label{Fig:imrho_ij_case:iii}
\begin{minipage}[b]{0.45\textwidth}
\centering \includegraphics[width=1\textwidth]{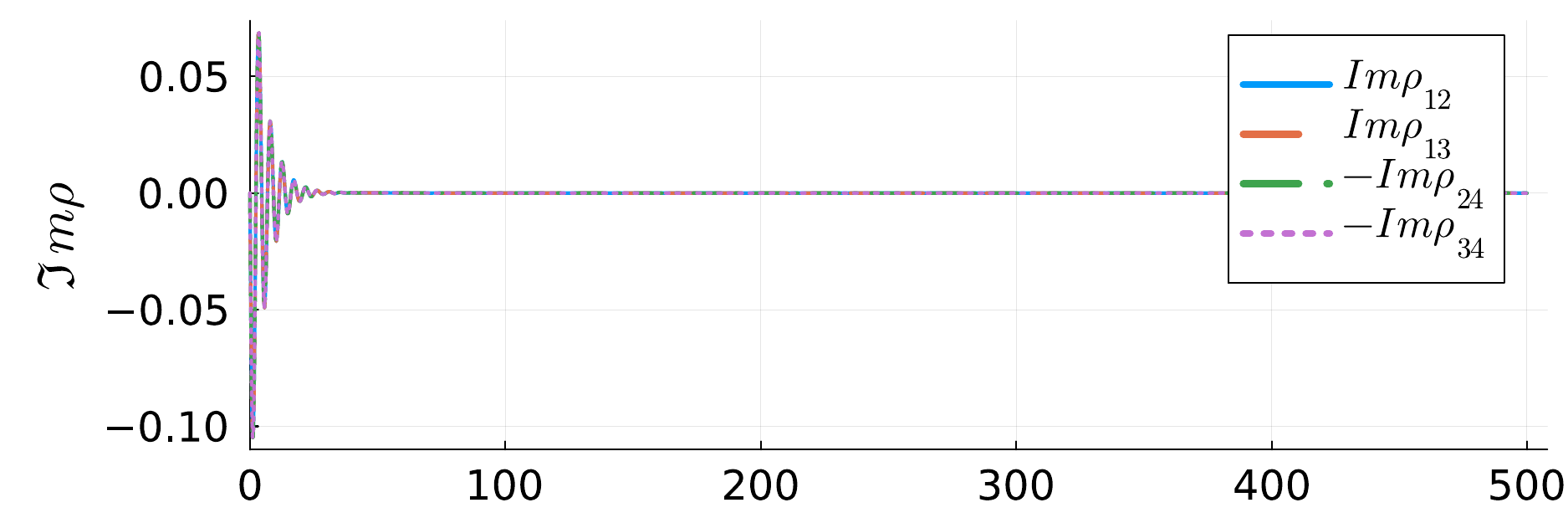}
\end{minipage}}
\hspace{0.025\textwidth}
\subfigure[$(0.5,1.0) \rightarrow (0.5,4.0)$]{
\label{Fig:imrho_ij_case:iv}
\begin{minipage}[b]{0.45\textwidth}
\centering \includegraphics[width=1\textwidth]{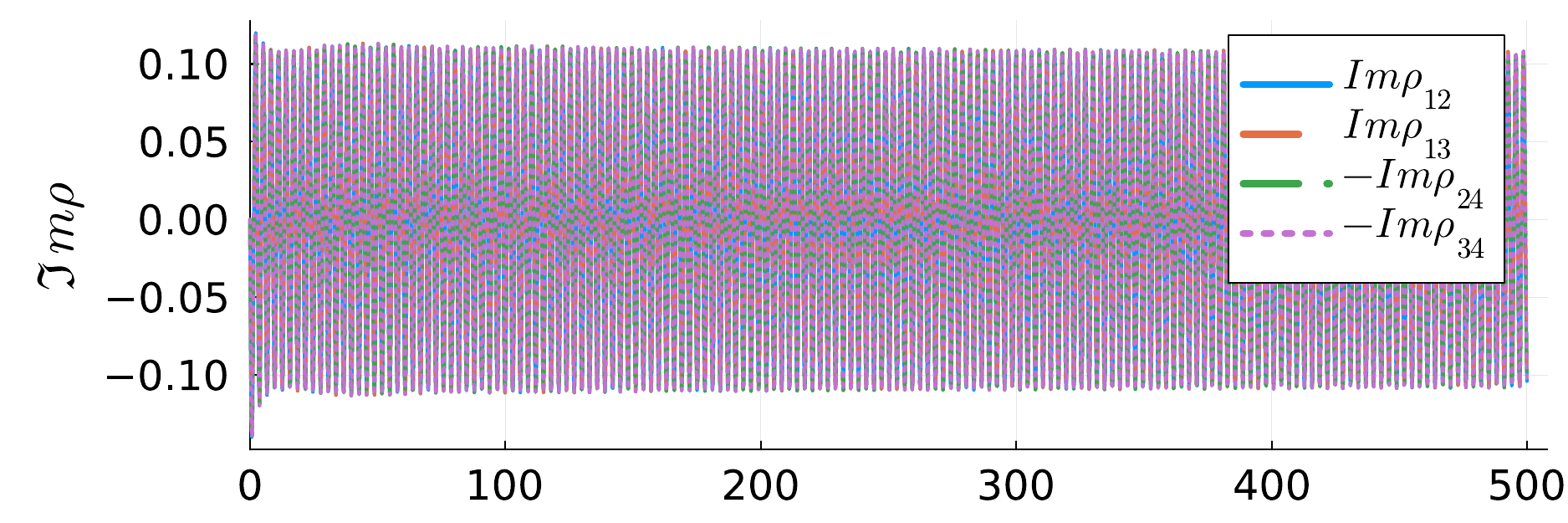}
\end{minipage}}
\caption{Imaginary part of selected off-diagonal elements of the density matrix $\rho_{ij}(t)$ as a function of time $t$ for the four different quenches (i)-(iv) $(V_{in}, U_{in})\rightarrow(V_{fi}, U_{fi})$ discussed in Sec.~\ref{Results}.}
\label{Fig:Imrho_ij_t}
\end{figure*}
\end{center}

\begin{center}
\begin{figure*}[t!]
\subfigure[$(0.5,1.0) \rightarrow (0.5,0.3)$]{
\label{Fig:Cij_case:i}
\begin{minipage}[b]{0.45\textwidth}
\centering \includegraphics[width=1\textwidth]{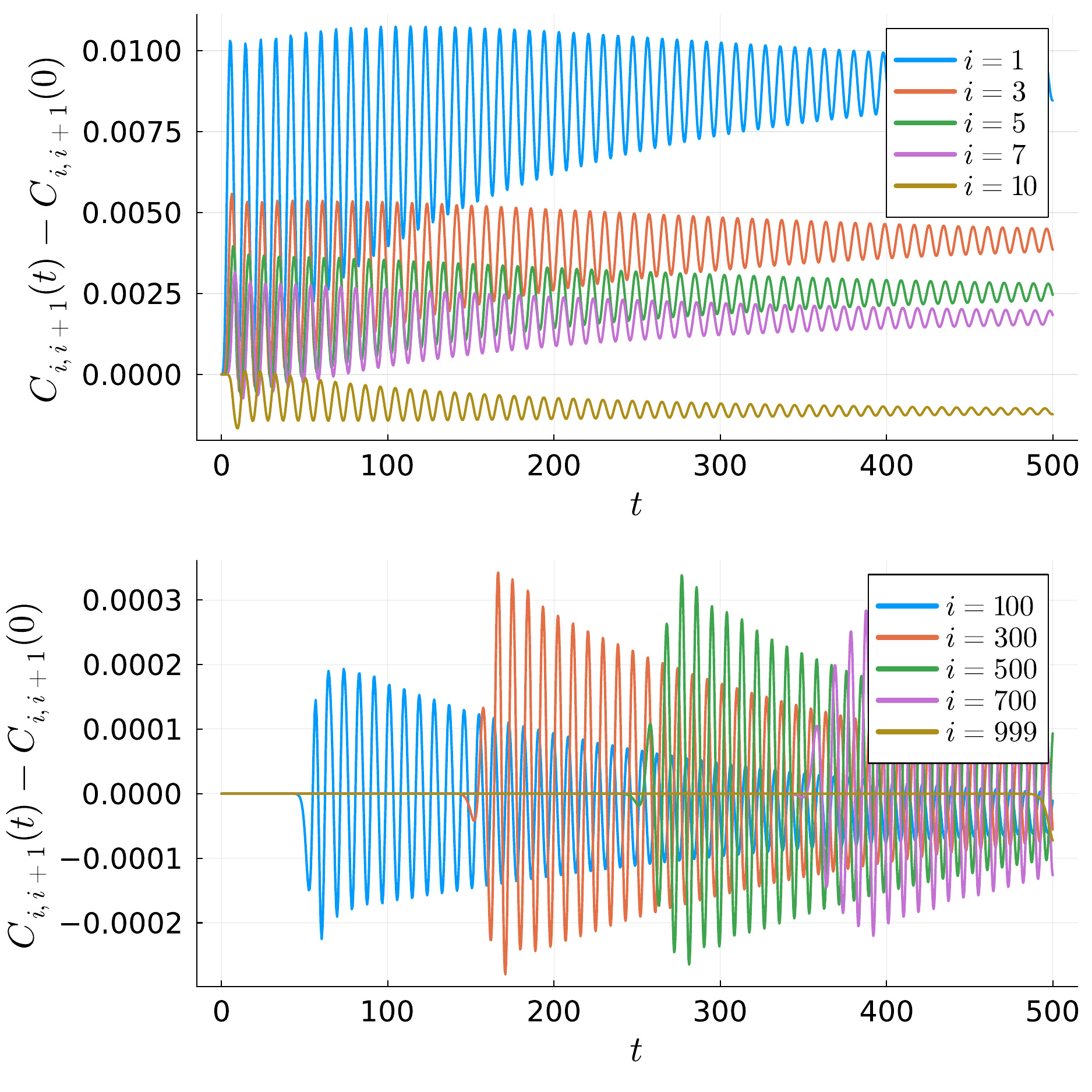}
\end{minipage}}
\hspace{0.025\textwidth}
\subfigure[$(0.5,1.0) \rightarrow (0.5,0.7) $]{
\label{Fig:Cij_case:ii}
\begin{minipage}[b]{0.45\textwidth}
\centering \includegraphics[width=1\textwidth]{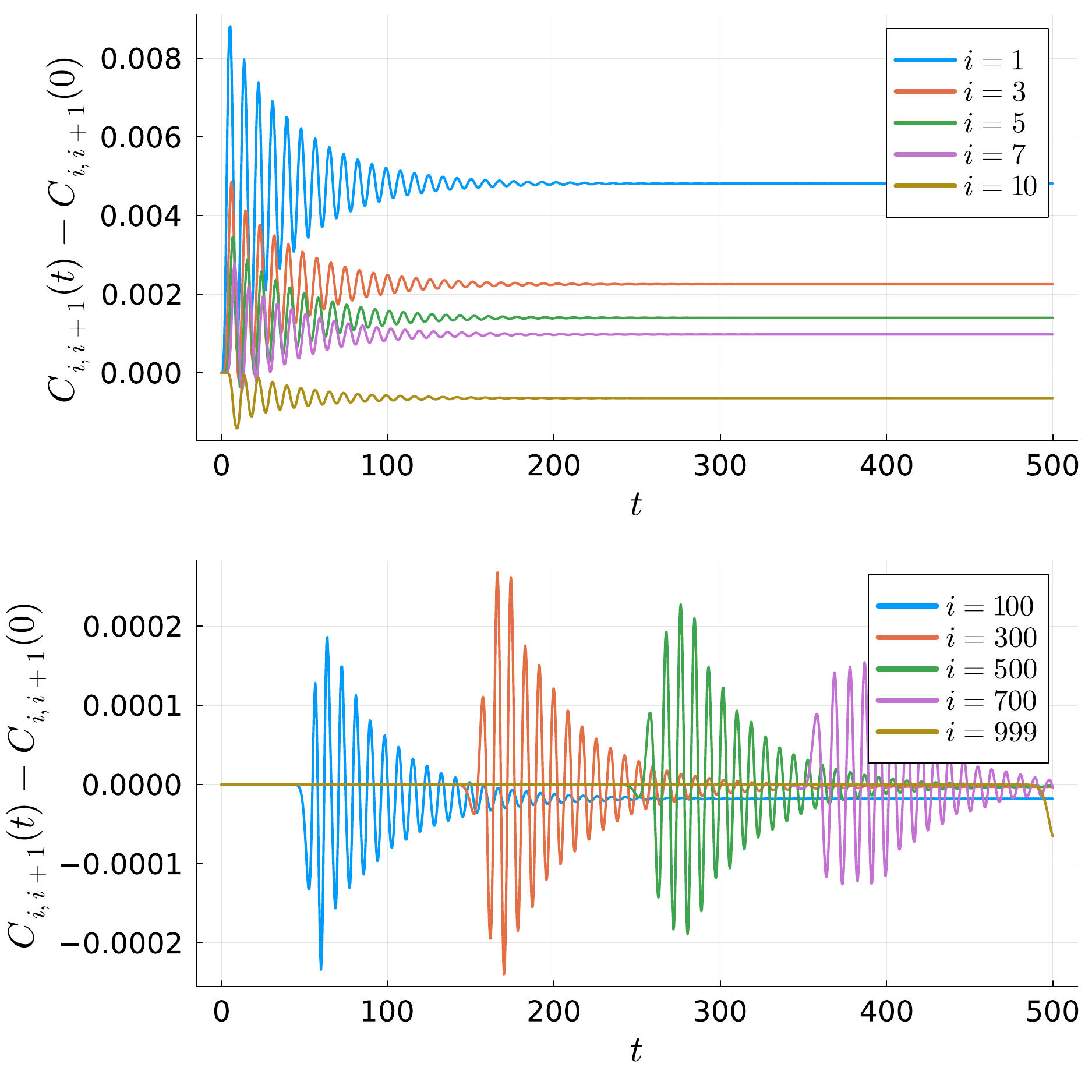}
\end{minipage}}
\\
\subfigure[$(0.5,1.0) \rightarrow (0.5,2.5)$]{
\label{Fig:Cij_case:iii}
\begin{minipage}[b]{0.45\textwidth}
\centering \includegraphics[width=1\textwidth]{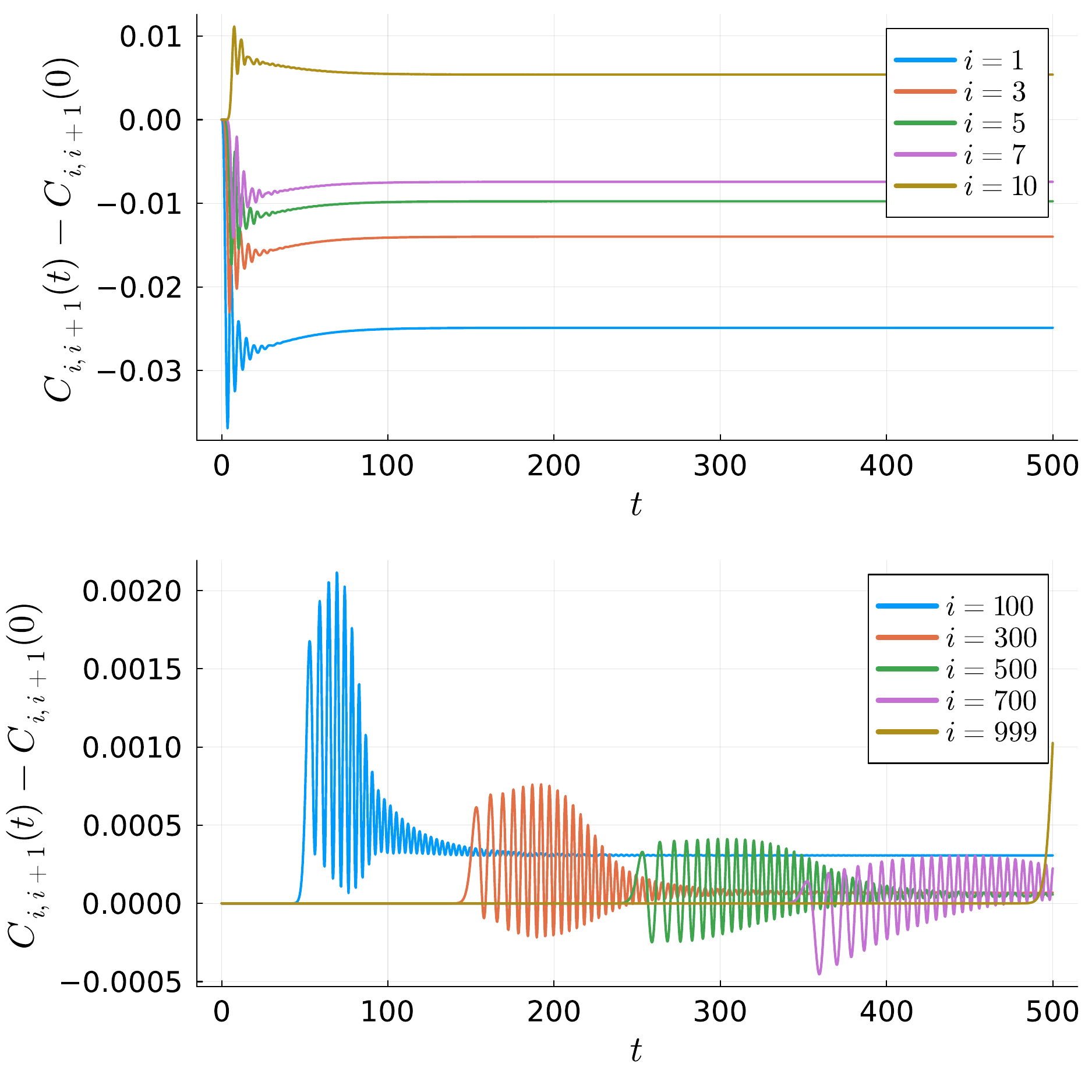}
\end{minipage}}
\hspace{0.025\textwidth}
\subfigure[$(0.5,1.0) \rightarrow (0.5,4.0)$]{
\label{Fig:Cij_case:iv}
\begin{minipage}[b]{0.45\textwidth}
\centering \includegraphics[width=1\textwidth]{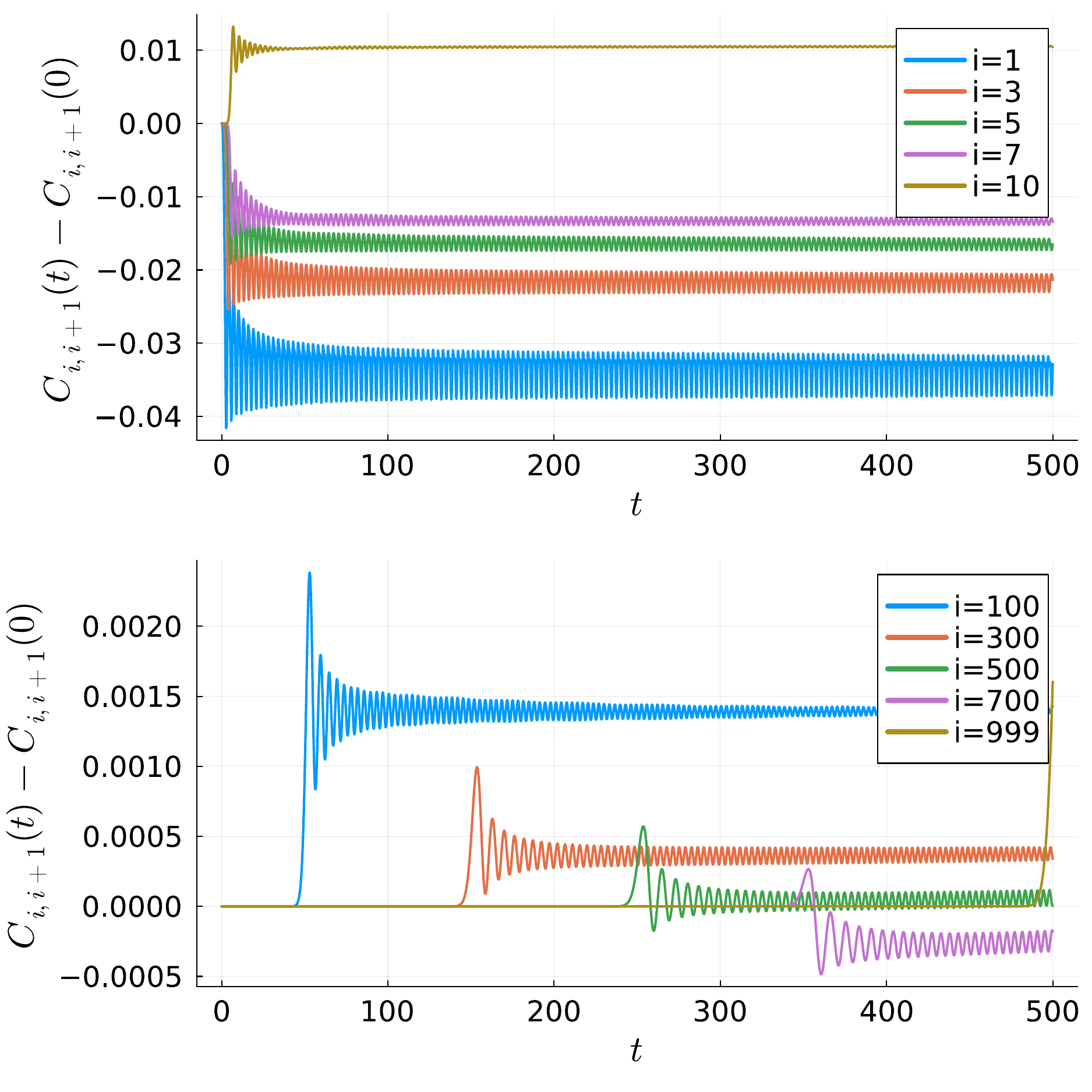}
\end{minipage}}
\caption{Times evolution of the nearest-neighbor correlators $C_{i(i+1)}(t)-C_{i(i+1)}(0)$ for the four different quenches (i)-(iv) discussed in Sec.~\ref{Results} and presented in Fig.~\ref{Fig:Cij_color_t}
at selected chain sites $i$.}
\label{Fig:Cij_color_t_slice}
\end{figure*}
\end{center}

\begin{figure}[t!]
\centering \includegraphics[width=0.45\textwidth]{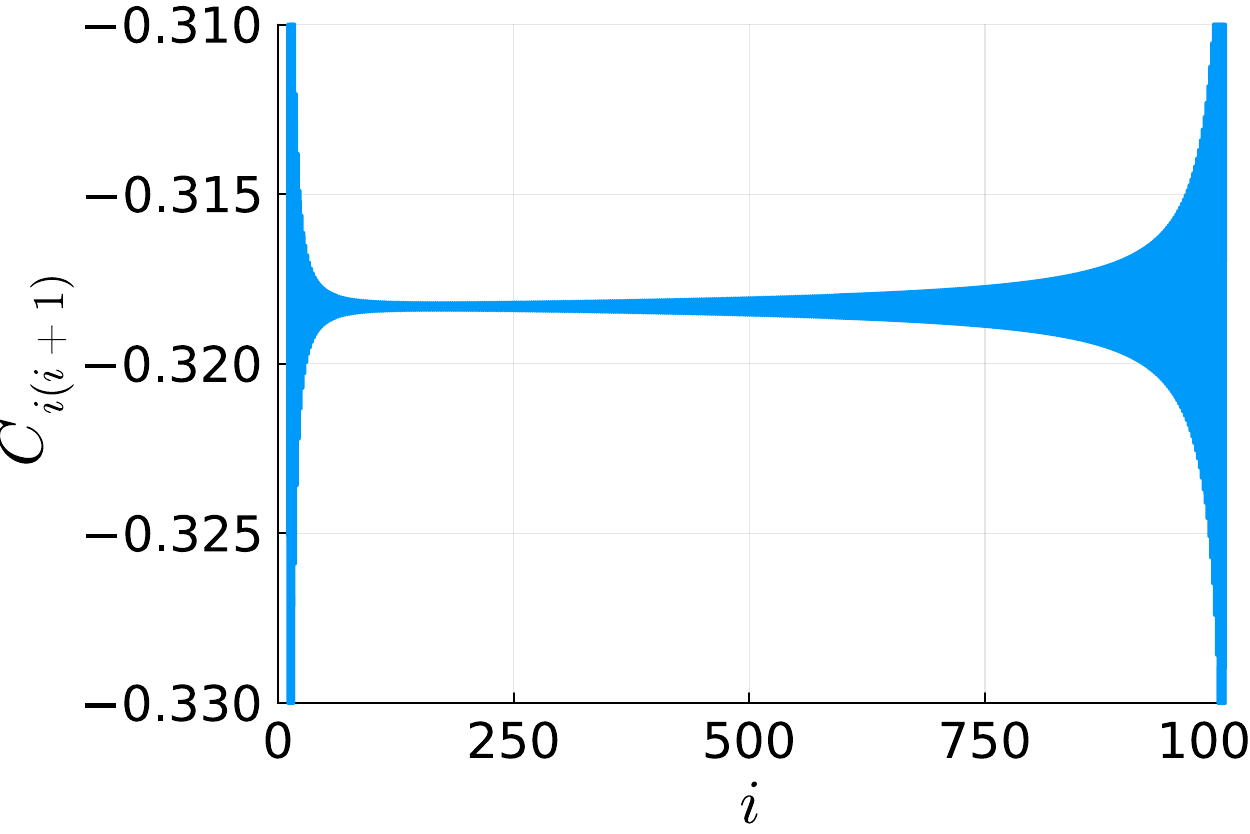} \\
\caption{$C_{i(i+1)}$ correlator at equilibrium for $V=0.5$ and $U=1.0$. 
The even and odd sites correspond to the lower and upper envelope of the function respectively.
}
\label{Fig:Cij_equilibrium}
\end{figure}

\begin{figure*}[t!]
\includegraphics[width=0.48\textwidth]{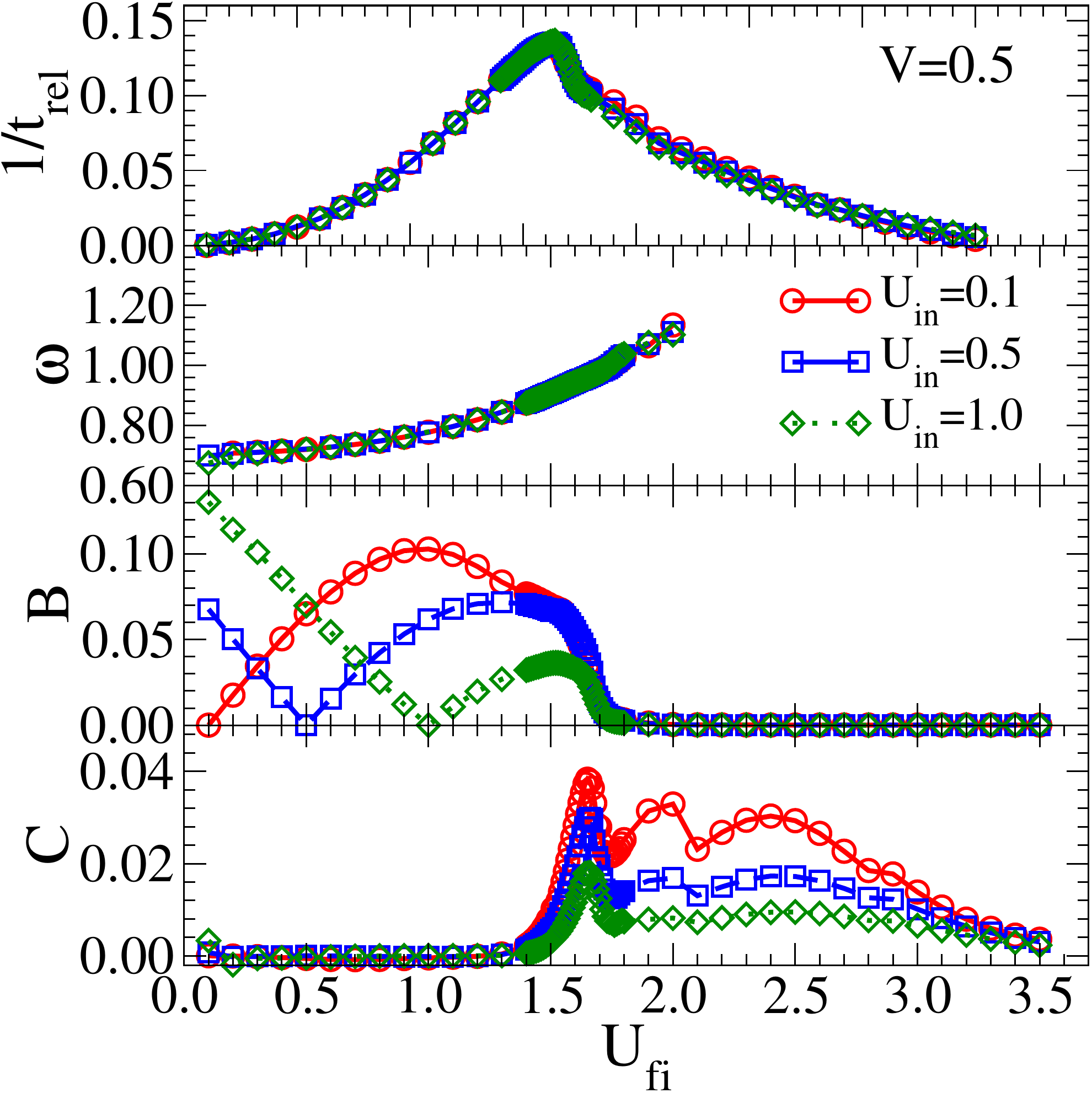}
\includegraphics[width=0.48\textwidth]{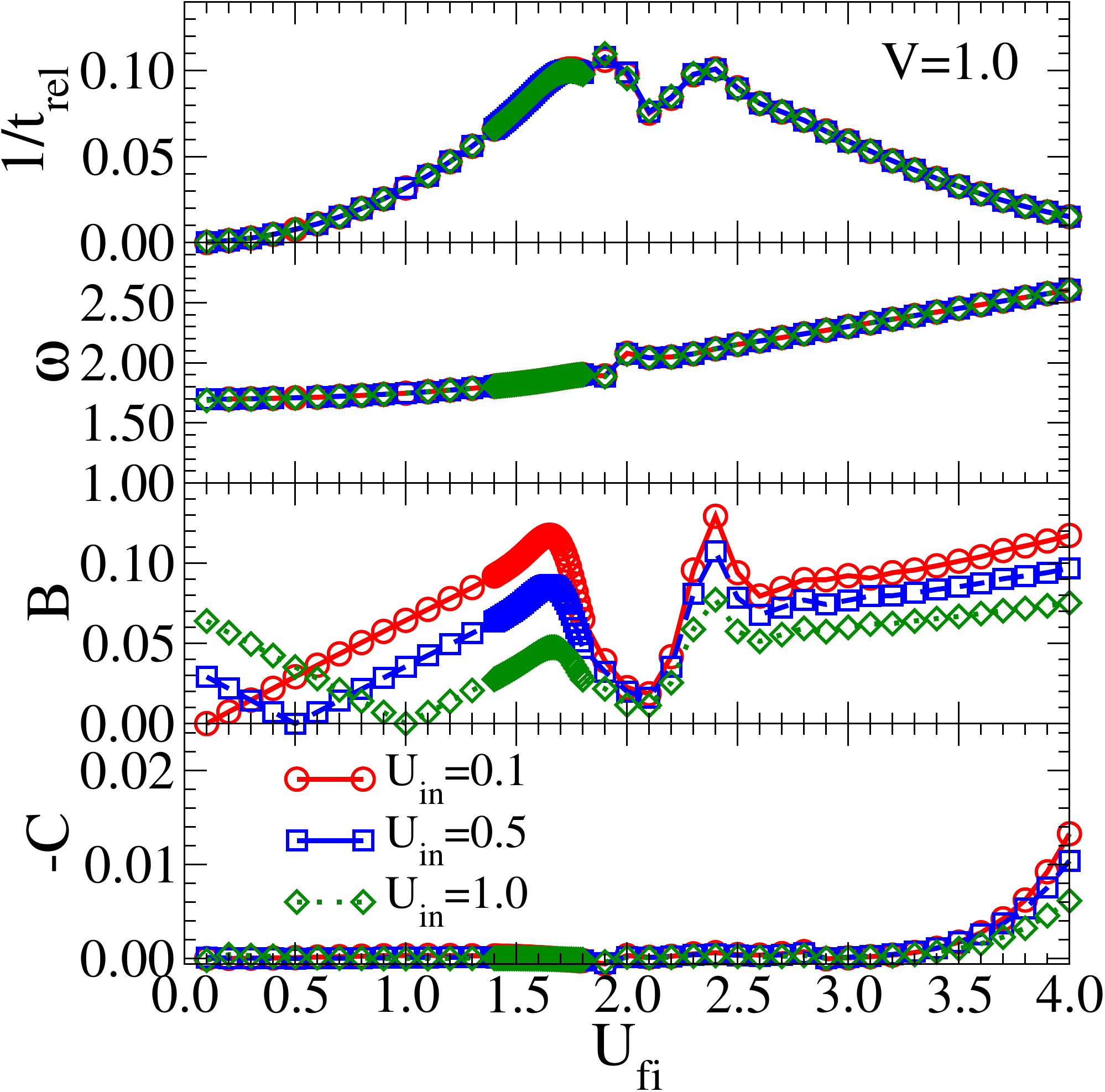}
\caption{Inverse relaxation time $1/t_{\rm rel}$, oscillation frequency $\omega$, oscillation amplitude $B$ and non-oscillating decay parameter $C$ as a function of the final interaction $U_{\rm fi}$ for different initial interactions $U_{\rm in}$ and fixed hybridization $V_{\rm in}=V_{\rm fi}=0.5$ (left) and $V_{\rm in}=V_{\rm fi}=1.0$ (right), respectively. All parameters are obtained by fitting the numerical data for the mean field parameter $\rho_{11}$ to Eq.~\eqref{Fitting}. 
}
\label{Fig:Fitting_rho11}
\end{figure*}

\begin{figure*}[t!]
\includegraphics[width=\textwidth]{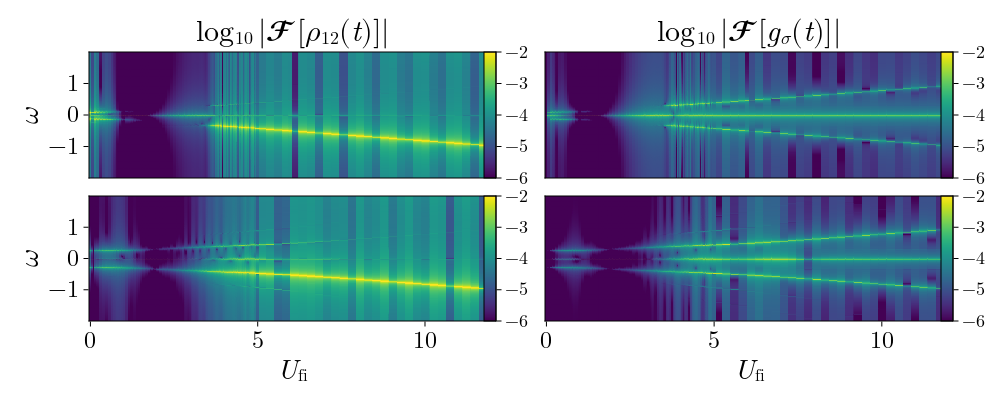}
%
\caption{Discrete Fourier transform of $\rho_{12}(t)$ [left] and $g(t)$ [right] for $(V_{\rm in},U_{\rm in})\!=\!(0.5,1.0)$ [top] and $(V_{\rm in},U_{\rm in})\!=\!(1.0,0.1)$ [bottom], respectively, as a function of $U_{\rm fi}$ (horizontal axis) and $\omega$ (vertical axis). Data are shown as a two dimensional color map on a logarithmic scale.}
\label{Fig:Fourier_g_rho12}
\end{figure*}

\section{Time evolution}
\label{App:mean field_dynamics}

\subsection{mean field dynamics of the impurity}
\label{App:mean field_dynamics_of_the_impurity}

To study the time evolution of the impurity we are using the equation of motion for the density operator
\begin{equation}
\label{App:equation_of_motion_Density_Operator}
\dot{\hat \rho}(t)= -i \left[\hat{\cal H}_\mathrm{imp}, \hat\rho(t)\right] \,.
\end{equation}
This leads to the following differential equations for the components of the density matrix
\begin{subequations}
\label{App:Term_by_term_equation_of_motion_Density_Operator}
\begin{align}
\label{rho11}
\dot\rho_{11}(t) &=-2\Im\left[V_\downarrow g_\downarrow^{*}(t)\rho_{12}(t)
+ V_\uparrow g_\uparrow^{*}(t) \rho_{13}(t) \right]
\\
\label{rho12}
\dot\rho_{12}(t) &=-i \Biggl(\Bigl(U-\mu_{\downarrow}+\varepsilon_{d,\downarrow}\Bigl)\rho_{12}(t)
+V_\downarrow g_\downarrow^{\phantom*}(t)\rho_{22}(t)
+V_\uparrow g_\uparrow^{\phantom*}(t)\rho_{23}^{*}(t)
-V_\downarrow g_\downarrow^{\phantom*}(t)\rho_{11}(t)- V_\uparrow g_\uparrow^{*}(t) \rho_{14}(t) \Biggl)
\\
\label{rho13}
\dot\rho_{13}(t) &=-i \Biggl(
 \Bigl(U-\mu_{\uparrow}+\varepsilon_{d,\uparrow}\Bigl)\rho_{13}(t)
+V_\downarrow g_\downarrow^{\phantom*}(t)\rho_{23}(t)
+V_\uparrow g_\uparrow^{\phantom*}(t)\rho_{33}(t)
- V_\uparrow g_\uparrow^{\phantom*}(t) \rho_{11}(t)
- V_\downarrow g_\downarrow^{*}(t) \rho_{14}(t) \Biggl)
\\
\label{rho14}
\dot\rho_{14}(t) &=-i \Biggl(
 \Bigl(U-\sum\limits_\sigma(\mu_{\sigma}-\varepsilon_{d,\sigma})\Bigl)\rho_{14}(t)
+V_\downarrow g_\downarrow^{\phantom*}(t)\rho_{24}(t)
+V_\uparrow g_\uparrow^{\phantom*}(t)\rho_{34}(t)
- V_\uparrow g_\uparrow^{\phantom*}(t) \rho_{12}(t)
- V_\downarrow g_\downarrow^{\phantom*}(t) \rho_{13}(t)
\Biggl)
\end{align}
\begin{align}
\label{rho22}
\dot\rho_{22}(t) &=2\Im\left[V_\downarrow g_\downarrow^{*}(t)\rho_{12}(t) - V_\uparrow g_\uparrow^{*}(t) \rho_{24}(t)\right]
\\
\label{rho23}
\dot\rho_{23}(t) &=-i \Biggl(
 V_\downarrow g_\downarrow^{*}(t)\rho_{13}(t)
-\left(2h-\varepsilon_{d,\uparrow}+\varepsilon_{d,\downarrow}\right)\rho_{23}(t)
+V_\uparrow g_\uparrow^{\phantom*}(t)\rho_{34}^{*}(t)
- V_\uparrow g_\uparrow^{\phantom*}(t) \rho_{12}^{*}(t)
- V_\downarrow g_\downarrow^{*}(t) \rho_{24}(t)
\Biggl)
\\
\label{rho24}
\dot\rho_{24}(t) &=-i \Biggl(
 V_\downarrow g_\downarrow^{*}(t)\rho_{14}(t)
-\left(\mu_{\uparrow}-\varepsilon_{d,\uparrow}\right)\rho_{24}(t)
+V_\uparrow g_\uparrow^{\phantom*}(t)\rho_{44}(t)
- V_\uparrow g_\uparrow^{\phantom*}(t) \rho_{22}(t)
- V_\downarrow g_\downarrow^{\phantom*}(t) \rho_{23}(t)
\Biggl)
\\
\label{rho33}
\dot\rho_{33}(t) &=2\Im\left[V_\uparrow g_\uparrow^{*}(t)\rho_{13}(t)
- V_\downarrow g_\downarrow^{*}(t) \rho_{34}(t) \right]
\\
\dot\rho_{34}(t) &=-i \Biggl(
 V_\uparrow g_\uparrow^{*}(t)\rho_{14}(t)
-\left(\mu_{\downarrow}-\varepsilon_{d,\downarrow}\right)\rho_{34}(t)
+V_\downarrow g_\downarrow^{\phantom*}(t)\rho_{44}(t)
- V_\uparrow g_\uparrow^{\phantom*}(t) \rho_{23}^*(t)
- V_\downarrow g_\downarrow^{\phantom*}(t) \rho_{33}(t)
\Biggl)
\\
\label{rho44}
\dot\rho_{44}(t) &=2\Im\left[V_\uparrow g_\uparrow^{*}(t)\rho_{24}(t)
+V_\downarrow g_\downarrow^{*}(t)\rho_{34}(t)\right]
\end{align}
\end{subequations}
Here we define $h=\frac{1}{2}(\mu_{\uparrow}-\mu_{\downarrow})$. We note that $\hat\rho$ is Hermitian matrix and  $\rho_{ji}^{\phantom*}(t)=\rho_{ij}^*(t)$.

Detailed derivation of these equations see in the supplemental material (see Sec. S1 in Ref.\cite{Supplemental_Material}).

\subsection{Mean field dynamics of the chain}
\label{App:mean field_dynamics_of_the_chain}

To study the dynamics of the chain, we use the equation of motion
\begin{subequations}
\label{App:equation_of_motion}
\begin{align}
&\frac{d}{dt}C_{ij,\sigma}(t) = i \langle [\hat{\cal H}_{\mathrm{chain},\sigma}, \hat C_{ij,\sigma}]\rangle(t)
\\
&\frac{d}{dt}K_{ij,\sigma}(t) = i \langle [\hat{\cal H}_{\mathrm{chain},\sigma}, \hat K_{ij,\sigma}]\rangle(t)
\end{align}
\end{subequations}
We obtain
\begin{subequations}
\label{App:Time_evolution_of_correlators_for_Chain}
\begin{itemize}
\item {\boldmath$i\neq 0, r$ , $j \neq 0, r$}
\begin{align}
\label{C_inot0r_jnot0r}
{\dot C}_{ij,\sigma}(t) &=i\Bigl(J_{i-1,\sigma} C_{(i-1)j,\sigma}(t)
+ J_{i,\sigma} C_{(i+1)j,\sigma}(t)
- J_{j-1,\sigma} C_{i(j-1),\sigma}(t)
- J_{j,\sigma} C_{i(j+1),\sigma}(t)
+\left(\varepsilon_{i,\sigma} - \varepsilon_{j,\sigma}\right)C_{ij,\sigma}(t)\Bigl)
\\
\label{K_inot0r_jnot0r}
{\dot K}_{ij,\sigma}(t) &=
-i\Bigl(J_{i-1,\sigma} K_{(i-1)j,\sigma}(t)+J_{i,\sigma} K_{(i+1)j,\sigma}(t)
+J_{j-1,\sigma} K_{i(j-1),\sigma}(t)
+J_{j,\sigma} K_{i(j+1),\sigma}(t)
\nonumber \\
&+\left(\varepsilon_{i,\sigma} + \varepsilon_{j,\sigma}-2\mu_{\sigma}(t)\right)K_{ij,\sigma}(t)
\Bigl)
\\
\label{K_inot0r}
K_{ii,\sigma}(t) &=0
\end{align}

\item {\boldmath$i=r$, $j \neq 0, r$}
\begin{align}
\label{C_ir_jnot0r}
{\dot C}_{rj,\sigma}(t) &=i\Bigl(J_{r-1,\sigma} C_{(r-1)j,\sigma}(t)
+ J_{r,\sigma} C_{(r+1)j,\sigma}(t)
- J_{j-1,\sigma} C_{r(j-1),\sigma}(t)
- J_{j,\sigma} C_{r(j+1),\sigma}(t)
\nonumber \\
&+\left(\varepsilon_{r,\sigma} - \varepsilon_{j,\sigma}\right)C_{rj,\sigma}(t)
+ V_\sigma f_\sigma^{*} (t) C_{0j,\sigma}(t)
\Bigl)
\\
\label{K_ir_jnot0r}
{\dot K}_{rj,\sigma}(t) &=-i\Bigl(
J_{r-1,\sigma} K_{(r-1)j,\sigma}(t)
+J_{r,\sigma} K_{(r+1)j,\sigma}(t)
+J_{j-1,\sigma} K_{r(j-1),\sigma}(t)
+J_{j,\sigma} K_{r(j+1),\sigma}(t)
\nonumber \\
&+\left(\varepsilon_{r,\sigma} + \varepsilon_{j,\sigma}-2\mu_{\sigma}(t)\right)K_{rj,\sigma}
+ V_\sigma f_\sigma^{\phantom*} (t)K_{0j,\sigma}(t)
\Bigl)
\end{align}

\item  {\boldmath$i=j = r$}
\begin{align}
\label{C_ir_jr}
{\dot C}_{rr,\sigma}(t) &=-2\Im \Bigl(J_{r-1,\sigma}C_{(r-1)r,\sigma}(t)
- J_{r,\sigma} C_{r(r+1),\sigma}(t)
+ V_\sigma f_\sigma^{*} (t) C_{0r,\sigma}(t)
\Bigl)
\\
\label{K_ir_jr}
K_{rr,\sigma}(t) &=0
\end{align}

\item {\boldmath$i=0$ , $j \neq 0, r$}
\begin{align}
\label{C_i0_jnot0r}
{\dot C}_{0j,\sigma} (t)&=-i\Bigl(J_{j-1,\sigma} C_{0(j-1),\sigma}(t)
+ J_{j,\sigma} C_{0(j+1),\sigma}(t)
- (\mu_{\sigma}(t)-\varepsilon_{j,\sigma}) C_{0j,\sigma}(t)
-2 V_\sigma f_\sigma^{\phantom*} (t) C_{rj,\sigma}(t)
+2 V_\sigma f_\sigma^{*} (t) K_{rj,\sigma}(t)
\Bigl)
\end{align}

\item {\boldmath$i=0$ , $j= r$}
\begin{align}
\label{C_i0_jr}
{\dot C}_{0r,\sigma}(t) &=-i\Bigl(J_{r-1,\sigma} C_{0(r-1),\sigma}(t)
+ J_{r,\sigma} C_{0(r+1),\sigma}(t)
- (\mu_{\sigma}(t)-\varepsilon_{r,\sigma}) C_{0r,\sigma}(t)
- 2 V_\sigma f_\sigma^{\phantom*} (t) C_{rr,\sigma}(t)
+ V_\sigma f_\sigma^{\phantom*} (t)\Bigl)
\end{align}

\item {\boldmath$i=j= 0$}
\begin{align}
\label{C_i0_j0}
C_{00,\sigma} (t) &=1
\end{align}
\end{itemize}
\end{subequations}
Here we note that for $x=1$ we have that $(x-1) \rightarrow L$ and for $x=L$ we have that $(x+1) \rightarrow 1$.
For PBC $J_{L,\sigma}$ is a finite value, while for OBC $J_{L,\sigma}=0$, therefore corresponding terms will be not there.

Finally, we would like to note that we have ${L(L+1)}$ differential equations for the chain.
There are ${L(L+3)/2}$ equations for $C_{ij}$ and ${L(L-1)/2}$ for $K_{ij}$. Since the diagonal elements of the correlation function $K_{ij}$ are identical to zero and ${C_{00,\sigma} \equiv 1}$, we do not write differential equations for them.  

A detailed derivation of these equations is in the supplemental material (see Sec. S2 in Ref.\cite{Supplemental_Material}).

\subsection{Beyond the half-filling}
\label{App:mean_field_dynamics_beyond_the_half_filling}

Similar to equilibrium, during the time evolution, the half-filling can be ensured by taking ${\mu_\sigma=0}$ and ${\varepsilon_{d}=-U/2}$. 
Away from the half-filling, in addition to the differential Eqs.~\eqref{App:Term_by_term_equation_of_motion_Density_Operator} and~\eqref{App:Time_evolution_of_correlators_for_Chain}, we need other equations to adjust the chemical potential so that the correct filling $N_\sigma$ is present at each time step. For this purpose, we again need to use Eq.~\eqref{App:Fixing_filling}, but now $\rho_{ii}$ and $C_{ii}$ are functions of $t$.
{Since this procedure is numerically expensive, in this manuscript, we restrict our calculations to half-filled systems, where adjusting the chemical potential is unnecessary, as the condition $\mu_\sigma = 0$ already ensures half-filling, as mentioned above.}

\subsection{Resonance level model}
\label{App:Resonance_level_model}

The Hamiltonian for the resonance level model (RLM) is similar to the Hamiltonian $\hat{\cal H}_{\mathrm{chain},\sigma}$, which we consider in our method to describe the behavior of the chain. 
The difference is that it does not contain a coupling to the Majorana fermion.
On this basis, the equations describing the time evolution of the RLM are also similar to those presented in Sec.~\ref{App:mean field_dynamics_of_the_chain}. 
The only difference is that the coupling to the Majorana fermion should be neglected and the first site should be considered as an impurity. 
Accordingly, then $\varepsilon_{1,\sigma}$ corresponds to the local energy of the impurity, and $J_{1,\sigma}$ corresponds to the hybridization.

\section{Additional numerical results}
\label{App:Additional_numerical_results}

In this Appendix, we present some additional results which we do not address in the main text. In particular, we present:
the time evolution of selected off-diagonal density matrix elements $\rho_{i\ne j}(t)$, nearest-neighbor chain correlators $C_{i(i+1)}(t)$, fitting parameters for $\rho_{11}(t)$, and Fourier transform analysis of $g(t)$ and $\rho_{12}$.   

\subsection{\texorpdfstring{Off-diagonal density matrix $\rho_{i\ne j}(t)$}{Off-diagonal density matrix rho\_ij(t)}}
\label{App:Rho_offdiagonal}
In this section, we complement our analysis of the time evolution of the density matrix $\rho_{ij}(t)$ by presenting selected off-diagonal elements $\rho_{i\ne j}(t)$.
As already pointed out in the main text, for the half-filled and SU($2$) symmetric system only the imaginary parts of $\rho_{12}(t)\!=\!\rho_{13}(t)\!=\!\rho^*_{24}(t)\!=\!\rho^*_{34}(t)$ carry additional information independent of the mean field parameter $f(t)$ and the diagonal density matrix elements $\rho_{ii}(t)$ (which are discussed in detail in Sec.~\ref{Results}).
Hence, only the imaginary parts of these components are presented in Fig.~\ref{Fig:Imrho_ij_t}. 

For quenches to the parameter regimes (i)-(iii), where the real parts of all observables decay to their equilibrium values defined by the final parameters $(V_{\rm fi},U_{\rm fi})$ after the quench, the imaginary parts converge to zero [see Figs.~\ref{Fig:imrho_ij_case:i}, \ref{Fig:imrho_ij_case:ii} and \ref{Fig:imrho_ij_case:iii}].
Interestingly, non-oscillatory contributions are absent for $\Im \rho_{i\ne j}(t)$ in regimes (i) and (iii) [Figs.~\ref{Fig:imrho_ij_case:i} and \ref{Fig:imrho_ij_case:iii}] in contrast to the corresponding results for $f(t)$ and $g(t)$ in Figs.~\ref{Fig:Case:i} and \ref{Fig:Case:iii} where such non-oscillatory drift can be observed.
In particular, for regime (iii) where oscillations are absent in $f(t)$, $g(t)$ and $\rho_{ii}(t)$ and only the non-oscillatory term remains, the imaginary part of $\rho_{i\ne j}(t)$ vanishes immediately after the settling time $t_{\rm S}$ as can be seen in Fig.~\ref{Fig:imrho_ij_case:iii}.
This can be understood by expressing the full density matrix $\rho_{ij}(t)$ in terms of the fitting function in Eq.~\eqref{Fitting} where $\cos(\omega t+\phi)$ is replaced by $e^{i(\omega t+\phi)}$ keeping all fitting parameters real.

In regime (iv), where $f(t)$, $g(t)$, $\rho_{ii}(t)$ do not decay to their equilibrium values but exhibit undamped oscillatory behavior at large times, also the imaginary part of the off-diagonal elements of the density matrix features persistent oscillations and does not decay to zero, as be seen in Fig.~\ref{Fig:imrho_ij_case:iv}. However, these persistent oscillations are centered around zero so that a time average about an oscillation period vanishes.

\subsection{\texorpdfstring{Nearest-neighbor correlators $C_{i(i+1)}(t)$}{Nearest-neighbor correlators C\_i(i+1)(t)}}
\label{App:Correlator}

In Sec.~\ref{Results} we have presented color maps of $C_{i(i+1)}(t)$ as a function of the time $t$ and the chain site $i$ for four different quenches $(V_{\rm in},U_{\rm in})\!\rightarrow\!V_{\rm fi},U_{\rm fi})$ [see Fig.~\ref{Fig:Cij_color_t}]. 
In this section, we complement our results by depicting the time evolution of $C_{i(i+1)}(t)$ for selected chain sites $i$ in Fig.~\ref{Fig:Cij_color_t_slice}. This allows us to better recognize the four different regimes (i)-(iv) of long-time behavior discussed in Sec.~\ref{Results} and, in particular, to estimate the size of oscillations at a given chain site $i$. 
We observe that overall the amplitude of the oscillations decays with increasing distance from the impurity (consider the different scales on the $y$-axis in the upper and lower plot in each of the four panels of Fig.~\ref{Fig:Cij_color_t_slice}). 
This is the expected behavior as the effect of a quench at the impurity site should indeed be less pronounced for chain sites far away from the impurity. However, for $i\!\gtrsim\!100$ this hierarchy of oscillations amplitudes is reversed for quenches (i) and (ii). In particular, in Figs.~\ref{Fig:Cij_case:i} and \ref{Fig:Cij_case:ii} the oscillation amplitude at $i\!=\!100$ is smaller than the ones at $i=300$ and $i=500$.
This can be explained by the observation that for the equilibrium we are starting from, the distance between the lower and upper envelope of $C_{i(i+1)}$, i.e. $\Delta C_{i} = \lvert C_{(i+1)(i+2)} - C_{i(i+1)} \rvert$ has a minimum at $i\!=\!160$, as can be seen in Fig.~\ref{Fig:Cij_equilibrium}. Since, for quenches to small values of $U_{\rm fi}$ the energy scale of the chain is comparable (or even larger) than the one of the impurity we can indeed expect that the size of the oscillations is mainly governed by the initial chain correlator. On the contrary, for larger values of $U_{\rm fi}$ (see Figs.~\ref{Fig:Cij_case:iii} and \ref{Fig:Cij_case:iv}) the dominant energy scale is the Hubbard interaction at the impurity and the initial values for the nearest-neighbor correlators play only a secondary role.

\subsection{\texorpdfstring{Fitting parameters for $\rho_{11}(t)$}{Fitting parameters for rho11(t)}}
\label{App:Fitting}

In this Appendix, we briefly discuss the evolution of the fitting parameters $1/t_{\rm rel}$, $\omega$, $B$ and $C$ obtained by fitting $\rho_{11}(t)$ to Eq.~\eqref{Fitting} as a function of $U_{\rm fi}$.
Data are presented in Fig.~\ref{Fig:Fitting_rho11} for $V_{\rm in}\!=\!0.5$ (left panel) and $V_{\rm in }\!=\!1.0$ (right panel) for three different values of initial interaction $U_{\rm in}$ given in the figures.
For the fitting parameters $1/t_{\rm rel}$ and $\omega$ we obtained the same results as for the fitting of the mean field parameters $f(t)$ in Fig.~\ref{Fig:Fitting_f}.
As expected, the oscillation amplitude $\omega$ and the inverse relaxation time $1/t_{\rm rel}$ are the same for all components of the density matrix $\rho_{ij}$ and the mean field parameters $f$ and $g$. 
Regarding the fitting parameters $B$ and $C$, we get a slightly different behavior. 
For very small values of $U_{\rm fi}$ regime (i) defined by $C\!\ne\!0$ for $f(t)$ is absent in the fits of $\rho_{11}(t)$, i.e., no non-oscillatory contribution can be observed in $\rho_{11}(t)$ for $U_{\rm fi}\!\lll\!1$.
Moreover, for these small values of $U_{\rm fi}$ $B$ decays linearly in contrast to a non-monotonic behavior observed for a fit of $f(t)$.
Apart from these small differences, the behavior of the fitting parameters is the same as in Fig.\ref{Fig:Fitting_f}).

\subsection{\texorpdfstring{Fourier transform of $g(t)$ and $\rho_{12}(t)$}{Fourier transform of g(t) and rho12(t)}}
\label{App:Fourier}

In this section, we extend our discussion of the Fourier transform of $\rho_{11}(t)$ and $f(t)$ Fig.~\ref{Fig:Fourier} to $g(t)$ and $\rho_{12}(t)$ in the right and left panel of Fig.~\ref{Fig:Fourier_g_rho12}, respectively.
In the upper and lower panels, we show results for two different sets of initial values specified in the caption of the figure.
Data is presented in the form of a color map as a function of $U_{\rm fi}$ ($x$-axis) and the frequency $\omega$ ($y$-axis).
The results for $g(t)$ in the right panels are completely analogous to the results for $f(t)$ in the right panels of Fig.~\ref{Fig:Fourier} and, hence, do not need further discussion.
For the Fourier transform of $\rho_{12}(t)$ we observe a bright feature for $U_{\rm fi}\!>\!3.75$ only at negative $\omega$ in contrast to $\rho_{11}(t)$ where this feature appears to be symmetric with respect to $\omega\!=\!0$.
This behavior originates from the imaginary part which emerges in $\rho_{12}(t)$ at large final interactions while $\rho_{11}(t)$ is purely real in the entire parameter regime.
The complex conjugate of $\rho_{12}(t)$, which is equivalent to $\rho_{34}(t)$ exhibits a bright feature at positive $\omega$.
Considering Eq.~\eqref{Define_f_via_rho_down} the combination of these two density elements leads to the two bright features of $f(t)$ which are symmetric with respect to $\omega\!=\!0$.


\end{appendix}


\begin{thebibliography}{100}
\bibitem{Quenching}
In experimental setups, the Hubbard interaction $U$ in systems like ultracold
  atoms or condensed matter systems can be manipulated in various ways. The key
  factor is the ratio of the on-site interaction $U$ to the hopping term $J$,
  i.e., $U/J$. This ratio can be controlled through the following techniques.
  In ultracold atomic gases: (i) Feshbach Resonances allow the control of the
  scattering length between atoms, which directly changes the value of $U$;
  (ii) the depth of the optical lattice can be adjusted, altering the hopping
  amplitude $J$, which consequently changes the $U/J$ ratio. Similarly, the
  hybridization $V$ can be controlled to keep $V /J$ constant. On the other
  hand in materials and heterostructures in condensed matter: (i) applying
  mechanical strain to a material can change the electron overlap between
  neighboring atoms, modifying the hopping term $J$. Changing the distance
  between the impurity and the chain can alter the hybridization $V$ , allowing
  control over the $U/J$ ratio while keeping $V /J$ constant; (ii) gate
  voltages can be applied to change the electronic properties, including the
  hopping and interaction terms. In some cases, gate voltages can also tune the
  Hubbard interaction $U$ by adjusting the carrier density.
\bibitem{Mean-field-artifact}
One example is the two impurity Anderson model. As hybridization decreases,
  this model undergoes a crossover between the Kondo regime observed at large
  values of hybridization and the Ruderman-Kittel-Kasuya-Yosida (RKKY) regime
  for small values of hybridization. This crossover is well described by the
  density matrix renormalization group (DMRG). On the other hand, if we try to
  address the problem using dynamicalo mean field theory (DMFT), we obtain a
  phase transition. In fact, comparison of the local and nonlocal magnetic
  susceptibilities for the impurities obtained by DMRG and DMFT shows good
  agreement for large values of hybridization when the system is in the Kondo
  regime, but for small values of hybridization the magnetic susceptibilities
  obtained by DMFT diverge, indicating a phase transition, while the
  susceptibilities obtained by DMGR continue to increase, indicating a
  crossover to the RKKY regime.\cite{ti.sc.12}.
\bibitem{NRG}
In standard numerical renormalization group (NRG) calculations, the impurity is
  coupled to a non-interacting fermionic bath described by a flat band. In this
  case, the hopping coefficients for the NRG chain are given by an analytical
  expression\cite{wils.75}. To make the comparison with our mean-field
  calculations more straightforward, we consider mapping to an infinite linear
  chain, unlike standard NRG calculations. In this scenario, the hopping
  coefficients are no longer determined analytically but must be calculated
  numerically (see details in Ref.~\cite{hofs.00}). Here are some technical
  details: we use the NRG parameters $\Lambda=2.0$, $N_{\rm iter}=60$, and
  $N=1000$ kept states (using number of up and down fermions as a good quantum
  number).
\bibitem{expectation_for_f}
The expectation value for $f$ in Eq.~\eqref{fgexpansionf} can be, in fact,
  calculated exactly since the corresponding Hilbert space is only four
  dimensional. The expansion in $g$ can be then obtained from this exact
  solution by a standard Talyor series expansion.
\bibitem{Neglecting_Constant}
This constant term must also be taken into account in equilibrium calculations
  when comparing the energies of two states during hysteresis.
\bibitem{Schroedinger_equation}
In fact, for the linear Schr\"odinger equation
  ${i\frac{d}{dt}\lvert{\psi}(t)\rangle\!=\!H\lvert\psi(t)\rangle}$ with the
  initial condition ${\lvert\psi(t\!=\!t_0)\rangle\!=\!\lvert\psi_0\rangle}$ we
  can express the solution as
  $\lvert\psi(t)\rangle\!=\!S(t,t_0)\lvert\psi_0\rangle$ where the time
  evolution operator $S(t,t_0)$ satisfies the linear differential equation
  ${\frac{d}{dt}{S}(t,t_0)\!=\!HS(t,t_0)}$ with the initial condition
  $S(t_0,t_0)\!=\!\mathds{1}$. The entire dynamics characterized by oscillation
  frequencies and decay rates is then determined by the time evolution operator
  $S(t,t_0)$ which is completely independent of the initial conditions and,
  hence, $t_{\rm rel}$ and $\omega$ cannot depend on $(V_{\rm in},U_{\rm in})$.
  Let us point out that the situation is not so clear in our case as we are
  concerned with a {\em nonlinear} system of differential equations, but our
  numerical results demonstrate that this feature is indeed preserved.
\end{thebibliography}
\bibliographystyle{aipnum4-1}

\pagebreak
\begin{center}
\textbf{\large Supplemental Materials: Mean Field Decoupling of Single Impurity Anderson Model through Auxiliary Majorana Fermions.}
\end{center}

\setcounter{section}{0}
\renewcommand{\thesection}{S\arabic{section}}
\renewcommand{\theequation}{S.\arabic{equation}}

\section{Derivation of the expressions for the time evolution for impurities}
\label{S:Derivation_for_time_evolution_Impurity}

The equation of motion for the impurity reads
\begin{equation}
\label{S:equation_of_motion_Density_Operator}
\dot{\hat \rho}(t)= -i \left[\hat{\cal H}_{imp}, \hat\rho(t)\right] \,.
\end{equation}
This leads to the following differential equations for the components of the density matrix:
\begin{subequations}
\label{S:Term_by_term_equation_of_motion_Density_Operator}
\begin{align}
\label{S:rho11}
\dot\rho_{11}(t)&=-i \Bigl(
 {\cal H}_{11}\rho_{11}(t)
+{\cal H}_{12}\rho_{21}(t)
+{\cal H}_{13}\rho_{31}(t)
+{\cal H}_{14}\rho_{41}(t)
-\rho_{11}(t){\cal H}_{11}
-\rho_{12}(t){\cal H}_{21}
-\rho_{13}(t){\cal H}_{31}
-\rho_{14}(t){\cal H}_{41}
\Bigl)
\nonumber\\
&=-i\Biggl(g_\downarrow^{\phantom*}(t)\rho_{21}(t) +g_\uparrow^{\phantom*}(t)\rho_{31}(t)
-\rho_{12}(t)g_\downarrow^{*}(t)
-\rho_{13}(t)g_\uparrow^{*}(t)
\Biggl)
=-2\Im\left[g_\downarrow^{*}(t)\rho_{12}(t)
+ g_\uparrow^{*}(t) \rho_{13}(t) \right]
\\
\label{S:rho12}
\dot\rho_{12}(t)&=-i \Bigl(
 {\cal H}_{11}\rho_{12}(t)
+{\cal H}_{12}\rho_{22}(t)
+{\cal H}_{13}\rho_{32}(t)
+{\cal H}_{14}\rho_{42}(t)
-\rho_{11}(t){\cal H}_{12}
-\rho_{12}(t){\cal H}_{22}
-\rho_{13}(t){\cal H}_{32}
-\rho_{14}(t){\cal H}_{42}
\Bigl)
\nonumber\\
&=-i \Biggl(\Bigl(U+\sum\limits_\sigma(\mu_{\sigma}-\varepsilon_{d,\sigma})\Bigl)\rho_{12}(t)
+g_\downarrow^{\phantom*}(t)\rho_{22}(t)
+g_\uparrow^{\phantom*}(t)\rho_{32}(t)
-\rho_{11}(t)g_\downarrow^{\phantom*}(t)
-\rho_{12}(t)\left(\mu_{\uparrow}-\varepsilon_{d,\uparrow}\right)
-\rho_{14}(t)g_\uparrow^{*}(t)
\Biggl)
\nonumber \\
&=-i \Biggl(\Bigl(U+\mu_{\downarrow}-\varepsilon_{d,\downarrow}\Bigl)\rho_{12}(t)
+g_\downarrow^{\phantom*}(t)\rho_{22}(t)
+g_\uparrow^{\phantom*}(t)\rho_{23}^{*}(t)
-g_\downarrow^{\phantom*}(t)\rho_{11}(t)- g_\uparrow^{*}(t) \rho_{14}(t) \Biggl)
\\
\label{S:rho13}
\dot\rho_{13}(t)&=-i \Bigl(
 {\cal H}_{11}\rho_{13}(t)
+{\cal H}_{12}\rho_{23}(t)
+{\cal H}_{13}\rho_{33}(t)
+{\cal H}_{14}\rho_{43}(t)
-\rho_{11}(t){\cal H}_{13}
-\rho_{12}(t){\cal H}_{23}
-\rho_{13}(t){\cal H}_{33}
-\rho_{14}(t){\cal H}_{43}
\Bigl)
\nonumber\\
&=-i \Biggl(
 \Bigl(U+\sum\limits_\sigma(\mu_{\sigma}-\varepsilon_{d,\sigma})\Bigl)\rho_{13}(t)
+g_\downarrow^{\phantom*}(t)\rho_{23}(t)
+g_\uparrow^{\phantom*}(t)\rho_{33}(t)
-\rho_{11}(t)g_\uparrow^{\phantom*}(t)
-\rho_{13}(t)\left(\mu_{\downarrow}-\varepsilon_{d,\downarrow}\right)
-\rho_{14}(t)g_\downarrow^{*}(t) \Biggl)
\nonumber\\
&=-i \Biggl(
 \Bigl(U+\mu_{\uparrow}-\varepsilon_{d,\uparrow}\Bigl)\rho_{13}(t)
+g_\downarrow^{\phantom*}(t)\rho_{23}(t)
+g_\uparrow^{\phantom*}(t)\rho_{33}(t)
- g_\uparrow^{\phantom*}(t) \rho_{11}(t)
- g_\downarrow^{*}(t) \rho_{14}(t) \Biggl)
\\
\label{S:rho14}
\dot\rho_{14}(t)&=-i \Bigl(
 {\cal H}_{11}\rho_{14}(t)
+{\cal H}_{12}\rho_{24}(t)
+{\cal H}_{13}\rho_{34}(t)
+{\cal H}_{14}\rho_{44}(t)
-\rho_{11}(t){\cal H}_{14}
-\rho_{12}(t){\cal H}_{24}
-\rho_{13}(t){\cal H}_{34}
-\rho_{14}(t){\cal H}_{44}
\Bigl)
\nonumber\\
&=-i \Biggl(
 \Bigl(U+\sum\limits_\sigma(\mu_{\sigma}-\varepsilon_{d,\sigma})\Bigl)\rho_{14}(t)
+g_\downarrow^{\phantom*}(t)\rho_{24}(t)
+g_\uparrow^{\phantom*}(t)\rho_{34}(t)
-\rho_{12}(t)g_\uparrow^{\phantom*}(t)
-\rho_{13}(t)g_\downarrow^{\phantom*}(t)
\Biggl)
\\
\label{S:rho22}
\dot\rho_{22}(t)&=-i \Bigl(
 {\cal H}_{21}\rho_{12}(t)
+{\cal H}_{22}\rho_{22}(t)
+{\cal H}_{23}\rho_{32}(t)
+{\cal H}_{24}\rho_{42}(t)
-\rho_{21}(t){\cal H}_{12}
-\rho_{22}(t){\cal H}_{22}
-\rho_{23}(t){\cal H}_{32}
-\rho_{24}(t){\cal H}_{42}
\Bigl)
\nonumber\\
&=-i \Biggl(
 g_\downarrow^{*}(t)\rho_{12}(t)
+g_\uparrow^{\phantom*}(t)\rho_{42}(t)
-\rho_{21}(t)g_\downarrow^{\phantom*}(t)
-\rho_{24}(t)g_\uparrow^{*}(t)
\Biggl)
=2\Im\left[g_\downarrow^{*}(t)\rho_{12}(t) - g_\uparrow^{*}(t) \rho_{24}(t)\right]
\\
\label{S:rho23}
\dot\rho_{23}(t)&=-i \Bigl(
 {\cal H}_{21}\rho_{13}(t)
+{\cal H}_{22}\rho_{23}(t)
+{\cal H}_{23}\rho_{33}(t)
+{\cal H}_{24}\rho_{43}(t)
-\rho_{21}(t){\cal H}_{13}
-\rho_{22}(t){\cal H}_{23}
-\rho_{23}(t){\cal H}_{33}
-\rho_{24}(t){\cal H}_{43}
\Bigl)
\nonumber\\
&=-i \Biggl(
 g_\downarrow^{*}(t)\rho_{13}(t)
+\left(\mu_{\uparrow}-\varepsilon_{d,\uparrow}\right)\rho_{23}(t)
+g_\uparrow^{\phantom*}(t)\rho_{43}(t)
-\rho_{21}(t)g_\uparrow^{\phantom*}(t)
-\rho_{23}(t)\left(\mu_{\downarrow}-\varepsilon_{d,\downarrow}\right)
-\rho_{24}(t)g_\downarrow^{*}(t)
\Biggl)
\nonumber\\
&=-i \Biggl(
 g_\downarrow^{*}(t)\rho_{13}(t)
+\left(2h-\varepsilon_{d,\uparrow}+\varepsilon_{d,\downarrow}\right)\rho_{23}(t)
+g_\uparrow^{\phantom*}(t)\rho_{34}^{*}(t)
- g_\uparrow^{\phantom*}(t) \rho_{12}^{*}(t)
- g_\downarrow^{*}(t) \rho_{24}(t)
\Biggl)
\end{align}
\begin{align}
\label{S:rho24}
\dot\rho_{24}(t)&=-i \Bigl(
 {\cal H}_{21}\rho_{14}(t)
+{\cal H}_{22}\rho_{24}(t)
+{\cal H}_{23}\rho_{34}(t)
+{\cal H}_{24}\rho_{44}(t)
-\rho_{21}(t){\cal H}_{14}
-\rho_{22}(t){\cal H}_{24}
-\rho_{23}(t){\cal H}_{34}
-\rho_{24}(t){\cal H}_{44}
\Bigl)
\nonumber\\
&=-i \Biggl(
 g_\downarrow^{*}(t)\rho_{14}(t)
+\left(\mu_{\uparrow}-\varepsilon_{d,\uparrow}\right)\rho_{24}(t)
+g_\uparrow^{\phantom*}(t)\rho_{44}(t)
-\rho_{22}(t)g_\uparrow^{\phantom*}(t)
-\rho_{23}(t)g_\downarrow^{\phantom*}(t)
\Biggl)
\\
\label{S:rho33}
\dot\rho_{33}(t)&=-i \Bigl(
 {\cal H}_{31}\rho_{13}(t)
+{\cal H}_{32}\rho_{23}(t)
+{\cal H}_{33}\rho_{33}(t)
+{\cal H}_{34}\rho_{43}(t)
-\rho_{31}(t){\cal H}_{13}
-\rho_{32}(t){\cal H}_{23}
-\rho_{33}(t){\cal H}_{33}
-\rho_{34}(t){\cal H}_{43}
\Bigl)
\nonumber\\
&=-i \Biggl(
 g_\uparrow^{*}(t)\rho_{13}(t)
+g_\downarrow^{\phantom*}(t)\rho_{43}(t)
-\rho_{31}(t)g_\uparrow^{\phantom*}(t)
-\rho_{34}(t)g_\downarrow^{*}(t)
\Biggl)
=2\Im\left[g_\uparrow^{*}(t)\rho_{13}(t)
- g_\downarrow^{*}(t) \rho_{34}(t) \right]
\\
\label{S:rho34}
\dot\rho_{34}(t)&=-i \Bigl(
 {\cal H}_{31}\rho_{14}(t)
+{\cal H}_{32}\rho_{24}(t)
+{\cal H}_{33}\rho_{34}(t)
+{\cal H}_{34}\rho_{44}(t)
-\rho_{31}(t){\cal H}_{14}
-\rho_{32}(t){\cal H}_{24}
-\rho_{33}(t){\cal H}_{34}
-\rho_{34}(t){\cal H}_{44}
\Bigl)
\nonumber\\
&=-i \Biggl(
 g_\uparrow^{*}(t)\rho_{14}(t)
+\left(\mu_{\downarrow}-\varepsilon_{d,\downarrow}\right)\rho_{34}(t)
+g_\downarrow^{\phantom*}(t)\rho_{44}(t)
-\rho_{32}(t)g_\uparrow^{\phantom*}(t)
-\rho_{33}(t)g_\downarrow^{\phantom*}(t)
\Biggl)
\\
\label{S:rho44}
\dot\rho_{44}(t)&=-i \Bigl(
 {\cal H}_{41}\rho_{14}(t)
+{\cal H}_{42}\rho_{24}(t)
+{\cal H}_{43}\rho_{34}(t)
+{\cal H}_{44}\rho_{44}(t)
-\rho_{41}(t){\cal H}_{14}
-\rho_{42}(t){\cal H}_{24}
-\rho_{43}(t){\cal H}_{34}
-\rho_{44}(t){\cal H}_{44}
\Bigl)
\nonumber\\
&=-i \Biggl(
g_\uparrow^{*}(t)\rho_{24}(t)
+g_\downarrow^{*}(t)\rho_{34}(t)
-\rho_{42}(t)g_\uparrow^{\phantom*}(t)
-\rho_{43}(t)g_\downarrow^{\phantom*}(t)
\Biggl)
=2\Im\left[g_\uparrow^{*}(t)\rho_{24}(t)
+g_\downarrow^{*}(t)\rho_{34}(t)\right]
\end{align}
\end{subequations}

Here we used the fact that $\hat\rho$ is Hermitian matrix and  $\rho_{ji}^{\phantom*}(t)=\rho_{ij}^*(t)$.

\section{Derivation of the expressions for the time evolution for chain}
\label{S:Derivation_for_time_evolution_Chain}

The equation of motion for the chain reads
\begin{subequations}
    \label{S:equation_of_motion}
    \begin{align}
    \label{S:equation_of_motion_C}
    &\frac{\dd}{\dd t}C_{ij,\sigma}(t) 
        = i \langle [\hat{\cal H}_{\chainLabelMFD,\sigma}, \hat C_{ij,\sigma}]\rangle(t)
\\
    \label{S:equation_of_motion_K}
    &\frac{\dd}{\dd t}K_{ij,\sigma}(t) 
        = i \langle [\hat{\cal H}_{\chainLabelMFD,\sigma}, \hat K_{ij,\sigma}]\rangle(t).
\end{align}
\end{subequations}
Here, $\hat C_{ij,\sigma}$ and $\hat K_{ij,\sigma}$ are given by the the following expressions 
\begin{subequations}
\label{S:Cij_and_Kij}
\begin{align}
    \label{S:Cij_correlator}
    &\hat C_{ij,\sigma} =\left\{
    \begin{array}{ccc}
            \hat c_{i,\sigma}^{\dagger} \hat{c}_{j,\sigma}^{\phantom\dagger} &
            \quad & 
            1 \leq i,j \leq L
        \\
            \hat\gamma_{\sigma}^{\phantom\dagger} \hat{c}_{j,\sigma}^{\phantom\dagger} &
            \quad & 
            i=0,\,1 \leq j \leq L
        \\
            \hat c_{i,\sigma}^{\dagger} \hat\gamma_{\sigma}^{\phantom\dagger} &
            \quad &
            1 \leq i \leq L,\,  j=0
        \\
        \hat{\mathbb{1}} &
        \quad & 
        i=j=0
    \end{array}
    \right.
\\
    \label{S:Cij_matrix}
    & C_{ij,\sigma} (t)
        = \langle \hat C_{ij,\sigma} \rangle (t)
\\
    \label{S:Kij_correlator}
    & \hat K_{ij,\sigma} =
\left. 
\begin{array}{ccc}
    \hat c_{i,\sigma}^{\phantom\dagger} \hat c_{j,\sigma}^{\phantom\dagger} & \quad & 1 \leq i , j \leq L
\end{array}
\right.
\\
\label{S:Kij_matrix}
&K_{ij,\sigma} (t) = \langle \hat K_{ij,\sigma} \rangle (t).
\end{align}
\end{subequations}



\begin{subequations}
\label{S:cumutator}
For Eq. \eqref{S:equation_of_motion_C} we have
\begin{align}
\label{S:cumutator_cdagger_c}
[\hat{\cal H}_{chain,\sigma}, \hat c_{i,\sigma}^{\dagger} \hat c_{j,\sigma}^{\phantom\dagger}] &=
\sum_{l=1}^L \Bigl( J_{l,\sigma} [\hat c_{l,\sigma}^{\dagger}\hat c_{l+1,\sigma}^{\phantom\dagger},\hat c_{i,\sigma}^{\dagger} \hat c_{j,\sigma}^{\phantom\dagger}]
+J_{l,\sigma}[\hat c_{l+1,\sigma}^{\dagger}\hat c_{l,\sigma}^{\phantom\dagger},\hat c_{i,\sigma}^{\dagger} \hat c_{j,\sigma}^{\phantom\dagger}]
+(\mu_{\sigma} -\varepsilon_l)[\hat n_{l,\sigma},\hat c_{i,\sigma}^{\dagger} \hat c_{j,\sigma}^{\phantom\dagger}] \Bigl)
\nonumber \\
&+f_\sigma^{\phantom*}(t) [\hat c_{r,\sigma}^{\dagger}\hat c_{0,\sigma}^{\phantom\dagger},\hat c_{i,\sigma}^{\dagger} \hat c_{j,\sigma}^{\phantom\dagger}]
+ f_\sigma^{*}(t) [\hat c_{0,\sigma}^{\dagger} \hat c_{r,\sigma}^{\phantom\dagger},\hat c_{i,\sigma}^{\dagger} \hat c_{j,\sigma}^{\phantom\dagger}]
\end{align}
and for Eq. \eqref{S:equation_of_motion_K} we obtain
\begin{align}
\label{S:cumutator_c_c}
[\hat{\cal H}_{chain,\sigma}, \hat c_{i,\sigma}^{\phantom\dagger} \hat c_{j,\sigma}^{\phantom\dagger}] &=
\sum_{l=1}^L \Bigl( J_{l,\sigma} [\hat c_{l,\sigma}^{\dagger}\hat c_{l+1,\sigma}^{\phantom\dagger},\hat c_{i,\sigma}^{\phantom\dagger} \hat c_{j,\sigma}^{\phantom\dagger}]
+J_{l,\sigma}[\hat c_{l+1,\sigma}^{\dagger}\hat c_{l,\sigma}^{\phantom\dagger},\hat c_{i,\sigma}^{\phantom\dagger} \hat c_{j,\sigma}^{\phantom\dagger}]
+(\mu_{\sigma} -\varepsilon_l)[\hat n_{l,\sigma},\hat c_{i,\sigma}^{\phantom\dagger} \hat c_{j,\sigma}^{\phantom\dagger}] \Bigl)
\nonumber \\
&+f_\sigma^{\phantom*}(t) [\hat c_{r,\sigma}^{\dagger}\hat c_{0,\sigma}^{\phantom\dagger},\hat c_{i,\sigma}^{\phantom\dagger} \hat c_{j,\sigma}^{\phantom\dagger}]
+ f_\sigma^{*}(t) [\hat c_{0,\sigma}^{\dagger} \hat c_{r,\sigma}^{\phantom\dagger},\hat c_{i,\sigma}^{\phantom\dagger} \hat c_{j,\sigma}^{\phantom\dagger}]
\end{align}
\end{subequations}
Here index $0$ indicates that we deal with Majorana fermions, i.e. $\hat c_{0,\sigma}^{\phantom\dagger} \equiv \gamma_{\sigma}$ and therefore $\hat c_{0,\sigma}^{\phantom\dagger}=\hat c_{0,\sigma}^{\dagger} $ and correspondingly $\{\hat c_{0,\sigma}^{\phantom\dagger}, \hat c_{i,\sigma'}^{\phantom\dagger}\} = 2\delta_{\sigma\sigma'} \delta_{0i}$. 
%
To calculate the commutation relations in Eqs. \eqref{S:cumutator}, we use the following formula
\begin{equation}
\label{S:cumutator_4operars}
[\hat A\hat B, \hat C \hat D]=A\{\hat B, \hat C\}\hat D
- \{\hat A, \hat C\}\hat B \hat D +\hat C \hat A \{\hat B,\hat D\}  - \hat C \{\hat A, \hat D\}\hat B \,.
\end{equation}
We obtain

\begin{subequations}
\label{S:comutator_tertm-to-term}
\begin{align}
[\hat c_{l,\sigma}^{\dagger}\hat c_{l+1,\sigma}^{\phantom\dagger},\hat c_{i,\sigma}^{\dagger} \hat c_{j,\sigma}^{\phantom\dagger}]
&=\hat c_{l,\sigma}^{\dagger}\{\hat c_{l+1,\sigma}^{\phantom\dagger},\hat c_{i,\sigma}^{\dagger}\} \hat c_{j,\sigma}^{\phantom\dagger}
- \hat c_{i,\sigma}^{\dagger} \{\hat c_{l,\sigma}^{\dagger}, \hat c_{j,\sigma}^{\phantom\dagger}\}\hat c_{l+1,\sigma}^{\phantom\dagger}
=\delta_{l,i-1}\hat c_{i-1,\sigma}^{\dagger} \hat c_{j,\sigma}^{\phantom\dagger}
- \delta_{l,j}\hat c_{i,\sigma}^{\dagger} \hat c_{j+1,\sigma}^{\phantom\dagger}
\\
[\hat c_{l+1,\sigma}^{\dagger}\hat c_{l,\sigma}^{\phantom\dagger},\hat c_{i,\sigma}^{\dagger} \hat c_{j,\sigma}^{\phantom\dagger}]
&=\hat c_{l+1,\sigma}^{\dagger}\{\hat c_{l,\sigma}^{\phantom\dagger},\hat c_{i,\sigma}^{\dagger}\} \hat c_{j,\sigma}^{\phantom\dagger}
- \hat c_{i,\sigma}^{\dagger}\{\hat c_{l+1,\sigma}^{\dagger},\hat c_{j,\sigma}^{\phantom\dagger}\}\hat c_{l,\sigma}^{\phantom\dagger}
=\delta_{l,i}\hat c_{i+1,\sigma}^{\dagger} \hat c_{j,\sigma}^{\phantom\dagger}
- \delta_{l,j-1}  \hat c_{i,\sigma}^{\dagger} \hat c_{j-1,\sigma}^{\phantom\dagger}
\\
[\hat n_{l,\sigma},\hat c_{i,\sigma}^{\dagger} \hat c_{j,\sigma}^{\phantom\dagger}]
&=\hat c_{l,\sigma}^{\dagger}\{\hat c_{l,\sigma}^{\phantom\dagger},\hat c_{i,\sigma}^{\dagger}\} \hat c_{j,\sigma}^{\phantom\dagger}
- \hat c_{i,\sigma}^{\dagger}\{\hat c_{l,\sigma}^{\dagger},\hat c_{j,\sigma}^{\phantom\dagger}\}\hat c_{l,\sigma}^{\phantom\dagger}
=\delta_{l,i}\hat c_{i,\sigma}^{\dagger} \hat c_{j,\sigma}^{\phantom\dagger}
- \delta_{l,j}  \hat c_{i,\sigma}^{\dagger} \hat c_{j,\sigma}^{\phantom\dagger}
=\left(\delta_{l,i}- \delta_{l,j}\right) \hat c_{i,\sigma}^{\dagger} \hat c_{j,\sigma}^{\phantom\dagger}
\\
[\hat c_{r,\sigma}^{\dagger}\hat c_{0,\sigma}^{\phantom\dagger},\hat c_{i,\sigma}^{\dagger} \hat c_{j,\sigma}^{\phantom\dagger}]
&=\hat c_{r,\sigma}^{\dagger}\{\hat c_{0,\sigma}^{\phantom\dagger},\hat c_{i,\sigma}^{\dagger}\} \hat c_{j,\sigma}^{\phantom\dagger}
- \hat c_{i,\sigma}^{\dagger}\{\hat c_{r,\sigma}^{\dagger},\hat c_{j,\sigma}^{\phantom\dagger}\}\hat c_{0,\sigma}^{\phantom\dagger}
+\hat c_{i,\sigma}^{\dagger}\hat c_{r,\sigma}^{\dagger} \{\hat c_{0,\sigma}^{\phantom\dagger}, \hat c_{j,\sigma}^{\phantom\dagger}\}
\nonumber
\\
&=2\delta_{0,i}\hat c_{r,\sigma}^{\dagger} \hat c_{j,\sigma}^{\phantom\dagger}
+2\delta_{0j} \hat c_{i,\sigma}^{\dagger}\hat c_{r,\sigma}^{\dagger}
- \delta_{r,j}  \hat c_{i,\sigma}^{\dagger} \hat c_{0,\sigma}^{\phantom\dagger}
\\
[\hat c_{0,\sigma}^{\dagger} \hat c_{r,\sigma}^{\phantom\dagger},\hat c_{i,\sigma}^{\dagger} \hat c_{j,\sigma}^{\phantom\dagger}]
&=\hat c_{0,\sigma}^{\dagger}\{\hat c_{r,\sigma}^{\phantom\dagger},\hat c_{i,\sigma}^{\dagger}\} \hat c_{j,\sigma}^{\phantom\dagger}
- \hat c_{i,\sigma}^{\dagger}\{\hat c_{0,\sigma}^{\dagger},\hat c_{j,\sigma}^{\phantom\dagger}\}\hat c_{r,\sigma}^{\phantom\dagger}
-\{\hat c_{0,\sigma}^{\dagger}, \hat c_{i,\sigma}^{\dagger}\}\hat c_{r,\sigma}^{\phantom\dagger}\hat c_{j,\sigma}^{\phantom\dagger}
\nonumber
\\
&=\delta_{r,i}\hat c_{0,\sigma}^{\dagger} \hat c_{j,\sigma}^{\phantom\dagger}
- 2\delta_{0,j}  \hat c_{i,\sigma}^{\dagger} \hat c_{r,\sigma}^{\phantom\dagger}
-2\delta_{0i}\hat c_{r,\sigma}^{\phantom\dagger}\hat c_{j,\sigma}^{\phantom\dagger}
\\
%
[\hat c_{l,\sigma}^{\dagger}\hat c_{l+1,\sigma}^{\phantom\dagger},\hat c_{i,\sigma}^{\phantom\dagger} \hat c_{j,\sigma}^{\phantom\dagger}]
&=
- \{\hat c_{l,\sigma}^{\dagger}, \hat c_{i,\sigma}^{\phantom\dagger}\} \hat c_{l+1,\sigma}^{\phantom\dagger} \hat c_{j,\sigma}^{\phantom\dagger}
- \hat c_{i,\sigma}^{\phantom\dagger}
\{\hat c_{l,\sigma}^{\dagger}, \hat c_{j,\sigma}^{\phantom\dagger}\}\hat c_{l+1,\sigma}^{\phantom\dagger}
=- \delta_{l,i}\hat c_{i+1,\sigma}^{\phantom\dagger} \hat c_{j,\sigma}^{\phantom\dagger}
- \delta_{l,j}\hat c_{i,\sigma}^{\phantom\dagger} \hat c_{j+1,\sigma}^{\phantom\dagger}
\\
[\hat c_{l+1,\sigma}^{\dagger}\hat c_{l,\sigma}^{\phantom\dagger},\hat c_{i,\sigma}^{\phantom\dagger} \hat c_{j,\sigma}^{\phantom\dagger}]
&=- \{\hat c_{l+1,\sigma}^{\dagger}, \hat c_{i,\sigma}^{\phantom\dagger}\} \hat c_{l,\sigma}^{\phantom\dagger} \hat c_{j,\sigma}^{\phantom\dagger}
- \hat c_{i,\sigma}^{\phantom\dagger}
\{\hat c_{l+1,\sigma}^{\dagger}, \hat c_{j,\sigma}^{\phantom\dagger}\}\hat c_{l,\sigma}^{\phantom\dagger}
=- \delta_{l,i-1}\hat c_{i-1,\sigma}^{\phantom\dagger} \hat c_{j,\sigma}^{\phantom\dagger}
- \delta_{l,j-1}\hat c_{i,\sigma}^{\phantom\dagger} \hat c_{j-1,\sigma}^{\phantom\dagger}
\\
[\hat n_{l,\sigma}^{\phantom\dagger},\hat c_{i,\sigma}^{\phantom\dagger} \hat c_{j,\sigma}^{\phantom\dagger}]
&= - \{\hat c_{l,\sigma}^{\dagger}, \hat c_{i,\sigma}^{\phantom\dagger}\} \hat c_{l,\sigma}^{\phantom\dagger} \hat c_{j,\sigma}^{\phantom\dagger}
- \hat c_{i,\sigma}^{\phantom\dagger}
\{\hat c_{l,\sigma}^{\dagger}, \hat c_{j,\sigma}^{\phantom\dagger}\}\hat c_{l,\sigma}^{\phantom\dagger}
=- \delta_{l,i}\hat c_{i,\sigma}^{\phantom\dagger} \hat c_{j,\sigma}^{\phantom\dagger}
- \delta_{l,j}\hat c_{i,\sigma}^{\phantom\dagger} \hat c_{j,\sigma}^{\phantom\dagger}
=-\left(\delta_{li}+\delta_{lj}\right) \hat c_{i,\sigma}^{\phantom\dagger} \hat c_{j,\sigma}^{\phantom\dagger}
\\
[\hat c_{r,\sigma}^{\dagger}\hat c_{0,\sigma}^{\phantom\dagger},\hat c_{i,\sigma}^{\phantom\dagger} \hat c_{j,\sigma}^{\phantom\dagger}]
&=\hat c_{r,\sigma}^{\dagger}\{\hat c_{0,\sigma}^{\phantom\dagger},\hat c_{i,\sigma}^{\phantom\dagger}\} \hat c_{j,\sigma}^{\phantom\dagger}
-\{\hat c_{r,\sigma}^{\dagger} , \hat c_{i,\sigma}^{\phantom\dagger}\} \hat c_{0,\sigma}^{\phantom\dagger}\hat c_{j,\sigma}^{\phantom\dagger}
+\hat c_{i,\sigma}^{\phantom\dagger}\hat c_{r,\sigma}^{\dagger} \{\hat c_{0,\sigma}^{\phantom\dagger}, \hat c_{j,\sigma}^{\phantom\dagger}\}
- \hat c_{i,\sigma}^{\phantom\dagger}\{\hat c_{r,\sigma}^{\dagger},\hat c_{j,\sigma}^{\phantom\dagger}\}\hat c_{0,\sigma}^{\phantom\dagger}
\nonumber
\\
&=2\delta_{0,i}\hat c_{r,\sigma}^{\dagger} \hat c_{j,\sigma}^{\phantom\dagger}
+ 2\delta_{0,j} \hat c_{i,\sigma}^{\phantom\dagger} \hat c_{r,\sigma}^{\dagger}
- \delta_{r,i}  \hat c_{0,\sigma}^{\phantom\dagger} \hat c_{j,\sigma}^{\phantom\dagger}
- \delta_{r,j}  \hat c_{i,\sigma}^{\phantom\dagger} \hat c_{0,\sigma}^{\phantom\dagger}
\\
[\hat c_{0,\sigma}^{\dagger}\hat c_{r,\sigma}^{\phantom\dagger},\hat c_{i,\sigma}^{\phantom\dagger} \hat c_{j,\sigma}^{\phantom\dagger}]
&=
-\{\hat c_{0,\sigma}^{\dagger} , \hat c_{i,\sigma}^{\phantom\dagger}\} \hat c_{r,\sigma}^{\phantom\dagger}\hat c_{j,\sigma}^{\phantom\dagger}
- \hat c_{i,\sigma}^{\phantom\dagger}\{\hat c_{0,\sigma}^{\dagger},\hat c_{j,\sigma}^{\phantom\dagger}\}\hat c_{r,\sigma}^{\phantom\dagger}
=-2\delta_{0i} \hat c_{r,\sigma}^{\phantom\dagger}\hat c_{j,\sigma}^{\phantom\dagger}
- \delta_{0j} \hat c_{i,\sigma}^{\phantom\dagger}\hat c_{r,\sigma}^{\phantom\dagger}
\end{align}
\end{subequations}

After implementing Eq.~\eqref{S:comutator_tertm-to-term} in Eq.~\eqref{S:cumutator}, based on equation of mouton \eqref{S:equation_of_motion} we obtain

\begin{subequations}
\label{S:Time_evolution_of_correlators_for_Chain}

\begin{itemize}
\item {\boldmath$i\neq 0, r$ , $j \neq 0, r$}
\begin{align}
\label{S:C_inot0r_jnot0r}
{\dot C}_{ij,\sigma} &=i\Bigl(J_{i-1,\sigma} C_{(i-1)j,\sigma}
+ J_{i,\sigma} C_{(i+1)j,\sigma} - J_{j-1,\sigma} C_{i(j-1),\sigma}
- J_{j,\sigma} C_{i(j+1),\sigma}
-\left(\varepsilon_{i,\sigma} - \varepsilon_{j,\sigma}\right)C_{ij,\sigma}\Bigl)
\\
\label{S:K_inot0r_jnot0r}
{\dot K}_{ij,\sigma} &=
-i\Bigl(J_{i-1,\sigma} K_{(i-1)j,\sigma}
+J_{i,\sigma} K_{(i+1)j,\sigma}
+J_{j-1,\sigma} K_{i(j-1),\sigma}
+J_{j,\sigma} K_{i(j+1),\sigma}
-\left(\varepsilon_{i,\sigma} + \varepsilon_{j,\sigma}-2\mu_{\sigma}\right)K_{ij,\sigma}
\Bigl)
\\
\label{S:K_inot0r}
K_{ii,\sigma} &=0
\end{align}

\item {\boldmath$i=r$ , $j \neq 0, r$}  and {\boldmath$i\neq 0, r$ , $j = r$}
\begin{align}
\label{S:C_ir_jnot0r}
{\dot C}_{rj,\sigma} &=i\Bigl(J_{r-1,\sigma} C_{(r-1)j,\sigma}
+ J_{r,\sigma} C_{(r+1)j,\sigma} - J_{j-1,\sigma} C_{r(j-1),\sigma}
- J_{j,\sigma} C_{r(j+1),\sigma}
-\left(\varepsilon_{r,\sigma} - \varepsilon_{j,\sigma}\right)C_{rj,\sigma}
+ f_\sigma^{*} (t) C_{0j,\sigma}
\Bigl)
\\
\label{S:C_inot0r_jr}
{\dot C}_{ir,\sigma} &=i\Bigl(J_{i-1,\sigma} C_{(i-1)r,\sigma}
+ J_{i,\sigma} C_{(i+1)r,\sigma} - J_{r-1,\sigma} C_{i(r-1),\sigma}
- J_{r,\sigma} C_{i(r+1),\sigma}
-\left(\varepsilon_{i,\sigma} - \varepsilon_{r,\sigma}\right)C_{ir,\sigma}
- f_\sigma^{\phantom*} (t) C_{i0,\sigma}
\Bigl)
\\
\label{S:K_ir_jnot0r}
{\dot K}_{rj,\sigma} &=-i\Bigl(
J_{r-1,\sigma} K_{(r-1)j,\sigma}
+J_{r,\sigma} K_{(r+1)j,\sigma}
+J_{j-1,\sigma} K_{r(j-1),\sigma}
+J_{j,\sigma} K_{r(j+1),\sigma}
-\left(\varepsilon_{r,\sigma} + \varepsilon_{j,\sigma}-2\mu_{\sigma}\right)K_{rj,\sigma}
\nonumber\\
&+ f_\sigma^{\phantom*} (t)K_{0j,\sigma}
\Bigl)
\\
\label{S:K_inot0r_jr}
{\dot K}_{ir,\sigma} &=-i\Bigl(
J_{i-1,\sigma} K_{(i-1)r,\sigma}
+J_{i,\sigma} K_{(i+1)r,\sigma}
+J_{r-1,\sigma} K_{i(r-1),\sigma}
+J_{r,\sigma} K_{i(r+1),\sigma}
-\left(\varepsilon_{i,\sigma} + \varepsilon_{r,\sigma}-2\mu_{\sigma}\right)K_{ir,\sigma}
\nonumber\\
&+ f_\sigma^{\phantom*} (t)K_{j0,\sigma}
\Bigl)
\end{align}

\item  {\boldmath$i=r$ , $j = r$}
\begin{align}
\label{S:C_ir_jr}
{\dot C}_{rr,\sigma} &=i\Bigl(J_{r-1,\sigma} C_{(r-1)r,\sigma}
+ J_{r,\sigma} C_{(r+1)r,\sigma} - J_{r-1,\sigma} C_{r(r-1),\sigma}
- J_{r,\sigma} C_{r(r+1),\sigma}
+ f_\sigma^{*} (t) C_{0r,\sigma}
- f_\sigma^{\phantom*} (t) C_{r0,\sigma}
\Bigl)
\\
\label{S:K_ir_jr}
K_{rr,\sigma} &=0
\end{align}

\item {\boldmath$i=0$ , $j \neq 0, r$} and {\boldmath$i\neq 0, r$ , $j=0 $}
\begin{align}
\label{S:C_i0_jnot0r}
{\dot C}_{0j,\sigma} &=i\left(- J_{j-1,\sigma} C_{0(j-1),\sigma}
- J_{j,\sigma} C_{0(j+1),\sigma} -(\mu_{\sigma}-\varepsilon_{j,\sigma}) C_{0j,\sigma}
+2 f_\sigma^{\phantom*} (t) C_{rj,\sigma}
-2 f_\sigma^{*} (t) K_{rj,\sigma}
\right)
\\
\label{S:C_inot0r_j0}
{\dot C}_{i0,\sigma} &=i\Bigl(J_{i-1,\sigma} C_{(i-1)0,\sigma}
+ J_{i,\sigma} C_{(i+1)0,\sigma} + (\mu_{\sigma}-\varepsilon_{i,\sigma}) C_{i0,\sigma}
-2 f_\sigma^{*} (t) C_{ir,\sigma}
+2 f_\sigma^{\phantom*} (t) K_{ir,\sigma}^{*}
\Bigl)
\end{align}

\item {\boldmath$i=0$ , $j= r$} and {\boldmath$i=r$ , $j= 0$}
\begin{align}
\label{S:C_i0_jr}
{\dot C}_{0r,\sigma} &=i\Bigl(- J_{r-1,\sigma} C_{0(r-1),\sigma}
- J_{r,\sigma} C_{0(r+1),\sigma} -(\mu_{\sigma}-\varepsilon_{r,\sigma}) C_{0r,\sigma}
+2 f_\sigma^{\phantom*} (t) C_{rr,\sigma}
-f_\sigma^{\phantom*} (t) C_{00,\sigma}
\Bigl)
\\
\label{S:C_ir_j0}
{\dot C}_{r0,\sigma} &=i\Bigl(J_{r-1,\sigma} C_{(r-1)0,\sigma}
+ J_{r,\sigma} C_{(r+1)0,\sigma} + (\mu_{\sigma}-\varepsilon_{r,\sigma}) C_{r0,\sigma}
-2 f_\sigma^{*} (t) C_{rr,\sigma}
+ f_\sigma^{\phantom*} (t) C_{00,\sigma}^{*}
\Bigl)
\end{align}

\item {\boldmath$i=0$ , $j= 0$}
\begin{align}
\label{S:C_i0_j0_der}
{\dot C}_{00,\sigma} &=i\Bigl(
2 f_\sigma^{\phantom*} (t) C_{r0,\sigma}
+2 f_\sigma^{\phantom*} (t) K_{r0,\sigma}^*
-2f_\sigma^{*} (t) C_{0r,\sigma}-2f_\sigma^{*} (t) K_{r0,\sigma}
\Bigl)
=2 if_\sigma^{\phantom*} (t)\left(C_{r0,\sigma} -
K_{0r,\sigma}^*\right) \nonumber\\
&- 2i f_\sigma^{*} (t)(\left(C_{0r,\sigma} -
C_{r0,\sigma}^*\right) =
2i f_\sigma^{\phantom*} (t)\left(C_{r0,\sigma} -
K_{0r,\sigma}^*\right)
- 2 if_\sigma^{*} (t)(\left(C_{0r,\sigma} -
C_{r0,\sigma}^*\right) =0
\\
\label{S:C_i0_j0}
C_{00,\sigma} &=1
\end{align}

\end{itemize}

\end{subequations}

For $x=1$ we have that $(x-1) \rightarrow L$ and for $x=L$ we have that $(x+1) \rightarrow 1$.  For PBC $J_{L,\sigma}$ is a finite value, while for OBC $J_{L,\sigma}=0$, therefore corresponding terms will be not there.

\section{\texorpdfstring{Detailed derivation for expressions for $a$ and $b$}{Detailed derivation for expressions for a and b}}
\label{S:Derivation_a_and_b}

Here we derive expressions for $a$ and $b$.

\begin{align*}
|\Psi_{\pm}\rangle &=|\Psi_{0,\pm}^{(0)} \rangle
- \frac{f V}{|J|} \sum_{l \neq 0}  \frac{A_{l0}^{\phantom*}+A_{0l}^{*}}
{\Delta \tilde E_{l0}}|\Psi_{l,\pm}^{(0)}\rangle
+\frac{f^2 V^2}{J^2} \Biggl[\sum_{l, l' \neq 0}
\frac{\left(A_{ll'}^{\phantom*}+A_{l'l}^{*}\right)\left(A_{l'0}^{\phantom*}+A_{0l'}^{*}\right)}
{\Delta \tilde E_{l0} \Delta \tilde E_{l'0}}
|\Psi_{l,\pm}^{(0)}\rangle
-\frac{1}{2}|\Psi_{0,\pm}^{(0)}\rangle \sum_{l \neq 0}
\frac{|A_{l0}^{\phantom*}+A_{0l}^{*}|^2}{\Delta \tilde E_{l0}^2}
\Biggl]
\end{align*}
Here
\begin{align*}
A_{l_1l_2} = \langle \Psi_{l_1,\pm}^{(0)}|\gamma c_r |\Psi_{l_2,\pm}^{(0)}\rangle
\end{align*}

We obtain
\begin{align*}
g& = \frac{\langle \Psi_{\pm}| \gamma c_r |\Psi_{\pm}\rangle}{\langle \Psi_{\pm}|\Psi_{\pm}\rangle}
=-\frac{f V}{|J|}\sum_{l\neq 0}\Biggl(\frac{A_{l0}^{\phantom*}+A_{0l}^{*}}{\Delta \tilde E_{l0}}A_{0l}^{\phantom*}
+\frac{A_{l0}^{*}+A_{0l}^{\phantom*}}{\Delta \tilde E_{l0}}A_{l0}^{\phantom*}
\Biggl)
+\frac{f^2 V^2}{J^2}\sum_{l_1,l_2 \neq 0}\Biggl(
\frac{A_{l_1 0}^{\phantom*}+A_{0l_1}^{*}}{\Delta \tilde E_{l_1 0}}
\frac{A_{l_2 0}^{*}+A_{0l_2}^{\phantom*}}{\Delta \tilde E_{l_2 0}}A_{l_1l2}^{\phantom*}
\\
&+\frac{\left(A_{l_1l_2}^{\phantom*}+A_{l_2l_1}^{*}\right)\left(A_{l_20}^{\phantom*}+A_{0l_2}^{*}\right)}
{\Delta \tilde E_{l_10} \Delta \tilde E_{l_20}}A_{0 l_1}^{\phantom*}
+\frac{\left(A_{l_1l_2}^{*}+A_{l_2l_1}^{\phantom*}\right)\left(A_{l_20}^{*}+A_{0l_2}^{\phantom*}\right)}
{\Delta \tilde E_{l_10} \Delta \tilde E_{l_20}}A_{l_1 0}^{\phantom*}
\Biggl)
+ {\cal O}(f^3)
= -\frac{a V}{|J|}f + \frac{b V^2}{J^2}f^2 + {\cal O}(f^3)
\end{align*}
So we obtain
\begin{align*}
a &= \sum_{l\neq 0}\Biggl(\frac{A_{l0}^{\phantom*}+A_{0l}^{*}}{\Delta \tilde E_{l0}}A_{0l}^{\phantom*}
+\frac{A_{l0}^{*}+A_{0l}^{\phantom*}}{\Delta \tilde E_{l0}}A_{l0}^{\phantom*}\Biggl)
=\sum_{l\neq 0}\frac{A_{l0}^{\phantom*}A_{0l}^{\phantom*} + |A_{0l}^{\phantom*}|^2
+|A_{l0}^{\phantom*}|^2+A_{0l}^{\phantom*}A_{l0}^{\phantom*}
}{\Delta \tilde E_{l0}}
=\sum_{l\neq 0}\frac{|A_{0l}^{\phantom*}|^2+|A_{l0}^{\phantom*}|^2}{\Delta \tilde E_{l0}}
\end{align*}
Here we use the fact that $A_{l_1 l_2}^{\phantom*}A_{l_2 l_1}^{\phantom*}=0$, because the condition that $A_{l_1 l_2}^{\phantom*}\neq 0$ requires that state $|\Psi_{l_2,\pm}^{(0)}\rangle$ have one more Dirac fermion than state  $|\Psi_{l_1,\pm}^{(0)}\rangle$, while  the condition that $A_{l_2 l_1}^{\phantom*} \neq 0$ requires that state $|\Psi_{l_1,\pm}^{(0)}\rangle$ have one more Dirac fermion than state  $|\Psi_{l_2,\pm}^{(0)}\rangle$. Obviously, both conditions can not be fulfilled.

\begin{align*}
b&=\sum_{l_1,l_2 \neq 0}\Biggl(
\frac{A_{l_1 0}^{\phantom*}+A_{0l_1}^{*}}{\Delta \tilde E_{l_1 0}}
\frac{A_{l_2 0}^{*}+A_{0l_2}^{\phantom*}}{\Delta \tilde E_{l_2 0}}A_{l_1l2}^{\phantom*}
+\frac{\left(A_{l_1l_2}^{\phantom*}+A_{l_2l_1}^{*}\right)\left(A_{l_20}^{\phantom*}+A_{0l_2}^{*}\right)}
{\Delta \tilde E_{l_10} \Delta \tilde E_{l_20}}A_{0 l_1}^{\phantom*}
+\frac{\left(A_{l_1l_2}^{*}+A_{l_2l_1}^{\phantom*}\right)\left(A_{l_20}^{*}+A_{0l_2}^{\phantom*}\right)}
{\Delta \tilde E_{l_10} \Delta \tilde E_{l_20}}A_{l_1 0}^{\phantom*}
\Biggl)
\\
&=\sum_{l_1,l_2 \neq 0}\Biggl(\frac{
\left(A_{l_1 0}^{\phantom*} + A_{0l_1}^{\phantom*}\right)
\left(A_{l_2 0}^{\phantom*} + A_{0l_2}^{\phantom*}\right)}
{\Delta \tilde E_{l_10} \Delta \tilde E_{l_20}}
A_{l_1l_2}^{\phantom*}
+\frac{\left(A_{l_1l_2}^{\phantom*}+A_{l_2l_1}^{\phantom*}\right)\left(A_{l_20}^{\phantom*}+A_{0l_2}^{\phantom*}\right)\left(A_{l_1 0}^{\phantom*} + A_{0 l_1}^{\phantom*}\right)}
{\Delta \tilde E_{l_10} \Delta \tilde E_{l_20}}
\Biggl)
\\
&=\sum_{l_1,l_2 \neq 0}\Biggl(\frac{
\left(A_{l_1 0}^{\phantom*} + A_{0l_1}^{\phantom*}\right)
\left(A_{l_2 0}^{\phantom*} + A_{0l_2}^{\phantom*}\right)}
{\Delta \tilde E_{l_10} \Delta \tilde E_{l_20}}
A_{l_1l_2}^{\phantom*}
+\frac{\left(A_{l_20}^{\phantom*}+A_{0l_2}^{\phantom*}\right)\left(A_{l_1 0}^{\phantom*} + A_{0 l_1}^{\phantom*}\right)}
{\Delta \tilde E_{l_10} \Delta \tilde E_{l_20}} A_{l_1l_2}^{\phantom*}
+\frac{\left(A_{l_20}^{\phantom*}+A_{0l_2}^{\phantom*}\right)\left(A_{l_1 0}^{\phantom*} + A_{0 l_1}^{\phantom*}\right)}
{\Delta \tilde E_{l_10} \Delta \tilde E_{l_20}} A_{l_2l_1}^{\phantom*}
\Biggl)
\\
&=3\sum_{l_1,l_2 \neq 0}\frac{
\left(A_{l_1 0}^{\phantom*} + A_{0l_1}^{\phantom*}\right)
\left(A_{l_2 0}^{\phantom*} + A_{0l_2}^{\phantom*}\right)}
{\Delta \tilde E_{l_10} \Delta \tilde E_{l_20}}A_{l_1l_2}^{\phantom*}
\end{align*}
Here we use the fact that by construction $A_{l_1l_2}^{\phantom*}$ are real numbers.

\end{document}